\documentclass[a4paper,11pt]{article}
\usepackage{jheppub} 
\usepackage{lineno}
\newcommand{\dd}{\mathrm{d}}
\newcommand{\bfp}{\mathbf{p}}
\newcommand{\bfP}{\mathbf{P}}
\newcommand{\bfq}{\mathbf{q}}
\newcommand{\bfk}{\mathbf{k}}

\newcommand{\bfw}{\mathbf{w}}
\newcommand{\bft}{\vec{\theta}}
\newcommand{\bfQ}{\mathbf{Q}}
\usepackage{subfigure}
\usepackage{titlesec}
\titlespacing*{\section}{0pt}{*1}{0.5em}
\title{\boldmath Renormalization Group Evolution for In-medium Energy Correlators}

\author{Weiyao Ke$^{a,b}$, }
\author{Bianka Me\c caj$^b$, }
\author{Ivan Vitev$^b$}
\affiliation{$^a$Key Laboratory of Quark and Lepton Physics (MOE) \& Institute of Particle Physics, Central\\
China Normal University, Wuhan 430079, China}
\affiliation{$^b$Theoretical Division T-2, Los Alamos National Laboratory, Los Alamos, NM 87545, USA}

\emailAdd{weiyaoke@ccnu.edu.cn, bmecaj@lanl.gov, ivitev@lanl.gov}

\abstract{
We present a first-principles analysis of the renormalization group (RG) evolution of the two-point energy-energy correlator (EEC) in light-quark and gluon jets propagating through nuclear matter. Our work focuses on the analytic structure of the RG equations in the thin-medium regime, highlighting how collinear emissions in the presence of a dense QCD medium reshape the EEC observables. We work in the opacity expansion of the SCET$_{\rm G}$ formalism, where the propagating quarks and gluons interact with the medium via Glauber gluon exchanges.  We compute the corresponding one-loop jet functions using the medium-induced splitting kernels at first order in opacity and perform  resummation at leading logarithmic (LL) order. In particular, we identify an experimentally accessible regime of jet energies and EEC angles where one can directly investigate the medium-induced scale evolution and extract the corresponding opacity-one correction to the anomalous dimensions. Furthermore, we demonstrate analytically, using the method of regions, the Coulomb-logarithmic enhancement regulated by plasma screening for EEC. We compare our theoretical predictions with experimental data in $p$-Pb collisions and make projections for O-O collisions to test whether energy correlators could serve as sensitive probe of the quark-gluon plasma (QGP) dynamics in small collision systems, offering a robust and model-independent avenue for constraining jet evolution in QCD matter.}     
\begin{document}

\noindent\makebox[\dimexpr\linewidth-1cm\relax][r]{LA-UR-25-24936}

\maketitle
\flushbottom

\section{Introduction}
Hadronic jets~\cite{Feynman:1978dt,UA2:1982nvv,Ellis:2007ib}, abundantly produced in high energy collisions at particle colliders, naturally offer an excellent opportunity to study the microscopic properties of Quantum Chromodynamics (QCD) in different medium environments~\cite{Vitev:2008rz}. Much of the interplay of the perturbative radiation, hadronization, and possible modifications from nuclear matter is encoded in the substructure of the jet~\cite{Seymour:1997kj,ZEUS:2004gcp,Ellis:2009me}. 
For instance, when high energy jets travel through Quark-Gluon Plasma (QGP)~\cite{Gyulassy:2004zy} they interact with the medium through multiple parton scatterings, leading to loss of energy and modifications in hadron~\cite{Wang:1992qdg} and jet~\cite{Vitev:2008rz} cross-sections, the jet's internal structure~\cite{Vitev:2009rd}, as well as collective response of the medium~\cite{Cao:2020wlm}. These phenomena, collectively referred to as jet quenching, were unambiguously experimentally established in heavy ion collisions~\cite{PHENIX:2004vcz,CMS:2011iwn} and one of the central goals of heavy ion phenomenology is to further understand such modifications in jet substructure from first principles in Quantum Field Theory. Therefore, defining jet substructure observables that are robust and enable detailed studies of how medium induced radiation and energy loss affect the internal dynamics of jets is essential.  

Among the broad class of jet substructure observables, the so-called energy-energy correlator (EEC) has emerged as a particularly clean probe~\cite{Chen:2020vvp}. Historically it was first introduced in the context of $e^+e^-$ annihilation as a global event shape, where it provided some of the earliest QCD precision tests~\cite{Basham:1978bw,Brown:1981jv}. These early theoretical studies established the calculability of the EEC in perturbation theory. More recently, they were shown to be sensitive to transverse momentum dependent parton distribution and fragmentation functions in deep inelastic scattering at the Electron-Ion Collider (EIC)~\cite{Li:2020bub,Li:2021txc}. The EEC for jet substructure is defined as the angular correlation of the energy flow restricted inside the jet and is sensitive to both vacuum and medium-induced radiation patterns. Renewed interest in the EECs has led to detailed analyses of the observable in the collinear limit (kinematic regime suitable for jet substructure), where factorization theorems and resummation techniques beyond leading logarithmic accuracy have been established for both light \cite{Dixon:2019uzg, Lee:2022uwt} and heavy flavor jets \cite{Craft:2022kdo, Holguin:2022epo}, yielding quantitative insight into mass effects and the dynamics of QCD radiation. Moreover, these theoretical advances have directly enabled a broad program of EEC measurements across multiple experiments \cite{CMS:2024mlf,ATLAS:2023tgo, ALICE:2024dfl, STAR:2025jut}, motivating further their role as powerful probes of QCD.


Much of the recent focus on the EECs as a jet substructure observable stems from their well-defined theoretical properties in the small-angle limit, yet at angles still large enough that non-perturbative effects remain subdominant. This is an infrared and collinear safe quantity that admits an operator language definition in terms of correlation functions of the energy flow operator \cite{Hofman:2008ar}              
\begin{align}
\label{EEC_operator}
EEC=\langle \Psi | \varepsilon (\vec{n}_1)\varepsilon (\vec{n}_2)|\Psi\rangle \,,
\end{align}
where $|\Psi\rangle$ describes the jet state where the EEC is measured and the $\varepsilon (\vec{n})$ is the energy flow operator defined in terms of the energy momentum tensor \cite{Sveshnikov:1995vi,Tkachov:1995kk}
\begin{align}
    \varepsilon(\vec{n})=\lim_{r\rightarrow \infty} \int_0^{\infty} dt \,  r^2 n^i T_{0i}(t,r\vec{{n}})\,.
\end{align}
This definition from first principles in field theory makes EEC a powerful tool for precision QCD measurements as it ensures calculability and predictability at higher orders in perturbative calculations. We want to emphasize here that, while in this work we focus on the two point energy-energy correlator defined in Eq.~(\ref{EEC_operator}), this approach can be extended to higher $n$-point correlators $\langle \Psi | \varepsilon (\vec{n}_1)\varepsilon(\vec{n_2})\cdots\varepsilon(\vec{n}_n)|\Psi\rangle$, which will correspond to $n$-insertions of the energy flow operator.                     


While the EEC has been extensively studied in proton–proton collisions, its behavior in hot and cold nuclear media has only recently begun to attract attention. Existing approaches typically rely on Monte-Carlo simulations or coupled Monte-Carlo--hydrodynamic frameworks to describe energy correlators in nuclear collisions~\cite{Yang:2023dwc,Barata:2023bhh,Bossi:2024qho,Xing:2024yrb,Chen:2024quk} or electron-ion collisions~\cite{Devereaux:2023vjz}. Semi-analytic studies usually include the fixed-order calculations of medium-induced corrections, either neglecting or incorporating the jet energy-loss effects ~\cite{PhysRevLett.130.262301,Andres:2023xwr,Barata:2023bhh,Andres:2024ksi,Barata:2024wsu,Fu:2024pic}. Other frameworks employ the twist expansions to analyze the scaling behavior of different components of medium corrections~\cite{Andres:2024xvk,Barata:2025fzd}, the application of Soft-Collinear Effective Theory (SCET) to study the factorization structure of EEC in the medium~\cite{Singh:2024vwb}, or using the transverse-momentum-dependent (TMD) framework with its in-medium non-perturbative corrections to study the collinear limit of EEC~\cite{Barata:2024wsu}.
Ref.~\cite{Barata:2023bhh} also starts to consider the medium correction to the EEC anomalous dimensions by implementing the phase-space veto to the vacuum calculation. 
Recent studies are also extending the analysis of EEC towards $N$-point correlators (ENC) in the medium~\cite{Bossi:2024qho,Barata:2025fzd}.
Currently, calculations or simulations with plasma effects in hadronic/nuclear colliders mostly often rely on the time-evolution picture using a transport equation~\cite{Yang:2023dwc,Bossi:2024qho,Xing:2024yrb}. It is therefore imperative and timely to investigate how medium effects modify the EEC using analytic methods from the perspective of a modified QCD scale evolution, which is grounded in first-principles theory. Furthermore, such approach might be more suitable for the case of a small and moderate opacities encountered in heavy-ion collisions, for example in $p$-Pb collisions~\cite{Liang-Gilman:2025gjl} and very recent light-ion systems (O-O, Ne-Ne) where a clear suppression in nuclear modification factor is observed~\cite{2969907}.

With this in mind, we present a first-principles analysis of the EEC observable in heavy-ion collisions using the framework of SCET extended to include Glauber gluon interactions. Glauber gluons represent off-shell transverse momentum exchanges between hard partons and the color charges in the QGP and provide a controlled way to describe screened-Coulomb-like scatterings between the jet and the medium. These momentum exchanges are responsible for transverse kicks that drive medium-induced broadening and radiation. To describe these interactions, we adopt the Glauber gluon formalism that couples a background medium and SCET degrees of freedoms developed in Ref.~\cite{Ovanesyan:2011xy}. In particular, we use the splitting functions derived at first order in opacity~\cite{Ovanesyan:2011kn,Ke:2024ytw} to compute the medium-modified quark and gluon jet functions that enter our factorized description of the in-medium EEC\footnote{In-medium splitting kernels have been obtained to higher orders in opacity, but show qualitatively similar behavior~\cite{Fickinger:2013xwa,Sievert:2019cwq}.}. This approach provides a rigorous and consistent framework for describing fixed order effects and QCD evolution in the presence of a strongly interacting medium~\cite{Kang:2017frl,Li:2017wwc,Li:2020rqj,Ke:2023ixa}. 

Starting with a factorization tailored for the EEC in heavy-ion collisions, we derive the renormalization group (RG) equation that governs its scale evolution. We begin by computing the medium-modified jet functions for both quark- and gluon-initiated jets within this framework. These jet functions capture the leading modifications to collinear radiation to first order in the opacity of the QGP and form a key ingredient in the resummation of large logarithms. We then carry out leading-logarithmic (LL) resummation and compare our theoretical predictions with experimental data from heavy-ion collisions. The results demonstrate that the EEC is a sensitive probe of medium-induced modifications, with the potential to discriminate between different underlying dynamics of parton energy loss and transport. By combining a rigorous effective field theory framework with phenomenologically relevant observables, this work provides a model-independent baseline for interpreting jet substructure measurements in heavy-ion collisions and opens the door to more precise constraints on QGP properties. 

These results are important for at least three reasons. First, in $e^+e^-$ and proton--proton collisions, the anomalous dimensions of the EEC have proven to be a robust feature, enabling the extraction of fundamental quantities such as the strong coupling constant. It is, therefore, natural to ask whether medium effects also induce corrections to the anomalous dimension, allowing the strength of jet--medium interactions to be extracted in a similarly robust manner. Second, the identification of a regime in which medium effects manifest solely as corrections to the anomalous dimension is of intrinsic interest, as it would indicate that the medium directly modifies the scale evolution of jets. This would provide a particularly transparent test of the picture of in-medium parton shower dynamics. Third, such a mechanism may also help explain the unexpectedly strong $p_T$ dependence observed in $p$--Pb data~\cite{Liang-Gilman:2025gjl}, thereby offering a new perspective on experimental observations. In the future, the in-medium EEC approach developed here can be extended to heavy flavor~\cite{Andronic:2015wma,Boer:2024ylx,Xing:2024yrb} for example including recent insights into quarkonium production~\cite{Copeland:2023wbu,Copeland:2025osx,Copeland:2025vop}.

The remainder of the paper is organized as follows. In Section~\ref{sec:eecvac}, we define the EEC observable and give key fixed order and resummed results for its evaluation in elementary collisions. Section~\ref{sec:eechi} discussed the EEC in the context of scales that emerge in QCD matter, introduces the SCET framework with Glauber gluons, and outlines the relevant factorization approach. We further give results for the collisional and radiative processes that are employed in the calculation of this observable in reactions with nuclei. Sections~\ref{sec:medium:exlcusive} and~\ref{sec:Medium:resummation}, respectively, are devoted to the evaluation of the medium-induced exclusive and semi-inclusive jet functions for the EEC. We further discuss renormalization and evolution in matters in Section~\ref{sec:Medium:resummation}, including the medium-induced contribution to  anomalous dimensions.  Comparison between theoretical results and $p$-Pb data from the Large Hadron Collider (LHC), as well as predictions for small collision systems are shown in Section~\ref{sec:pheno}. We conclude in Section~\ref{sec:conclude} with a discussion of implications and directions for future work. Some of the important technical derivations and results are placed in the Appendix for completeness.

\section{Effective Field Theory Framework for Energy Correlators}
\label{sec:eecvac}
The energy-energy correlator is a jet substructure observable that probes the angular distribution of energy flow within a final state. In this work, we consider the EEC observable defined within reconstructed jets in heavy-ion collisions. In the context of collider processes, the operator definition in Eq.~(\ref{EEC_operator}) can be reformulated as a weighted cross-section by the energy of particles inside the jet \cite{Belitsky:2013xxa}. Given a jet with transverse momentum $p_{T,\text{jet}}$, the EEC is constructed by summing over all pairs of constituents $i$ and $j$ within the jet 
\begin{equation}
\label{eq:EEC-cross-section}
\frac{1}{\sigma_{\text{jet}}}\,
\frac{d\Sigma}{d\theta\, dp_T\, dy}(p_T,y;R)
= \frac{1}{\sigma_{\text{jet}}}
\sum_{i,j \in \text{jet}}
\int d\sigma_{\text{jet}}\!\big(p_T,y;R;\{p_k\}\big)\,
\frac{p_{T,i} p_{T,j}}{p_{T,\text{jet}}^{2}}\,
\delta(\cos\vec{\theta} - \cos\vec{\theta}_{ij}) \, .
\end{equation}
Here, $d\sigma_{\text{jet}} \equiv \tfrac{d\sigma_{\text{jet}}}{dp_T\,dy}(p_T,y;R)$ 
denotes the inclusive jet cross-section differential in the jet transverse momentum $p_T$ 
and rapidity $y$ for a jet of radius $R$. The quantity 
$d\sigma_{\text{jet}}(p_T,y;R;\{p_k\})$ represents the differential jet cross-section 
further resolved in the constituent momenta $\{p_k\}$. The numerator $d\Sigma$ is the 
energy–energy weighted cross-section, obtained by summing over all pairs of jet 
constituents $i,j$, with weights $p_{T,i},\,  p_{T,j}$ and normalization  $p_{T,\text{jet}}^2$
and an angular delta function that fixes their opening angle $\theta_{ij}$. This normalization ensures that the EEC is 
defined as a probability distribution in $\cos\theta$ for a jet with given kinematics. 

To analyze the energy distribution within a jet using the EEC, we focus on the limit where the angular separation between pairs is small such that $\cos\theta_{ij} \approx 1$, but the characteristic scale $p_{T,\rm jet}\theta$ remains within the perturbative domain.     
In this regime, the EEC becomes sensitive to the structure of collinear radiation and the weighted cross-section in Eq.~(\ref{eq:EEC-cross-section}) admits a factorization within the SCET framework. For simplicity in notation we will refer to the $p_{T,\rm jet}$ as simply $p_T$ hereinafter.

\subsection{Energy Correlators for Semi-Inclusive Jets in the Vacuum}
Energy–energy correlators in reactions with nuclei build upon the baseline established in elementary collisions, which we discuss first. In proton–proton collisions, the factorization of the energy-weighted cross-section $\Sigma$ in the collinear limit takes the form
\begin{eqnarray}
\label{eq:EECfactorized-vac}
\frac{d\Sigma}{d\theta\, dp_T\, dy}
&=&
\sum_{a,b,c,X}\! \int\! dx_a\,dx_b\;
f_{a/A}(x_a,\mu)\,f_{b/B}(x_b,\mu)\; \nonumber  \\
&& \times  \int\! \frac{dz_J}{z_J}\; \mathcal{H}_{ab\to cX}\!\left(\tfrac{p_T}{z_J},\,y;\mu\right)\mathcal{J}_{\rm EEC,c}^{\rm vac}(\theta;\,z_J,\,p_T, R,\mu) \,.
\end{eqnarray}
The differential cross-section is expressed in a factorized form as the convolution of parton distribution functions $f_{a/A}(x_a,\mu)$ and $f_{b/B}(x_b,\mu)$, which govern the initial-state partonic composition, with the perturbative hard-scattering cross-section $\mathcal{H}_{ab\to cX}$ for the production of parton $c$, and the semi-inclusive energy correlator jet function $\mathcal{J}_{\rm EEC,c}^{\rm vac}(\theta;\,z_J,\,p_T, R,\mu)$. This jet function encapsulates the subsequent fragmentation of parton $c$ into a jet carrying a momentum fraction $z_J$, while simultaneously encoding the probability distribution for the energy-energy correlator observable measured at a characteristic angular separation $\theta$ within the jet.

In the collinear limit, the result for the semi-inclusive jet function in the vacuum up to next-to-leading order (NLO) in $\alpha_s$, expanded in $\theta/R$, contains the following contributions (we use a quark jet for definitiveness)
\begin{align}
\label{eq:NLO-vac-EEC}
\mathcal{J}_{\rm EEC,q}^{\rm vac}(\theta;\,z_J,\,p_T,\,R,\,\mu) = &\delta(1-z_J)\mathcal{J}_{\rm EEC,q}^{\rm vac,(0)}(\theta; p_T,\,R,\,\mu) \nonumber\\
&+ \delta(1-z_J)\mathcal{J}_{\rm EEC, q\rightarrow qg}^{\rm vac, (1)}(\theta; p_T,\,R,\,\mu) \nonumber\\
& + \mathcal{J}_{\rm EEC, q\rightarrow q(g)}^{\rm vac, (1)}(\theta; \,z_J,\,p_T,\,R,\,\mu) + \mathcal{J}_{\rm EEC,{q\rightarrow (q)g}}^{\rm vac, (1)}(\theta;\,z_J,\,p_T,\,R,\,\mu)\,.
\end{align}
The first line on the right hand side of the decomposition is the leading order (LO) result, where $\mathcal{J}_{\rm EEC, q}^{\rm vac, (0)}(\theta; p_T, R, \mu)$, without the $z_J$ dependence, is the LO vacuum exclusive jet function, This is the energy weighted probability to find two energy flows separated by angle $\theta$ for a quark initiated jet at the fixed transverse momentum $p_T$. Here, $\delta(1-z_J)$ is the LO matching coefficients from exclusive to semi-inclusive jet function.
In the absence of a perturbative splitting, the only mechanism to produce two energy flows separated by an angle $\theta$ is through non-perturbative physics at a characteristic scale $\Lambda_{\rm QCD}$. 

\begin{figure}
    \centering 
    \includegraphics[width=0.75\linewidth]{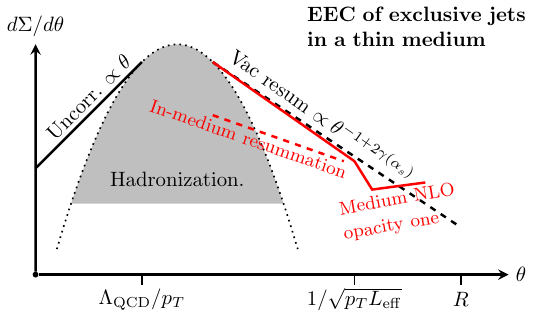}
    \caption{An illustration of the EEC of exclusive jets in the vacuum and after it is modified in a thin medium. In the asymptotic region, the EEC in the vacuum takes a power-law form with the time-like anomalous dimension $\gamma(\alpha_s) = \sum_{n=0} \left(\frac{\alpha_s}{4\pi}\right)^{n+1}\gamma_T^{{\rm vac}, (n)}$. It is also known that medium correction significantly
    modify the large-angle behavior of EEC, as shown by the red solid line. In this paper, we identify an important medium-induced resummation effect in the region $\Lambda_{\rm QCD}/p_T \ll \theta\ll \sqrt{8\pi/p_T L_{\rm eff}}$, resulting in a medium correction $\Delta\gamma(\alpha_s)$ to the vacuum $\gamma(\alpha_s)$. This is illustrated by the red dashed line.}
    \label{fig:illustration}
\end{figure}

This scale is responsible for the transition near the hadronization peak at $\theta \sim \Lambda_{\rm QCD}/p_T$, as illustrated in Figure~\ref{fig:illustration}. When $p_T\theta$ is large, this non-perturbative contribution is expected to be power-suppressed as $[\Lambda_{\rm QCD}/( p_T\theta)]^n$. For instance, a renormalon analysis in Ref.~\cite{Schindler:2023cww} finds $n=1$. Consequently, the importance of this term rapidly decreases at larger angles within the collinear limit, rendering it subleading compared to the NLO terms in the power expansion.

The second and third lines of Eq.~(\ref{eq:NLO-vac-EEC}) are NLO corrections. $\mathcal{J}_{\rm EEC, q\rightarrow qg}^{\rm vac, (1)}(\theta;\,p_T,\,R,\,\mu)$ is the NLO exclusive EEC jet function with both partons from the perturbative splitting inside the cone. The jet function $\mathcal{J}_{\rm EEC, i\rightarrow j(k)}^{\rm vac, (1)}(\theta; \,z_J,\,p_T,\,R,\,\mu)$ is the NLO semi-inclusive EEC jet function, where parton $j$ forms the jet and the parenthesized parton $(k)$ emitted outside of the cone. 

\subsubsection{Exclusive Energy-Energy Correlator Jet Function}
For $p_T\theta  \gg \Lambda_{\rm QCD}$, the NLO correction to the exclusive EEC jet function, originating from a perturbative splitting, decomposes into contact and non-contact components
\begin{align}
\mathcal{J}_{\rm EEC, q\rightarrow qg}^{\rm vac, (1)}(\theta;\,p_T,\,R,\,\mu) = \mathcal{J}_{\rm EEC,q\rightarrow qg}^{\rm c, vac, (1)}(\theta;\,p_T,\,R,\,\mu)
+ \mathcal{J}_{\rm EEC, q\rightarrow qg}^{\rm nc, vac, (1)}(\theta;\,p_T,\,R,\,\mu)\, .
\end{align}
The non-contact component is defined by diagrams where the two energy flow operators act on different daughter partons, in contrast to the contact component, where they act on the same one. We note that the distinction between contact and non-contact contributions disappears when $p_T\theta \sim \Lambda_{\rm QCD}$. 
Following this definition, the non-contact term reads
\begin{small}
\begin{align}
\label{eq:ncEEC_vac}
 \mathcal{J}_{\rm EEC, q\rightarrow qg}^{\rm nc, vac, (1)}(\theta; p_T, R, \mu) = &
  \frac{\alpha_s(\mu^2) C_F}{2\pi^2}\frac{\mu^{2\epsilon}e^{\epsilon\gamma_E}}{(4\pi)^{\epsilon} (2\pi)^{-2\epsilon}} \int_0^1 dx x(1-x)P_{qq, \epsilon}(x) \nonumber\\
 & \times \int\frac{d^{2-2\epsilon}\bfk}{\bfk^2}\delta\left(\theta^2-\frac{\bfk^2}{x^2(1-x)^2p_T^2}\right)\Theta_{\textrm{anti-}k_T}^{<R} \left[1+ \mathcal{O}\left(\frac{\Lambda^2}{ p_T^2\theta^2}\right)\right] \nonumber\\
 = &\frac{\alpha_s(\mu^2)}{2\pi}\left\{\frac{\Theta(R-\theta)}{[\theta^2]_+} \frac{3}{4}C_F -\delta(\theta^2)\left[\frac{3}{4}C_F\left(\frac{1}{\epsilon}+\ln\frac{\mu^2}{p_T^2\theta_{\rm max}^2}\right)+\frac{37}{12}\right]\right\}\,,
\end{align}
\end{small}
with $\Theta_{\textrm{anti-}k_T}^{\rm <R} = \Theta\left(2x(1-x)p_T \tan\frac{R}{2}-|\bfk|\right)$. Here $\bfk$ is the transverse momentum of the emitted parton and $x$ and $(1-x)$ are the momentum fractions carried by the quark and the gluon, respectively,  in the collinear splitting $q\rightarrow q g$. This expression contains an infrared divergence at $\theta = 0$, which in practice is regulated by non-perturbative effects. In the present analysis, the corresponding non-perturbative regulator has been power expanded away. In the asymptotic region $\Lambda_{\rm QCD}/p_T \ll \theta \ll R$, the calculation remains well separated from the pole at $\theta = 0$.
The contact term reads 
\begin{small}
\begin{align}
\label{eq:cEEC_vac}
\mathcal{J}_{\rm EEC, q\rightarrow qg}^{\rm c, vac, (1)}(\theta; p_T, R, \mu) = & \frac{\alpha_s(\mu^2) C_F}{2\pi^2}\frac{\mu^{2\epsilon}e^{\epsilon\gamma_E}}{(4\pi)^{\epsilon} (2\pi)^{-2\epsilon}} \int\frac{d^{2-2\epsilon}\bfk}{\bfk^2} \Theta_{\textrm{anti-}k_T}^{<R}\nonumber \\
&\times \int_0^1 dx P_{qq, \epsilon}(x) \left[\mathcal{J}_{\rm EEC, q}^{\rm vac, (0)}(\theta; xp_T)+\mathcal{J}_{\rm EEC, g}^{\rm vac, (0)}(\theta;(1-x)p_T)\right] \nonumber \\
= & \frac{\alpha_s(\mu^2) C_F}{2\pi} \left(-\frac{1}{\epsilon}\right)\left(\frac{\mu^2}{(2p_T)^2\tan^2\frac{R}{2}}\right)^\epsilon  \frac{e^{\epsilon\gamma_E}}{\Gamma(1-\epsilon)}\nonumber\\
&\times \int_0^1 dx \frac{P_{qq, \epsilon}(x)}{[x(1-x)]^{2\epsilon}} \left[x^2 \mathcal{J}_{\rm EEC, q}^{\rm vac, (0)}(\theta; p_T, R, \mu) \right . \nonumber\\
& \left. \qquad \qquad \qquad \qquad \qquad +(1-x)^2 \mathcal{J}_{\rm EEC, g}^{\rm vac, (0)}(\theta; p_T, R, \mu)\right] \,.
\end{align}
\end{small}
Note that the loop correction is dimensionless when power expanded in $\theta/R$, so it is already dropped in the equation. In the second line we use the LO scaling law of the EEC such that up to higher order corrections, we have $\Sigma(\theta, xp_T) = x^2 \Sigma(\theta, p_T)$. 

\subsubsection{Energy-Energy Correlator with Out-of-Cone Radiation}

In addition to the exclusive contribution discussed in the previous subsection, the EEC receives an equally important semi-inclusive correction when one of the daughter partons produced in a collinear splitting exits the jet cone. Physically, the contribution encodes the fact that an energetic splitting can transfer a non-negligible fraction of the parent parton’s transverse momentum outside the cone radius $R$, thereby modifying the normalization and shape of the energy distribution inside the jet. In SCET language, this corresponds to matching the exclusive EEC jet function onto a semi-inclusive jet function, with a hard-collinear coefficient capturing the effect of the phase-space region where one of the two daughters fails the anti-clustering condition.
The first out-of-cone configuration corresponds to the case where the quark stays inside the jet radius and the accompanying gluon is emitted outside, leading to
\begin{align}
\label{eq:EEC_sinc_vac1}
\mathcal{J}_{\rm EEC,q\rightarrow q(g)}^{\rm vac,(1)}(\theta; z_J, p_T, R, \mu) = &  \mathcal{J}_{\rm EEC, q}^{\rm vac,(0)}(\theta; p_T, R, \mu)\frac{\alpha_s(\mu^2) C_F}{2\pi^2}\frac{\mu^{2\epsilon}e^{\epsilon\gamma_E}}{(4\pi)^{\epsilon} (2\pi)^{-2\epsilon}}\nonumber\\
&\times \int_0^1 dx P_{qq, \epsilon}(x) \delta(z_J-x)\int\frac{d^{2-2\epsilon}\bfk}{\bfk^2} \Theta_{\textrm{anti-}k_T}^{>R}\nonumber\\
 = & \mathcal{J}_{\rm EEC, q}^{\rm vac, (0)}(\theta; p_T, R, \mu)\frac{\alpha_s(\mu^2) C_F}{2\pi} \frac{1}{\epsilon} \left(\frac{\mu^2}{(2p_T)^2 \tan^2 \frac{\mathcal{R}}{2} }\right)^\epsilon \nonumber\\
& \times\frac{e^{\epsilon\gamma_E}}{\Gamma(1-\epsilon)}\frac{P_{qq, \epsilon}(z_J)}{\left( 1-z_J \right)^{2\epsilon}} \, .
\end{align}
Similarly we compute the contribution with a gluon inside the jet, which reads
\begin{align}
\label{eq:EEC_sinc_vac2}
\mathcal{J}_{\rm EEC,q\rightarrow (q)g}^{\rm vac, (1)}(\theta; z_J, p_T, R, \mu) = &\mathcal{J}_{\rm EEC, g}^{\rm vac, (0)}(\theta; p_T, R, \mu)\frac{\alpha_s(\mu^2) C_F}{2\pi^2}\frac{\mu^{2\epsilon}e^{\epsilon\gamma_E}}{(4\pi)^{\epsilon} (2\pi)^{-2\epsilon}}\nonumber\\
&\times \int_0^1 dx P_{qq, \epsilon}(x) \delta(z_J-(1-x))\int\frac{d^{2-2\epsilon}\bfk}{\bfk^2} \Theta_{\textrm{anti-}k_T}^{\rm >R}\nonumber\\
= & \mathcal{J}_{\rm EEC, g}^{\rm vac, (0)}(\theta; p_T, R, \mu)\frac{\alpha_s(\mu^2) C_F}{2\pi} \frac{1}{\epsilon} \left(\frac{\mu^2}{(2p_T)^2 \tan^2 \frac{\mathcal{R}}{2} }\right)^\epsilon  \nonumber\\
&\times  \frac{e^{\epsilon\gamma_E}}{\Gamma(1-\epsilon)}\frac{P_{gq, \epsilon}(z_J)}{\left( 1-z_J \right)^{2\epsilon}}\, .
\end{align}
Note that $\Theta_{\textrm{anti-}k_T}^{\rm >R} = \Theta\left(|\bfk|-2(1-z_J)p_T \tan\frac{R}{2}\right)$ in both cases.
From a perturbative perspective, the out-of-cone piece is governed entirely by collinear physics, meaning the angular separation $\theta$ between the two energy-flow operators remains much smaller than the cone radius, while the transverse recoil $\bfk$ generated by the splitting determines whether the second parton remains inside or outside the jet cone. This leads to a characteristic dependence on the splitting variable $z_J$ and the constraint $\Theta_{\text{anti-}k_T}^{>R}$, which isolates emissions with $|\bfk| > 2(1 - z_J) p_T \tan\frac{R}{2}$.

\subsubsection{The Full Next-to-Leading Order Result for $\Sigma_q$ in the Vacuum}
Here we combine all the results from Eqs.~(\ref{eq:ncEEC_vac}), (\ref{eq:cEEC_vac}), (\ref{eq:EEC_sinc_vac1}), (\ref{eq:EEC_sinc_vac2}) we derived in the previous subsections for the exclusive and semi-inclusive contributions to obtain the full NLO result for the EEC measured on a quark-initiated jet in the vacuum
\begin{small}
\begin{align}
\label{eq:vac-full-LO+NLO}
\mathcal{J}_{\rm EEC, q}^{\rm vac, (0)+(1)}(\theta; z_J, p_T, R, \mu)  = & \delta(1-z_J)\left(\mathcal{J}_{\rm EEC, q}^{\rm vac,(0)}(\theta; p_T, R,\mu)+\mathcal{J}_{\rm EEC, q}^{\rm nc, vac,(1)}(\theta; p_T, R,\mu)\right)\nonumber\\
&  + \frac{\alpha_s(\mu^2)C_F}{2\pi} \frac{1}{\epsilon}\left(\frac{\mu^2}{(2p_T)^2 \tan^2 \frac{\mathcal{R}}{2} }\right)^\epsilon \frac{e^{\epsilon\gamma_E}}{\Gamma(1-\epsilon)} \nonumber\\
&  \times \left\{ \mathcal{J}_{\rm EEC, q}^{\rm vac,(0)}(\theta; p_T, R,\mu)\left[\frac{P_{qq, \epsilon}(z_J)}{(1-z_J)^{2\epsilon}}-\delta(1-z_J)\int_0^1dx \frac{P_{qq, \epsilon}(x)}{[x(1-x)]^{2\epsilon}}\right]\right. \nonumber\\
& \quad\quad+ \mathcal{J}_{\rm EEC, g}^{\rm vac,(0)}(\theta; p_T, R,\mu)\frac{P_{gq, \epsilon}(z_J)}{(1-z_J)^{2\epsilon}} \nonumber\\
& \quad\quad- \mathcal{J}_{\rm EEC, q}^{\rm vac,(0)}(\theta; p_T, R,\mu)\delta(1-z_J)\int_0^1dx (x^2-1) \frac{P_{qq, \epsilon}(x)}{[x(1-x)]^{2\epsilon}} \nonumber\\
& \quad\quad\left.- \mathcal{J}_{\rm EEC, g}^{\rm vac,(0)}(\theta; p_T, R,\mu)\delta(1-z_J)\int_0^1dxx^2\frac{P_{gq, \epsilon}(x)}{[x(1-x)]^{2\epsilon}}\right\}\,.
\end{align}    
\end{small}
The expression can be organized into the factorized production of semi-inclusive quark EEC jet function and the EEC of exclusive jets, up to differences of $\mathcal{O}(\alpha_s^2$), such that
\begin{align}
\mathcal{J}_{\rm EEC, q}^{\rm vac, (0)+(1)}(\theta; z_J, p_T, R, \mu) =&\left[\mathcal{H}_{ jq}^{\rm vac, (0)}(z_J, p_T, R,\mu)+\mathcal{H}_{jq}^{(1)}(z_J, p_T, R,\mu)\right]\nonumber\\
&\times \left[\mathcal{J}_{\rm EEC, j}^{\rm vac, (0)}(\theta;p_T,R,\mu)+\mathcal{J}_{\rm EEC, j}^{\rm vac, (1)}(\theta;p_T,R,\mu)\right] +\mathcal{O}(\alpha_s^2)\,.
\end{align}
The hard-collinear matching coefficients $\mathcal{H}_{ji}$ (not to be confused with the hard function $\mathcal{H}_{ab\rightarrow cX}$) at LO and NLO are the same as those for semi-inclusive jet function, and match the exclusive EEC jet function to the semi-inclusive ones. After performing the $\epsilon$ expansion, we find
\begin{align}
\mathcal{H}_{jq}^{(0)}(z_J, p_T, R,\mu) &= \delta_{jq}\delta(1-z_J)\,, \nonumber\\
\mathcal{H}_{qq}^{(1)}(z_J, p_T, R,\mu) &= \frac{\alpha_s(\mu^2)}{2\pi}\left[\left(\frac{1}{\epsilon}+\mathcal{L}\right) [C_Fp_{qq}(z_J)]_++\delta(1-z_J)d_{qq}\right.\nonumber\\
&\hspace{2cm}\left.-2C_F(1-z_J)p_{qq}(z_J)\left(\frac{\ln(1-z_J)}{1-z_J}\right)_+-C_F(1-z_J)\right]\,, \nonumber\\
\mathcal{H}_{gq}^{(1)}(z_J, p_T, R,\mu) &= \frac{\alpha_s(\mu^2)}{2\pi}\left[\left(\frac{1}{\epsilon}+\mathcal{L}\right) C_F p_{gq}(z_J)-2C_Fp_{gq}(z_J)\ln(1-z_J)-C_Fz_J\right]\,.
\end{align}
The logarithm reads $\mathcal{L} = \ln\frac{\mu^2}{(2p_T)^2\tan^2\frac{R}{2}}$ and the coefficient corresponding to the anti-$k_T$ jet is $d_{qq} = C_F\left(\frac{13}{2}-\frac{2}{3}\pi^2\right)$. The unrenomalized EEC of exclusive jet at NLO is
\begin{small}
\begin{align}
\mathcal{J}_{\rm EEC, j}^{\rm vac, (1)}(\theta; p_T,R,\mu)=&\frac{\alpha_s(\mu^2)}{4\pi} \left(\frac{1}{\epsilon}+\mathcal{L}\right) \left[\gamma_{qq}^{\rm vac}(3) \mathcal{J}_{\rm EEC, q}^{\rm vac,(0)}(\theta;p_T,R,\mu)+ \gamma_{gq}^{\rm vac}(3)\mathcal{J}_{\rm EEC, g}^{\rm vac,(0)}(\theta; p_T,R,\mu)\right] \nonumber\\
&+\frac{\alpha_s(\mu^2)}{2\pi}\left[\frac{\Theta(R-\theta)}{\theta^2} \frac{3C_F}{4}+\frac{295}{36}C_F \mathcal{J}_{\rm EEC, q}^{\rm vac,(0)}(\theta;p_T,R,\mu)\right. \nonumber \\
& \qquad \qquad\qquad+ \left.\frac{73}{36}C_F \mathcal{J}_{\rm EEC, g}^{\rm vac,(0)}(\theta;p_T,R,\mu)\right] \, ,
\end{align}
\end{small}
where the first line come from the contact contribution, and the second line are the non-contact expressions.
For the contact terms, we have defined the moments of LO QCD splitting functions as
\begin{align}
\label{eq:ENC-gamma-vac}
\gamma_{qq}^{\rm vac}(N) =& -2\int_0^1 z^{N-1} [C_F p_{qq}(z)]_+ dz\,, \quad \gamma_{qq}^{\rm vac}(3) = \frac{25}{6}C_F \, .\\
\gamma_{gg}^{\rm vac}(N) = & -2\int_0^1 z^{N-1} [P_{gg}(z)]_+ dz\,, \quad \gamma_{gg}^{\rm vac}(3) = \frac{14}{5}C_A+\frac{2}{3}N_f \, .\\
\gamma_{gq}^{\rm vac}(N) =& -2\int_0^1 z^{N-1} C_F p_{gq}(z) dz\,, \quad \gamma_{gq}^{\rm vac}(3) = -\frac{7}{6}C_F \, .\\
\gamma_{qg}^{\rm vac}(N) =& -2\int_0^1 z^{N-1} 2N_f T_F p_{qg}(z) dz\,, \quad \gamma_{qg}^{\rm vac}(3) = -\frac{14}{15}T_F N_f \, .
\end{align}
Similar expressions can be written down for $\Sigma_g$. Note that here we only show the results for the EEC, while the complete expressions for the semi-inclusive jet functions can be found in Ref.~\cite{Kang:2016mcy}.

\subsection{Renormalization Group Equations in Elementary Collisions} 
After the standard $\overline{\textrm{MS}}$ renormalization procedure, one arrives at the LL evolution equation for the exclusive EEC jet function
\begin{align}
\renewcommand*{\arraystretch}{1.5}
\frac{\partial }{\partial \ln \mu^2}\begin{bmatrix}
\mathcal{J}_{\rm EEC, q}^{\rm vac}(\theta; p_T,R,  \mu) \\
\mathcal{J}_{\rm EEC, g}^{\rm vac}(\theta; p_T, R,\mu)
\end{bmatrix}
= 
\frac{\alpha_s(\mu^2)}{4\pi}\begin{bmatrix}
\gamma_{qq}^{\rm vac}(3) & \gamma_{gq}^{\rm vac}(3)  \\
\gamma_{qg}^{\rm vac}(3) & \gamma_{gg}^{\rm vac}(3) 
\end{bmatrix}
\begin{bmatrix}
\mathcal{J}_{\rm EEC, q}^{\rm vac}(\theta; p_T,R,  \mu) \\
\mathcal{J}_{\rm EEC, g}^{\rm vac}(\theta; p_T,R, \mu) 
\end{bmatrix} \, ,
\end{align}
and for the semi-inclusive EEC jet function we have the following evolution equation
\begin{align}\label{eq:SIJFDGLAP} 
\frac{\partial}{\partial \ln \mu^2} 
\begin{bmatrix}
\mathcal{H}_{jq}(z_J,p_T,R,\mu)\\
\mathcal{H}_{jg}(z_J,p_T,R,\mu)
\end{bmatrix}
= \frac{\alpha_s(\mu^2)}{2 \pi}
\begin{bmatrix}
\hat{P}_{qq}(z) & \hat{P}_{gq}(z)\\
2N_f \hat{P}_{qg}(z) & \hat{P}_{gg}(z)
\end{bmatrix}
 \otimes_{z_J}
 \begin{bmatrix}
\mathcal{H}_{jq}(z_J,p_T,R,\mu)\\
\mathcal{H}_{jg}(z_J,p_T,R,\mu)
\end{bmatrix}\, ,
\end{align}
where the action is on the hard-collinear matching coefficients.
These are RG equations extracted from a calculation at the jet scale $2p_T\tan\frac{R}{2}$, as can be seen from the argument of the logarithm $\mathcal{L} = \ln\frac{\mu^2}{(2p_T)^2\tan^2\frac{R}{2}}$, where $\mu$ is an IR cut off. Therefore, the evolution for both $\mathcal{H}_{ji}(z_J, p_T, R, \mu)$ and $\mathcal{J}_{\rm EEC, i}^{\rm vac}(\theta; p_T, R, \mu)$ start at scale $\mu_R=2p_T\tan\frac{R}{2}$. Then, $\mathcal{J}_{\rm EEC, i}^{\rm vac}(\theta; p_T, R, \mu)$ is evolved to $\mu_\theta=\theta E$ to match the EEC measurement, while the hard-collinear functions $\mathcal{H}_{ji}$ are evolved to $\mu_H=p_T$ to match the hard function.
For, this reason, the above equation for $\mathcal{J}_{\rm EEC, i}^{\rm vac}$ differs from those in Ref~\cite{Dixon:2019uzg} by a minus sign, as the latter evolves $\mu$ as an UV cut off from $\mu_\theta$ to $\mu_R$. Finally, the boundary conditions of the exclusive jet function  evaluated at $\mu_\theta$ and accurate to NLO are
\begin{small}
\begin{align}
\label{eq:EEC-vac-initial-conditions}
\mathcal{J}_{\rm EEC, q}^{\rm vac}(\theta; p_T,R, \mu_\theta) =&  \mathcal{J}_{\rm EEC, q}^{\rm vac,(0)}(\theta; p_T,R,\mu_{\theta}) + \frac{\alpha_s(\mu_\theta^2) C_F}{2\pi}\left[\frac{\Theta(R-\theta)}{\theta^2} \frac{3}{4}+\frac{295}{36} \mathcal{J}_{\rm EEC, q}^{\rm vac,(0)}+\frac{73}{36} \mathcal{J}_{\rm EEC, g}^{\rm vac,(0)}\right]  \nonumber \\
\approx&  \frac{\Theta(R-\theta)}{\theta^2}\frac{\alpha_s(\mu_\theta^2) }{2\pi} \frac{3}{4}C_F\left[1+\mathcal{O}\left(\frac{\Lambda}{\theta E}\right)\right], \nonumber\\
\mathcal{J}_{\rm EEC, g}^{\rm vac}(\theta; p_T,R, \mu_\theta) \approx& \frac{\Theta(R-\theta)}{\theta^2}\frac{\alpha_s(\mu_\theta^2) }{2\pi}\left(\frac{7}{10}C_A+ \frac{1}{10} N_f T_f  \right)\left[1+\mathcal{O}\left(\frac{\Lambda}{\theta E}\right)\right], 
\end{align}
\end{small}
where we further use the condition that in the asymptotic region the contribution from the non-perturbative $\mathcal{J}_{\rm EEC, q}^{\rm vac,(0)}(\theta; p_T,R,\mu)$ is power suppressed compared to the  perturbative piece. For example, if we consider the non-perturbative power correction is of order $\frac{1}{\theta^2}\frac{\Lambda_{\rm QCD}}{\theta p_T}$, then we should define the lower bound of this asymptotic region by
\begin{align}
\frac{\Lambda_{\rm QCD}}{\alpha_s(\theta p_T)} \lesssim \theta p_T\,.
\end{align}
This means that, even though the perturbative evolution should still be working for $\theta p_T \gg \Lambda_{\rm QCD}$, the initial condition would have to include non-perturbative power corrections for $\Lambda_{\rm QCD} \ll \theta  p_T\lesssim \frac{\Lambda_{\rm QCD}}{\alpha_s(\theta p_T)}$.

\section{Factorization of Energy Correlators in Heavy-Ion Collisions}
\label{sec:eechi}
To analyze how the QCD medium affects each ingredient of the factorized formula for the EECs, we begin by identifying the relevant medium-induced energy scales that enter the problem and discuss their hierarchy relative to the hard and jet scales. We then introduce the theoretical framework of SCET$_{\rm G}$, which provides a systematic treatment of parton-medium interactions within the opacity expansion. In this context, we define the small expansion parameter that organizes the power counting of medium-induced corrections.

Two primary mechanisms govern the interaction between an energetic jet and the QGP: collisional energy loss due to elastic scatterings, and medium-induced parton radiation generated by transverse momentum broadening and color exchanges with the medium. The structure of these effects, and their interplay within the SCET$_{\rm G}$ formalism, will be discussed at the end of this Section, as they form essential inputs for computing the medium-modified EEC in Section~\ref{sec:medium:exlcusive}.

\subsection{Scales in Nuclear Matter and Modified Energy Correlator Factorization Formula}

In the heavy-ion environment, the jet propagates through a strongly interacting QCD medium, and the EEC becomes sensitive not only to vacuum QCD dynamics but also to medium-induced effects. These effects arise from multiple scales: the hard jet energy scale $Q \sim p_{T,\text{jet}}$, the collinear scale associated with small-angle emissions $Q\theta$, and the medium scale $\mu_{\text{med}} \sim T$, where $T$ is the temperature of the QGP. To capture this multi-scale behavior, we work within the SCET framework extended to include Glauber gluon interactions, which model transverse momentum exchanges between the jet and the medium~\cite{Ovanesyan:2011kn,Ovanesyan:2011xy}. These interactions are characterized by momentum transfers of the form $q^\mu \sim (0, 0, \vec{q}_\perp)$, and lead to medium-induced modifications of the jet function and the evolution of the observable. Last but not least, in reactions with nuclei there are geometry considerations, such as the size of the medium $L$ and its opacity $L/\lambda$ that become important.

The energy-energy correlator in heavy-ion collisions admits a factorized description
in terms of hard and collinear functions, valid in the regime of small opening angles
and for jets traversing a dilute QCD medium. Working in the opacity expansion of SCET with
Glauber interactions (SCET$_\mathrm{G}$), we can systematically separate the dynamics associated
with the hard production of the jet and its collinear in-medium evolution.
The factorization theorem relies on the separation of the following dynamical scales. In proton-proton collisions, apart from $\Lambda_{\rm QCD}$, most of the characteristic scales are introduced by the probes, such as the hard production scale $p_T$, the jet scale $p_TR$ and the jet substructure scale ($\theta p_T$ for EEC).
\begin{align}
Q &\sim p_{T}  && \text{hard production scale}, \nonumber \\
\mu_{R} &\sim p_{T} R && \text{hard-collinear scale, jet cone}, \nonumber \\
\mu_\theta &\sim \theta p_{T} && \text{collinear virtuality of EEC}\,.
\end{align}
In nuclear interactions, there are a new class of medium-related energy scales:
\begin{align}
M_{\rm LPM} & \sim \sqrt{2 p_{T}/L} &&  \text{LPM scale},\nonumber \\
\sqrt{\langle\delta k_T^2\rangle} &\sim \sqrt{g_s^4 T^3 L} && \textrm{medium $k_T$ broadening scale}, \nonumber\\
m_{\rm eff} &\sim g_s T,~~\Lambda_{\rm QCD} && \text{medium screening scale}, 
\end{align}
Here, $m_{\rm eff}$ denotes the effective screening mass of the gluon propagator in the medium. In a high-temperature plasma, it can be estimated by the leading-order plasma Debye mass $m_{\rm eff} = m_D\sim g_sT$. 
Multiple scattering leads to transverse momentum broadening scale $\langle\delta k_T^2\rangle$, controlled by the product of the length of the medium $L$ and the jet transport parameter $\hat{q}\sim g_s^4 T^3$ that characterizes the transverse momentum transfer squared per unit length.
In a thin medium, the Landau–Pomeranchuk–Migdal (LPM) effect~\cite{Landau:1953um,Migdal:1956tc} introduces a new semi-hard scale $M_{\rm LPM} \propto \sqrt{p_{T}/L}$~\cite{Ke:2023ixa,Ke:2024yzp}. This can be understood as the parametric scale of the typical virtuality of collinear fluctuations whose lifetime matches the medium length.

In this work, we consider the following hierarchy of scales
\begin{align}
\textrm{Scenario I:}~~&\Lambda_{\rm QCD} \lesssim m_{\rm eff} \lesssim \sqrt{\langle\delta k_T^2\rangle} \ll \mu_\theta \ll M_{\rm LPM} \ll \mu_R \lesssim Q \, , \nonumber\\
\textrm{Scenario II:}~~&\Lambda_{\rm QCD} \lesssim m_{\rm eff} \lesssim \sqrt{\langle\delta k_T^2\rangle} \ll M_{\rm LPM} \ll \mu_\theta \ll \mu_R \lesssim Q \, .
\label{eq:all-the-scales}
\end{align}
Both of these cases can be achieved for the propagation of high-energy small $R$ jets in a thin (short length $L$) medium, such that $M_{\rm LPM}$ is the hardest among all the medium-related energy scales. For later convenience, we also define an angle related to the LPM effect\footnote{Note that our estimates for $M_{\rm LPM}$ and $\theta_{\rm LPM}$ are parametric, where we drop the explicit $x(1-x)$ kinematic dependence. For $M_{\rm LPM}$, this adequately describes its power and logarithmic appearance. If one wishes to examine the in-medium EEC features in the phenomenological applications that follow, e.g. transitional behavior around $\theta_{\rm LPM}$, a more accurate estimate of $\theta_{\rm LPM}$ is desirable. From the first peak of the LPM interference factor $1-\cos\frac{\bfk^2 L}{2x(1-x)p_T}$, the argument should equal $\pi$. Taking the peak value of $x(1-x)=1/4$ in $\theta = |{\bf k}|/[x(1-x)p_T]$, its is easy to obtain Eq.~(\ref{eq:define-LPM-scale-and-angle}). },
\begin{align}
\label{eq:define-LPM-scale-and-angle}
\theta_{\rm LPM}  = \sqrt{\frac{8 \pi}{p_T L}}\,.
\end{align}
Conditions in Eq.~(\ref{eq:all-the-scales}) also ensure that the primary hard scattering remains unaffected by the medium, while medium-induced modifications enter at the collinear level. 
The medium effects correction to EEC can be analyzed in opacity expansion, where we will discuss the medium resummation effects for both cases of $M_{\rm LPM} \ll \mu_\theta$ and $\mu_\theta \ll M_{\rm LPM}$.
From Eq.~(\ref{eq:all-the-scales}), one can conclude that the semi-inclusive jet function is the same as the one for the vacuum. However, medium radiations causes collinear-soft radiations outside the jet cone and is also an effect at first order in opacity. Therefore, to consistently compute EEC in opacity expansion, the modification to semi-inclusive jet function is also needed, which is discussed in subsection~\ref{sec:medium:semi-inlcusive}.

In the following Section, we present the factorization that governs the EEC in the presence of a QGP and outline the details for calculating the quark and gluon jet functions. These ingredients will form the foundation for the renormalization group analysis and resummation presented in subsequent sections.

The normalized EEC cross-section can be written as 
\begin{equation}
\label{eq:EECfactorized}
\frac{d\Sigma}{d\theta\, dp_T\, dy}
=
\begin{aligned}[t]
& \sum_{a,b,c}\!\int\! dx_a\,dx_b\,dz_J\;
f_{a/A}(x_a,\mu)\, f_{b/B}(x_b,\mu)\;
\mathcal{H}_{ab\to c}\!\left(\tfrac{p_T}{z_J},\,y,\mu\right) \\
& \quad \times \Big[
\mathcal{J}_{\rm EEC, c}^{\rm vac}(\theta;\,z_J,\,p_T, R,\mu)
+ \mathcal{J}_{\rm EEC, c}^{\rm med}(\theta;\,z_J,\,p_T, R,\mu;\,L,m_{\rm eff})
\Big]\,,
\end{aligned}
\end{equation}
where $d\Sigma$ is the energy–energy weighted cross-section that depends on the kinematics and the medium parameters, and $f_{a/A}(x_a,\mu)$ and $f_{b/B}(x_b,\mu)$ are the parton distribution functions (or nuclear PDFs in reactions with nuclei) for partons $a$ and $b$ in the incoming beams $A$ and $B$. Here $z_J$ is the momentum fraction of the parent parton $c$ carried by the reconstructed jet and $\mathcal{H}_{ab\to c}(\hat p_T,y,\mu)$ is the hard function for producing a parton $c$ at transverse momentum $\hat p_T=p_T/z_J$ and rapidity $y$. The jet functions $\mathcal{J}_{\rm EEC, c}^{\rm vac}(\theta;z_J,p_T,R,\mu)$ and $\mathcal{J}_{\rm EEC, c}^{\rm med}(\theta;z_J,p_T,R,\mu;L,m_{\rm eff})$ are the vacuum jet function, which contains the EEC measurement through the angle $\theta$ between constituents inside the jet and the medium modification, depending on medium parameters such as the length $L$ and effective screening mass $m_{\rm eff}$. 

Equation~\eqref{eq:EECfactorized} encapsulates the physical picture that underlies our calculation: the hard scattering is unmodified,
the jet develops collinear structure through both vacuum and medium-induced branchings, and Glauber scatterings in the medium generate transverse broadening and color decoherence. The absence of a separate soft function reflects the fact that the EEC weighting suppresses soft contributions, leaving collinear radiation as the dominant source of logarithmic enhancements.

The form of $\mathcal{J}_{\rm EEC, c}^{\rm vac}(\theta;\,z_J,\,p_T,\,R,\,\mu)$ up to NLO in $\alpha_s$ is already introduced for quark jets in Eq.~(\ref{eq:vac-full-LO+NLO}). The medium modified part of the jet function follows the same structure as Eq.~(\ref{eq:vac-full-LO+NLO}), but with the vacuum splitting functions replaced by the medium-induced transverse momentum dependence splitting functions calculated at first order in opacity. The full EEC jet function $\mathcal{J}_{\rm EEC, c}=\mathcal{J}_{\rm EEC, c}^{\rm vac}+\mathcal{J}_{\rm EEC, c}^{\rm med}$ for any flavor $c$ can be written as a sum of various jet functions
\begin{small}
\begin{align}
\mathcal{J}_{\rm EEC, c}(\theta;z_J, p_T, R,\mu)=& \delta(1-z_J)\mathcal{J}_{\rm EEC,c}^{\rm vac(0)}(\theta;p_T,R,\mu) + \Delta z_{\rm coll}(-\delta'(1-z))\mathcal{J}_{\rm EEC, c}^{\rm vac, (0)}(\theta;p_T,R,\mu)\nonumber\\
& +\delta(1-z_J)\mathcal{J}_{\rm EEC, c}^{\rm nc, vac,(1)}(\theta;p_T,R,\mu)+\delta(1-z_J)\mathcal{J}_{\rm EEC, c}^{\rm nc, med, (1)}(\theta;p_T,R,\mu)\nonumber\\
& + \delta(1-z_J)\mathcal{J}_{\rm EEC, c}^{\rm c, vac, (1)}(\theta;p_T, R, \mu) +  \delta(1-z_J)\mathcal{J}_{\rm EEC, c}^{\rm c, med, (1)}(\theta;p_T, R, \mu) \nonumber\\
&  + \mathcal{J}_{\rm EEC, c}^{\rm vac, (1)}(\theta;z_J, p_T, R, \mu) + \mathcal{J}_{\rm EEC,c}^{\rm med,(1)}(\theta;z_J, p_T, R, \mu)\,.
\end{align}
\end{small}
The first correction comes from the collisional energy loss of the jet, which is often neglected in the formal collinear power counting of SCET$_G$. However, it can be important for jet phenomenology, so we include it here.
The corrections in the second and third line corresponds to the medium modification to exclusive jet function with non-contact or contact energy weightings.
Finally, the last line contains the correction to the semi-inclusive jet function, denoted by their additional dependence on $z_J$.

In this Section, we will first discuss the opacity expansion in subsection~\ref{sec:SCETG_opacity_expansion} and important features of the TMD splitting function at opacity order one in subsection~\ref{sec:SCETG_TMD_splitting_functions}, which will be used for all the remaining sections. As a reminder
Section~\ref{sec:medium:exlcusive} will discuss the correction to the exclusive jet function. Section~\ref{sec:Medium:resummation}  will discuss the semi-inclusive jet function in the medium and the final formula with renormalization and resummation.

\subsection{SCET$_\mathrm{G}$ Opacity Expansion} 
\label{sec:SCETG_opacity_expansion}
In its original form, which includes (ultra)soft and collinear modes, the SCET framework cannot describe the collisional and radiative processes induced
by the propagation of an energetic parton in strongly-interacting matter. To overcome this deficiency, it must be supplemented with off-shell modes that mediate the interactions between the collinear sector and the quasi-particles of the medium. Depending on their nature and the kinematics of scattering processes, there are two possible momentum scalings for the virtual gluon that go at most as:
\begin{eqnarray}
~  [ \lambda^2, \lambda^2, \lambda ], && \qquad {\rm Gauber \; gluon}, \\
 ~ [ \lambda, \lambda, \lambda],  && \qquad {\rm Coulomb \; gluon}, 
\end{eqnarray} 
in lightcone coordinates~\cite{Ovanesyan:2011xy,Makris:2019ttx}. For the case of energetic jets and and massive (or bound) scattering centers the former scenario applies and the Feynman rules for SCET$_\mathrm{G}$ for different gauges can be found in Ref.~\cite{Ovanesyan:2011xy}. While operator level expression for the Glauber gluon field  $A_{\rm G}$  can be written down for free sources, medium quasi-particles are strongly coupled. Consequently, a background field approach that relates its strength to a very small number of non-perturbative parameters is sufficient and, in fact, preferred to separate the perturbative and non-perturbative sectors of the theory. 

With SCET$_\mathrm{G}$ at hand, we can compute radiative processes in matter. Formation times of collinear splittings can  exceed the mean free paths of energetic partons in the nuclear medium, leading to coherent emission that changes the splitting fraction and angular dependence in the branching processes~\cite{Gyulassy:2003mc}. One way to study in-medium interactions is through the opacity expansion approach which evaluates quantities of interest as a series in the correlation between the multiple scattering centers (not to be confused with the average number of interactions determined by the size and transport properties of the nuclear medium). It was first introduced to describe soft gluon radiation~\cite{Gyulassy:2000er,Gyulassy:2000fs,Wiedemann:2000za} and later applied to collinear branching beyond the energy loss limit~\cite{Ovanesyan:2011xy,Ovanesyan:2011kn}. For heavy ion collisions it is the first order in opacity that captures the dominant modifications to parton branching due to the LPM effect. While higher orders can result in some quantitative differences, their effect is subleading, especially for thin media~\cite{Sievert:2019cwq,Sievert:2018imd}.       

\subsection{Collisional Energy Loss}
\label{sec:coll-eloss}
Modification to hadron and jet observables can also arise through tree-level interactions with the quasi-particles of the medium. We adopt the approach from Refs.~\cite{Ke:2022gkq,Li:2018xuv}. 
The correction to the jet function due to collisional energy loss of the jet is given by
\begin{align}
 & J_i^{(0)}(E, R, \theta)\sum_j \int dz^+ \rho_j\int d^2\bfq \frac{d\sigma_{ij}}{d^2\bfq} \left(\delta\left(1-z_J-\frac{\omega(\bfq, E)}{E}\right)-\delta(1-z_J) \right) \nonumber\\
&  \qquad \qquad \qquad \approx  \Delta z_{i, \rm coll}(E) (-\delta'(1-z_J))J^{(0)}(E,R,\theta) \, .
\end{align}
Here, $\frac{d\sigma_{ij}}{d^2\bfq}$ is the differential cross-section that describes the  collision between parton $i$ and $j$  and $\omega(\bfq, E)$ is energy carried away by the Glauber gluon. Note that $\omega$ is dropped in the Glauber gluon propagator because of the power counting. 
The collisional energy loss fraction is defined as
\begin{align}
\Delta z_{i, \rm coll}(E) = \frac{1}{E}\int dz^+ \rho_j\int d^2\bfq \frac{d\sigma_{ij}}{d^2\bfq} \omega(\bfq, E)\, .
\end{align}
In a thermal plasma, there are important plasma excitation effects that screen the infrared divergence in the exchanged gluon propagator. Using the hard thermal loop theory in a weakly-coupled thermal plasma, the collisional energy loss  of a quark per unit time is 
\begin{eqnarray}
\frac{d\Delta z_{i, \rm coll}}{dt} = v \frac{C_F}{4} \frac{\alpha_s(ET)}{E} m_D^2 \ln\left(\frac{E T}{m_D^2}\right)\left( \frac{1}{v} - \frac{1-v^2}{2v^2}\ln\frac{1+v}{1-v}\right) \, ,
\end{eqnarray}
where $v=p/E$ is the velocity of the parton, and $m_D$ is the plasma Debye screening mass 
\begin{align}
m_D^2 = \left(1+\frac{N_f}{6}\right) \left(g_s^{\rm med}T\right)^2\,.
\end{align}
$g_s^{\rm med}$ is supposed to be the running coupling at scale $m_D$. In a strongly coupled QGP, $m_D$ is not fully in the perturbative region, so we take $g_s^{\rm med}$ as a phenomenological parameter that quantify the jet-medium interaction strength.
Here, the energy are then carried away by the deflection of thermal partons or the excitation of plasma modes. This is always at a large angle compared to a moderate jet cone and causes energy loss outside of the cone~\cite{Neufeld:2011yh}. Even though collective or quasi-particle-like medium response can carry the energy back into the cone, but such contributions are suppressed by $R^2$ for small radii jets.

Summing over the elastic collisions from all orders of opacity, the result is an energy shift by the amount of the energy loss to the hard parton spectrum. Finally, the collisional energy loss of a gluon is $C_A/C_F$ times of that for a quark.

\subsection{TMD Splitting Functions at First Order in Opacity}
\label{sec:SCETG_TMD_splitting_functions}
As key inputs to the calculation of medium corrections to EEC, we summarize the transverse-momentum dependent splitting functions obtained in SCET$_\mathrm{G}$ at first order in opacity~\cite{Ke:2024ytw}. The collinear sector receives radiative corrections governed by medium-induced splitting kernels, and for a   quark jet
\begin{align}
P_{q\to qg}^{\rm med}(x,p_\perp) = 
 \mathcal{R}_{\rm RR} + \mathcal{R}_{\rm RV} + \mathcal{R}_{\rm VR} + \mathcal{R}_{\rm VV} \, ,
\end{align}
where the RR, RV, VR, VV channels denote real/virtual splitting combined with real/virtual Glauber scattering, see for example~\cite{Ke:2024yzp}. These contributions generate the medium-dependent anomalous dimensions that control the RG evolution of the EEC.

Consider an initial quark with large lightcone momentum $P^+$ produced with virtuality of order $P^2=Q^2\gg \Lambda_{\rm QCD}^2$ with a transverse momentum distribution $N_q(\bfP)$. The transverse direction is in $2-2\epsilon$ dimension.
At leading order $\mathcal{O}(\alpha^0_s)$ and opacity $n=0$, the final state quark has the same distribution
\begin{align}
\frac{dN_{qq}^{(0),n=0}}{dx d^{2-2\epsilon}\bfp} = \delta(1-x)N_q(\bfp) \, .
\end{align}
At next-to-leading order $\mathcal{O}(\alpha_s)$ and opacity $n=0$, one considers a gluon radiation with momentum $\bfk$ such that the quark distribution reads\footnote{Loop correction are scaleless in leading power of $\Lambda_{\rm QCD}^2/Q^2$.}
\begin{align}
\frac{dN_{qq}^{(1),n=0}}{dx d^{2-2\epsilon}\bfp} = \frac{\alpha_s C_F}{(2\pi)^3}P_{qq, \epsilon}(x)\int \frac{d^{2-2\epsilon}\bfk}{(2\pi)^{-2\epsilon}} \frac{1}{\bfk^2}N_q(\bfp+\bfk)\,,
\end{align}
where $P_{qq, \epsilon} = \frac{1+x^2}{1-x} - \epsilon(1-x)$ is the vacuum splitting function at leading order. In this case the virtuality cut off $P^2 = \frac{(x\bfk-(1-x)\bfp)^2}{x(1-x)} < Q^2$ will provide a natural upper cut to the $\bfk$ integration and in turn introduces $(1-x)^{2\epsilon}$ to regulate the soft divergence in $P_{qq}$.

\begin{figure}
    \centering
    \includegraphics[width=0.75\linewidth]{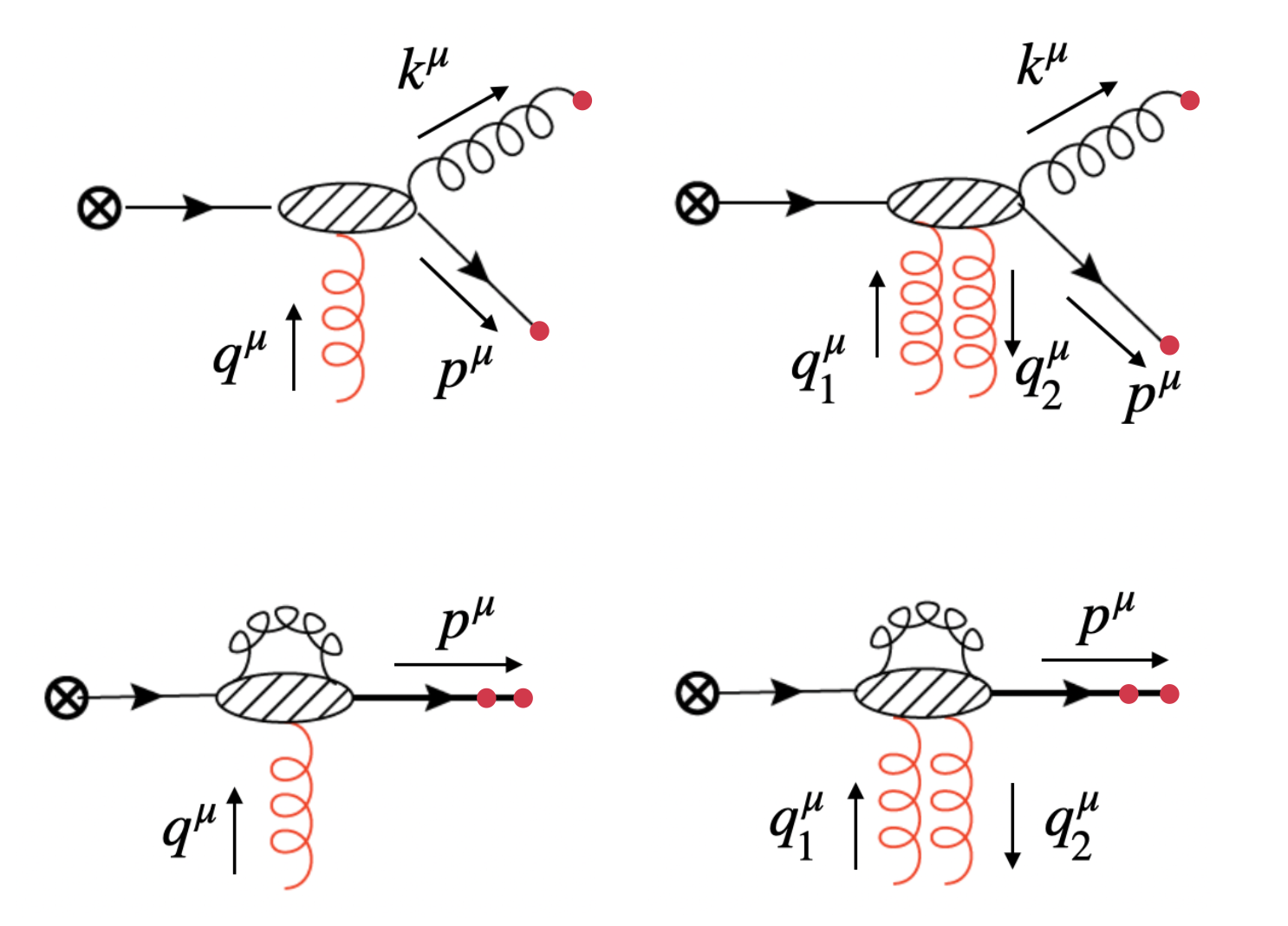}
    \caption{Feynman diagrams that contribute to the TMD splitting functions at the amplitude level that are relevant for the calculation at opacity $N=1$. The red gluons are the Glauber gluon medium interactions. }
    \label{fig:Jet-diagrams}
\end{figure}

At leading order in the coupling constant and opacity $n=1$, there are single-Glauber (squared amplitude) and double-Glauber (interference with 0-Glauber diagram, hereafter referred to as virtual Glauber exchange) contributions. The forward scattering with the medium, mediated by Glauber gluon, cannot change $x$ at leading power, but recoils the quark's transverse momentum
\begin{align}
\frac{dN_{qq}^{(0),n=1}}{dx d^{2-2\epsilon}\bfp} = \delta(1-x)\int_0^\infty dz^+ \rho_T^-(z^+) \int \frac{d^{2-2\epsilon}\bfq}{(2\pi)^{2-2\epsilon}}\frac{d\sigma_{FT}}{d^{2-2\epsilon}\bfq}
\left[N_q(\bfp-\bfq)-N_q(\bfp)\right] \, ,
\end{align}
where $\rho_T^-(z^+)$ is the density of the medium particles of color representation $T$ at a location with positive light-cone coordinate $z^+$.
$\frac{d\sigma_{FT}}{d^{2-2\epsilon}\bfq}$ is the screened forward scattering cross-section between the quark and the medium particle,
\begin{align}
\frac{d\sigma_{FT}}{d^{2}\bfq} = \frac{g_s^2 C_F g_s^2 C_T}{d_A}\frac{1}{(\bfq^2+m_{\rm eff}^2)^2}  \, .
\end{align} 
Here, $C_T$ is the quadratic Casimir of the target parton representation and $d_A = 8$ is the dimension of the adjoint representation. In a QGP medium, the effective screening mass $m_{\rm eff}$ is often taken to be the plasma Debye mass $m_D$, which can be perturbatively calculated at high temperature $m_{\rm eff}^2 = m_D^2$. In cold nuclear matter $m_{\rm eff}\sim \Lambda_{\rm QCD}$. 
If we power expand in $m_{\rm eff}^2/Q^2$ with $Q$ being a harder scale in the problem, then the second term in $\int \frac{d^{2-2\epsilon}\bfq}{(2\pi)^{2-2\epsilon}}\frac{d\sigma_{FT}}{d^{2-2\epsilon}\bfq}\left[N_q(\bfp-\bfq)-N_q(\bfp)\right]$ becomes a scaleless integral, while the first term is regulated in DimReg. 

At next-to-leading order $^{(1)}$ and opacity $n=1$. There are four types of contributions: 
\begin{align}
\frac{dN_{qq}^{(1),n=1}}{dx d^{2-2\epsilon}\bfp } = \frac{dN_{qq}^{RR}}{dx d^{2-2\epsilon}\bfp } + \frac{dN_{qq}^{RV}}{dx d^{2-2\epsilon}\bfp } + \frac{dN_{qq}^{VR}}{dx d^{2-2\epsilon}\bfp } + \frac{dN_{qq}^{VV}}{dx d^{2-2\epsilon}\bfp }
\, .
\end{align}
The diagrams and kinematic variables for the four types of contributions are shown in Figure~(\ref{fig:Jet-diagrams}). For brevity, it is useful to define linear combinations of transverse momenta such that
\begin{align}
\bfQ_1 &= x\bfk-(1-x)(\bfp-\bfq)\, ,\quad
\bfQ_2 = x\bfk-(1-x)\bfp\, , \nonumber\\
\bfQ_3 &= x(\bfk-\bfq)-(1-x)\bfp\, ,\quad
\bfQ_4 = x(\bfk-\bfq)-(1-x)(\bfq+\bfq)\, ,\nonumber\\
\bfQ_5 &= x\bfk-(1-x)(\bfp-\bfq-\bfk)\, ,\quad
\bfQ_6 = x\bfk-(1-x)(\bfp-\bfk)\, ,\nonumber\\
\bfQ_7 &= x(\bfk-\bfq)-(1-x)(\bfp-\bfk)\, ,\quad
\bfQ_8 = x(\bfk+\bfq)-(1-x)(\bfp-\bfk-\bfq)\, ,
\end{align}
with the interference factor
\begin{align}
\Phi(u) = 1-\cos(u) \, .
\end{align}
For simplicity for now on we will use the following short-hand notations
\begin{align}
\Phi_n = \Phi\left( \frac{\bfQ_n^2 z^+}{2x(1-x)P^+} \right)\,,\quad  \Phi_{n,m} = \Phi\left( \frac{(\bfQ_n^2-\bfQ_m^2) z^+}{2x(1-x)P^+} \right)\,.
\end{align}
With these definitions, we list below the contributions to the quark TMD splitting functions. \\
 $RR$: real splitting, real-Glauber interaction (squared single-Glauber exchange amplitude).
\begin{align}
\frac{dN_{qq}^{RR}}{dx dd^{2-2\epsilon}\bfp } =& \frac{g^2 C_F}{(2\pi)^3} P_{qq, \epsilon}(x) \int_0^\infty dz^+ \rho_T^-(z^+) \int \frac{d^{2-2\epsilon}\bfq}{(2\pi)^{2-2\epsilon}}\frac{1}{C_F}\frac{d\sigma_{FT}}{d^{2-2\epsilon}\bfq}
\int \frac{d^{2-2\epsilon}\bfk}{(2\pi)^{-2\epsilon}}   N_q(\bfp+\bfk-\bfq) \nonumber\\
&\times \left\{C_F \frac{1}{\bfQ_2^2} +
(2C_F-C_A) \left(\frac{1}{\bfQ_1^2}- \frac{\bfQ_1}{\bfQ_1^2}\cdot\frac{\bfQ_2}{\bfQ_2^2}\right)\Phi_1 + C_A \frac{\bfQ_1}{\bfQ_1^2}\cdot\frac{\bfQ_3}{\bfQ_3^2}\Phi_{1,3}\right.\nonumber\\
& \left.
+ C_A \left(\frac{1}{\bfQ_3^2}-\frac{\bfQ_2}{\bfQ_2^2}\cdot\frac{\bfQ_3}{\bfQ_3^2}\right)\Phi_3 + C_A \left(\frac{1}{\bfQ_3^2}-\frac{\bfQ_1}{\bfQ_1^2}\cdot\frac{\bfQ_3}{\bfQ_3^2}\right)\Phi_3  \right.\nonumber\\
& \left. + C_A \left(\frac{1}{\bfQ_1^2}-  \frac{\bfQ_1}{\bfQ_1^2}\cdot\frac{\bfQ_3}{\bfQ_3^2}\right)\Phi_1  \right\} \, .\label{eq:Pqq_RR}
\end{align}
 $RV$: real splitting, virtual-Glauber interaction (interference of double-Glauber exchange diagram and 0-Glauber diagram).
\begin{align}
\frac{dN^{RV}_{qq}}{dx d^{2-2\epsilon}\bfp} =&  \frac{g^2 C_F}{(2\pi)^3} P_{qq, \epsilon}(x) \int_0^\infty dz^+ \rho_T^-(z^+) \int \frac{d^{2-2\epsilon}\bfq}{(2\pi)^{2-2\epsilon}} \frac{1}{C_F}\frac{d\sigma_{FT}}{d^{2-2\epsilon}\bfq} \int \frac{d^{2-2\epsilon}\bfk}{(2\pi)^{-2\epsilon}}  N_q(\bfp+\bfk)\nonumber\\
&  \times \left\{-C_F\frac{1}{\bfQ_2^2} - C_A\frac{\bfQ_2}{\bfQ_2^2}\cdot\frac{\bfQ_4}{\bfQ_4^2} \Phi_{2,4} -  C_A \left(\frac{1}{\bfQ_2^2}-\frac{\bfQ_2}{\bfQ_2^2}\cdot\frac{\bfQ_4}{\bfQ_4^2}\right)\Phi_2 \right\} \, .
\label{eq:Pqq_RV}
\end{align}
 $VR$: loop splitting, real-Glauber interaction.
\begin{align}
\frac{dN^{VR}_{qq}}{dx d^{2-2\epsilon}\bfp}  =& \delta(1-x)  \int_0^\infty dz^+\rho_T^-(z^+)\int \frac{d^{2-2\epsilon}\bfq}{(2\pi)^{2-2\epsilon}}\frac{1}{C_F}\frac{d\sigma_{FT}}{d^{2-2\epsilon}\bfq} N_q(\bfp-\bfq)  \nonumber\\
& \times \int_0^1 dx' \frac{g^2 C_F}{(2\pi)^3}P_{qq, \epsilon}(x')\int \frac{d^{2-2\epsilon}\bfk}{(2\pi)^{-2\epsilon}}\left\{-2C_F \frac{1}{{\bfQ_5'}^2}\Phi_5' -  C_F\frac{1}{\bfk^2} \right.\nonumber\\
&\left.+(2C_F-C_A)\frac{\bfQ_5'}{{\bfQ_5'}^2}\cdot \frac{\bfQ_6'}{{\bfQ_6'}^2} \Phi_5'+ C_A\frac{\bfQ_6'}{{\bfQ_6'}^2}\cdot \frac{\bfQ_7'}{{\bfQ_7'}^2} \Phi_7' \right\} \, .\label{eq:Pqq_VR}
\end{align}
$VV$: loop correction, virtual-Glauber interaction.
In the $VR$ and $VV$ expressions, $\bfQ_n'$ are obtained by replacing $x$ by $x'$ in $\bfQ_n$.

\begin{align}
\frac{dN^{VV}_{qq}}{dx d^{2-2\epsilon}\bfp} =& \delta(1-x) N_q(\bfp) \int_0^\infty dz^+\rho_T^-(z^+)\int \frac{d^{2-2\epsilon}\bfq}{(2\pi)^{2-2\epsilon}}\frac{1}{C_F}\frac{d\sigma_{FT}}{d^{2-2\epsilon}\bfq}  \int_0^1 dx' \frac{g^2 C_F}{(2\pi)^3}P_{qq}(x') \nonumber\\
& \times \int \frac{d^{2-2\epsilon}\bfk}{(2\pi)^{-2\epsilon}} \left\{ C_A \frac{\bfQ_6'}{{\bfQ_6'}^2}\cdot\frac{\bfQ'_8}{{\bfQ_8'}^2} \Phi_8' - C_A\frac{1}{{\bfQ_6'}^2 } \Phi_6'  + C_F  \frac{1}{\bfk^2}   \right\}  \, .
\label{eq:Pqq_VV}
\end{align}

The contributions to the gluon splitting functions can be similarly written down.

\section{Medium-Modified Energy Correlator at NLO in Exclusive Jets}\label{sec:medium:exlcusive}

With the medium-modified QCD splitting functions, we first perform the NLO calculation of EEC at first order in opacity. Then, we will use dimensional regularization to extract the leading logarithmic divergence of EEC, and derive the medium-modified RG evolution equation. This Section will only summarize the key results that lead to the conclusion, for details, please refer to the Appendix.

\subsection{Non-Contact Energy-Energy Correlator at First Order in Opacity}
In the TMD splitting function, we have been defining the transverse momentum of a daughter parton relative to the direction of the mother parton before scattering and the splitting. Therefore, depending on whether the EEC comes from $RR$ or $RV$ type of contribution, the definition of the angle is different. In the process of $i\rightarrow j+k$ under the small-angle approximation $\theta_{jk} \ll 1$, 
\begin{align}
\label{eq:RR-angle}
\theta^{RR}_{jk} &= \left|\vec{\theta}_j-\vec{\theta}_k\right| =  \frac{|\bfk-(1-x)\bfq|}{x_jx_kE_i},~~\textrm{RR: real emission with real collision}\,,
\end{align}
and
\begin{align}
\label{eq:RV-angle}
\theta^{RV}_{jk} &= \left|\vec{\theta}_j-\vec{\theta}_k\right| =  \frac{|\bfk|}{x_jx_kE_i},~~\textrm{RV: real emission with virtual collision}\,.
\end{align}
Here $x_j = x$, $x_k = (1-x)$ are the light-cone momentum fraction of the two daughter partons, 
$\bfk$ is always understood as the transverse momentum of the parton $j$ relative to the mother parton $i$ and $\bfq$ is the net momentum transfer from medium to the splitting system at opacity one. Note that for $RV$ contribution, this is zero as imposed by the $\delta^{(2)}(\bfq)$, so it is removed from $\theta_{jk}^{RV}$.
With these definitions of angles, we can write down the general formula for the non-contact EEC at opacity one
\begin{align}
\label{eq:non-contact EEC}
& \frac{d\Sigma^i_{\rm non-contact}}{d\theta^2} = \sum_{(jk)}
\int_0^\infty dz^+ \int_0^1 dx \int d^2\bfk \int d^2\bfq \nonumber \\
& \, \, \times x_j x_k \left[ \delta\left(\bft^2-\left(\theta^{RR}_{jk}\right)^2\right) \frac{dN^{RR}_{i\rightarrow jk}}{dz^+ dx d^2\bfk d^2\bfq} +  \delta\left(\bft^2-\left(\theta^{RV}_{jk}\right)^2\right) \delta^{(2)}(\bfq)\frac{dN^{RV}_{i\rightarrow jk}}{dz^+ dx d^2\bfk}
\right] \, ,
\end{align}
where $i=q$, the summation goes over $(jk)=(qg)$ and for $i=g$, the summation goes over $(jk)=(gg),(q\bar{q})$ for a given number of active flavors. Note that there only $RR$ and $RV$ contributions, because the other two $VR$ and $VV$ do not correspond to a real perturbative splitting. The latter will play a role when we consider contact EEC contributions.

The needed formula for opacity-one real-emission function for all three channels $q\rightarrow q+g$, $g\rightarrow g+g$ and $g\rightarrow q+\bar{q}$, separated into $RR$ and $RV$, are listed below for completeness
\textbullet \ \  $q\rightarrow q+g$: real emission with real collision
\begin{align}
\label{eq:TMD-RR}
 \frac{dN^{RR}_{q\rightarrow q+g}}{dz^+ dx d^{2-2\epsilon}\bfk d^{2-2\epsilon}\bfq} 
= &\frac{g_s^2 C_F}{(2\pi)^{3-2\epsilon}} P_{qq,\epsilon}(x) \frac{1}{(2\pi)^{2-2\epsilon}} \frac{\sum_T\rho_T^-(z^+) g_s^2 C_T/d_A}{(\bfq^2+m_{\rm eff}^2)^2} \nonumber\\
&\times g_s^2 \left\{
\frac{C_F}{[\bfk-(1-x)\bfq]^2}\right.\nonumber\\
&  + (2C_F-C_A)\left[\frac{1}{\bfk^2}-\frac{\bfk\cdot(\bfk-(1-x)\bfq)}{\bfk^2 (\bfk-(1-x)\bfq)^2}\right]\Phi\left(\frac{\bfk^2 z^+}{2x(1-x)P^+}\right) \nonumber 
\\ 
&+ C_A\left[\frac{1}{(\bfk-\bfq)^2}-\frac{(\bfk-\bfq)\cdot(\bfk-(1-x)\bfq)}{(\bfk-\bfq)^2 (\bfk-(1-x)\bfq)^2}\right]\Phi\left(\frac{(\bfk-\bfq)^2z^+}{2x(1-x)P^+}\right) \nonumber\\
&+ C_A\left[\frac{1}{(\bfk-\bfq)^2}-\frac{\bfk\cdot(\bfk-\bfq)}{\bfk^2 (\bfk-\bfq)^2}\right]\Phi\left(\frac{(\bfk-\bfq)^2z^+}{2x(1-x)P^+}\right) \nonumber\\
& + C_A\left[\frac{1}{\bfk^2}-\frac{(\bfk-\bfq)\cdot \bfk}{ (\bfk-\bfq)^2 \bfk^2}\right]\Phi\left(\frac{\bfk^2 z^+}{2x(1-x)P^+}\right) \nonumber\\
&+\left.C_A \frac{\bfk\cdot(\bfk-\bfq)}{\bfk^2(\bfk-\bfq)^2} \Phi\left(\frac{[\bfk^2-(\bfk-\bfq)^2]z^+}{2x(1-x)P^+}\right)\right\}.
\end{align}
\textbullet \ \  $q\rightarrow q+g$, real emission with virtual collisions
\begin{align}
\label{eq:TMD-RV}
 \frac{dN^{RV}_{q\rightarrow q+g}}{dz^+ dx d^{2-2\epsilon}\bfk} 
= &\frac{g_s^2 C_F}{(2\pi)^{3-2\epsilon}} P_{qq,\epsilon}(x) \int \frac{d^{2-2\epsilon}\bfq}{(2\pi)^{2-2\epsilon}} \frac{ \sum_T \rho_T^-(z^+) g_s^2 C_T/d_A}{(\bfq^2+m_{\rm eff}^2)^2} \nonumber\\
& \times g_s^2\left\{
- \frac{C_F}{\bfk^2}-C_A \frac{\bfk\cdot\left( \bfk-\bfq\right)}{\bfk^2\left(\bfk-\bfq\right)^2} \Phi\left(\frac{(\bfk^2-(\bfk-\bfq)^2) z^+}{2x(1-x)P^+}\right) \right. \nonumber\\
&\left. -C_A\left( \frac{1}{\bfk^2}-\frac{\bfk\cdot(\bfk-\bfq)}{\bfk^2(\bfk-\bfq)^2}\right)\Phi\left(\frac{\bfk^2z^+}{2x(1-x)P^+}\right)\right\}\,.
\end{align}
\textbullet \ \  $g\rightarrow g+g$: real emission with real collisions
\begin{align}
\frac{dN^{RR}_{g\rightarrow g+g}}{dz^+ dx d^{2-2\epsilon}\bfk d^{2-2\epsilon}\bfq} 
= &\frac{g_s^2 C_A}{(2\pi)^{3-2\epsilon}} P_{gg}(x) \frac{1}{(2\pi)^{2-2\epsilon}} \frac{\sum_T\rho_T^-(z^+) g_s^2 C_T/d_A}{(\bfq^2+m_{\rm eff}^2)^2} \nonumber\\
& \times  g_s^2 \left\{
\frac{C_A}{(\bfk-x\bfq)^2} \right.\nonumber\\
&+ \frac{2C_A}{(\bfk-\bfq)^2}\Phi\left(\frac{(\bfk-\bfq)^2z^+}{2x(1-x)P^+}\right)
+\frac{2C_A}{\bfk^2} \Phi\left(\frac{\bfk^2z^+}{2x(1-x)P^+}\right) \nonumber\\
&
-C_A\frac{\bfk-x\bfq}{(\bfk-x\bfq)^2}\cdot\frac{\bfk}{\bfk^2}\Phi\left(\frac{\bfk^2z^+}{2x(1-x)P^+}\right) \nonumber\\
& -C_A\frac{\bfk-x\bfq}{(\bfk-x\bfq)^2}\cdot\frac{\bfk-\bfq}{(\bfk-\bfq)^2}\Phi\left(\frac{(\bfk-\bfq)^2z^+}{2x(1-x)P^+}\right) \nonumber\\
&  
- C_A\frac{\bfk}{\bfk^2}\cdot\frac{\bfk-\bfq}{(\bfk-\bfq)^2}\Phi\left(\frac{\bfk^2z^+}{2x(1-x)P^+}\right) \nonumber\\
& 
- C_A\frac{\bfk}{\bfk^2}\cdot\frac{\bfk-\bfq}{(\bfk-\bfq)^2}\Phi\left(\frac{(\bfk-\bfq)^2z^+}{2x(1-x)P^+}\right)\nonumber\\
&\left.
+ C_A\frac{\bfk}{\bfk^2}\cdot\frac{\bfk-\bfq}{(\bfk-\bfq)^2}\Phi\left(\frac{(\bfk^2-(\bfk-\bfq)^2)z^+}{2x(1-x)P^+}\right)
\right\}.
\end{align}
\textbullet \ \  $g\rightarrow g+g$: real emission with virtual collisions
\begin{align}
 \frac{dN^{RV}_{g\rightarrow g+g}}{dz^+ dx d^{2-2\epsilon}\bfk} 
= &\frac{g_s^2 C_A}{(2\pi)^{3-2\epsilon}} P_{gg}(x) \int \frac{d^{2-2\epsilon}\bfq}{(2\pi)^{2-2\epsilon}} \frac{ \sum_T \rho_T^-(z^+) g_s^2 C_T/d_A}{(\bfq^2+m_{\rm eff}^2)^2} \nonumber\\
& \times g_s^2\left\{
-\frac{C_A}{\bfQ_1^2} - C_A\frac{\bfk}{\bfk^2}\cdot \frac{\bfk-\bfq}{(\bfk-\bfq)^2}\Phi\left(\frac{(\bfk^2-(\bfk-\bfq)^2)z^+}{2x(1-x)P^+}\right)\nonumber\right.\\
&\left.- C_A\left(\frac{1}{\bfk^2}- \frac{\bfk}{\bfk^2}\cdot \frac{\bfk-\bfq}{(\bfk-\bfq)^2}\right)\Phi\left(\frac{\bfk^2 z^+}{2x(1-x)P^+}\right)
\right\}\,,
\end{align}
\textbullet \ \  $g\rightarrow q\bar{q}$: real emission with real collisions
\begin{align}
 \frac{dN^{RR}_{g\rightarrow q+\bar{q}}}{dz^+ dx d^{2-2\epsilon}\bfk d^{2-2\epsilon}\bfq} 
= &\frac{g_s^2 T_F}{(2\pi)^{3-2\epsilon}} P_{qg,\epsilon}(x) \frac{1}{(2\pi)^{2-2\epsilon}} \frac{\sum_T\rho_T^-(z^+) g_s^2 C_T/d_A}{(\bfq^2+m_{\rm eff}^2)^2} \nonumber\\
&  \times g_s^2 \left\{
\frac{C_F}{(\bfk-x\bfq)^2} \right.\nonumber\\
& + \frac{2C_F}{(\bfk-\bfq)^2}\Phi\left(\frac{(\bfk-\bfq)^2z^+}{2x(1-x)P^+}\right)
+\frac{2C_F}{\bfk^2} \Phi\left(\frac{\bfk^2z^+}{2x(1-x)P^+}\right) \nonumber\\
&
-C_A\frac{\bfk-x\bfq}{(\bfk-x\bfq)^2}\cdot\frac{\bfk}{\bfk^2}\Phi\left(\frac{\bfk^2z^+}{2x(1-x)P^+}\right) \nonumber\\
&
-C_A\frac{\bfk-x\bfq}{(\bfk-x\bfq)^2}\cdot\frac{\bfk-\bfq}{(\bfk-\bfq)^2}\Phi\left(\frac{(\bfk-\bfq)^2z^+}{2x(1-x)P^+}\right) \nonumber\\
&  
- (2C_F-C_A)\frac{\bfk}{\bfk^2}\cdot\frac{\bfk-\bfq}{(\bfk-\bfq)^2}\Phi\left(\frac{\bfk^2z^+}{2x(1-x)P^+}\right) \nonumber\\
&
- (2C_F-C_A)\frac{\bfk}{\bfk^2}\cdot\frac{\bfk-\bfq}{(\bfk-\bfq)^2}\Phi\left(\frac{(\bfk-\bfq)^2z^+}{2x(1-x)P^+}\right)\nonumber\\
&\left.
+ (2C_F-C_A)\frac{\bfk}{\bfk^2}\cdot\frac{\bfk-\bfq}{(\bfk-\bfq)^2}\Phi\left(\frac{(\bfk^2-(\bfk-\bfq)^2)z^+}{2x(1-x)P^+}\right)
\right\}.
\end{align}
\textbullet \ \  $g\rightarrow q\bar{q}$: real emission with virtual collisions
\begin{align}
\frac{dN^{RV}_{g\rightarrow q+\bar{q}}}{dz^+ dx d^{2-2\epsilon}\bfk} 
= &\frac{g_s^2 T_F}{(2\pi)^{3-2\epsilon}} P_{qg}(x,\epsilon) \int \frac{d^{2-2\epsilon}\bfq}{(2\pi)^{2-2\epsilon}} \frac{ \sum_T \rho_T^-(z^+) g_s^2 C_T/d_A}{(\bfq^2+m_{\rm eff}^2)^2} \nonumber\\
& \times g_s^2\left\{
-\frac{C_F}{\bfQ_1^2} - (2C_F-C_A)\frac{\bfk}{\bfk^2}\cdot \frac{\bfk-\bfq}{(\bfk-\bfq)^2}\Phi\left(\frac{(\bfk^2-(\bfk-\bfq)^2)z^+}{2x(1-x)P^+}\right)\nonumber\right.\\
&\left.- (2C_F-C_A)\left(\frac{1}{\bfk^2}- \frac{\bfk}{\bfk^2}\cdot \frac{\bfk-\bfq}{(\bfk-\bfq)^2}\right)\Phi\left(\frac{\bfk^2 z^+}{2x(1-x)P^+}\right)
\right\}\,.
\end{align}


\subsection{The Coulomb Logarithm in the Non-Contact Energy-Energy Correlator}
\label{sec:Coulomb_log}
In the phenomenological analysis, one can directly evaluate the expressions we obtain numerically once a medium density profile function is provided by independent simulations of QGP dynamics, such as hydrodynamic simulations~\cite{SHEN201661}. When using the equations presented in the last sections, the temperature profile along the path length are obtained from an ensemble average of the fluctuating medium path lengths $L$, by randomly sampling the initial production point according to the density of binary collision and azimuthal angle of jet propagation in the transverse plain of the interaction region. These details will be provided in subsection~\ref{sec:results-L-avg}.

Typically, such numerical evaluations require the introduction of a phase-space regulator. Here, we take the opportunity to examine this issue analytically in order to clarify the nature of the regulator: (1) which divergence it regulates and the corresponding logarithmic enhancement, and (2) whether a re-summation of this logarithm is required at the present order of the theory. 
From a method-of-regions analysis, it turns out that the logarithm originates from the Coulomb logarithm of the Glauber interaction. One only needs to introduce the IR regulator $m_{\rm eff}$ for the Glauber gluon propagator to ensure finiteness of the result. 
It is well known that in plasma physics, the re-summation of the Coulomb logarithm leads to the familiar plasma screening mass, but it does not generate the scale evolution of the EEC exclusive jet function, which instead arises from the collinear logarithm of radiation. Therefore, in our numerical evaluation we regulate this divergence using the LO QCD Debye mass $m_{\rm eff} = m_D$. 

The proof using the method of region method is not given for an arbitrary medium, but a simplified exponential medium profile
\begin{align}
\int_0^\infty \rho^-(z^+) \Phi\left(\frac{\bfQ_n^2 z^+}{2x(1-x)P^+}\right) dz^+  =& \int_0^\infty \rho_0^-e^{-z^+/L^+} \left[ 1-\cos\left(\frac{\bfQ_n^2 z^+}{2x(1-x)P^+}\right) \right] dz^+  \nonumber\\
=& \rho_0^-L^+\frac{\left(\bfQ_n^2\right)^2}{(\bfQ_n^2+i Q_{\rm LPM}^2)(\bfQ_n^2-i Q_{\rm LPM}^2)} \, , 
\end{align}
where $Q_{\rm LPM}= \sqrt{2x(1-x)P^+/L^+}$. This way, the LPM phase is just like two propagators in the calculation and we have plenty of techniques to manipulate them. From the method of region analysis of the integral, we can also identify the complete Coulomb log as $\ln\frac{P^+/L^+}{m_{\rm eff}^2}$. This phase-space region of this Coulomb logarithm is shown in Figure~\ref{fig:hierachy-of-scales}. Thus, we can identify that for exponential medium, the non-contact medium-induced EEC is
\begin{align}
& \frac{d\Sigma^{RR+RV}_{\rm non~contact}}{d\theta^2}\left(\theta\ll \theta_{\rm LPM}; \textrm{~in an exponential medium}\right) \nonumber\\
= &\frac{\left[\alpha_s(\mu^2)\right]^2C_F\rho_0 L^3}{32\pi}\Big\{\frac{3C_F+16C_A}{15}\ln\frac{p_T/L}{m_{\rm eff}^2} + \textrm{non-log enhanced terms}
\Big\}\, .
\end{align}

\begin{figure}
\centering
    \includegraphics[scale=0.8]{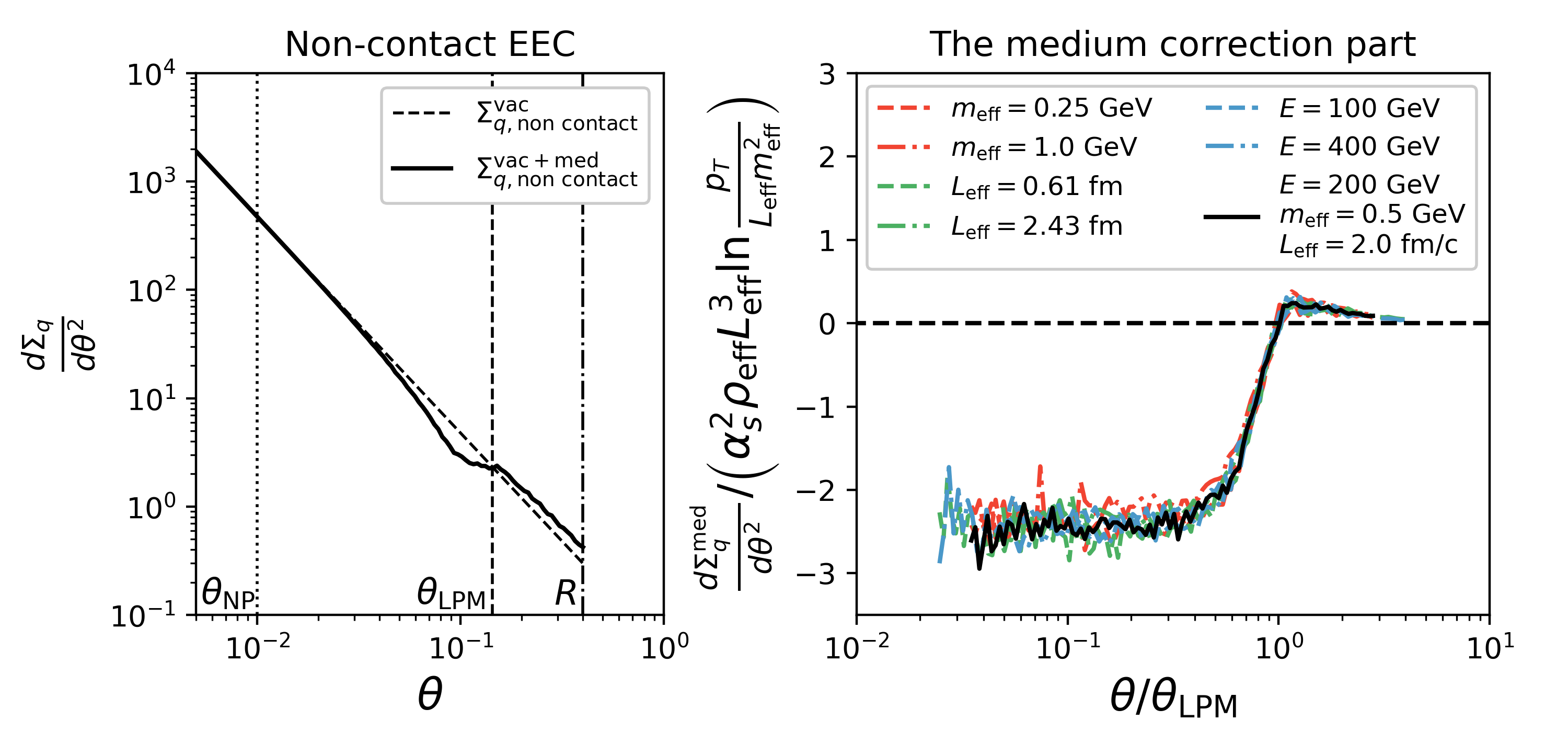}
    \caption{
    Left: numerical results for the non-contact contribution to the energy--energy correlator in the vacuum (black dashed line) and with medium correction at first order in opacity (black solid line) for $p_T = 200~\mathrm{GeV}$. The ``brick" medium has a constant temperature $T=0.4$~GeV with maximum length $2$~fm and the 
    medium-modified EEC shows the transition across the characteristic angular scales $\theta_{\mathrm{LPM}}$, which is defined in Eq.~(\ref{eq:define-LPM-scale-and-angle}).
    Right: only plotting the medium-induced non-contact component (RR + RV channels) normalized by $\alpha_s^2 \rho_{\mathrm{eff}}L_{\rm eff}^3 \ln \frac{p_T}{L_{\rm eff} m_{\rm eff}^2}$, as motivated by the analytic calculation of exponentially decaying medium. Different lines correspond to the variations of the jet $p_T$,  the effective path length, and the screening mass of the brick medium. They all collapse to a universal curve, corroborating the leading power dependence on medium parameters and  the presence of the Coulomb logarithm $\ln \frac{p_T}{L_{\rm eff}^+ m_{\rm eff}^2}$.}
    \label{fig:NonContactEEC}
\end{figure}

Even though this is here analytically demonstrated for an exponentially decaying medium. We can check the robustness of the Coulomb logarithm for other types of medium numerically. For example, in Figure~\ref{fig:NonContactEEC}, we numerically evaluate $d\Sigma^{RR+RV}/d\theta^2$ for $q\rightarrow q+g$ in a brick medium of constant temperature $T_0=0.4$ GeV with a fixed length. On the left panel, we combine it with the vacuum non-contact EEC to show the general form of medium correction. The opacity-one non-contact correction is an enhancement at angles larger than $\theta_{\rm LPM}$, while a depletion for $\theta<\theta_{\rm LPM}$. On the right panel, we only plot the medium-induced part and varies the jet $p_T$, effective path length $L_{\rm eff}$ and the effective screening mass $m_{\rm eff}$ by a factor of 4. The plotted results are rescaled by a form motivated by the method-of-region calculation, i.e., $\alpha_s^2 C_F \rho_{\rm eff} L_{\rm eff}^3 \ln\frac{p_T}{L_{\rm eff} m_{\rm eff}^2}$, while the angles are scaled by $\theta_{\rm LPM}$. It is clear that all lines collapse for $\theta<\theta_{\rm LPM}$. Indicating the form of the Coulomb logarithm is indeed also present for the brick medium. 

The appearance of a Coulomb logarithm at this order suggests that, once higher-order corrections are included, the phase space associated with soft-gluon emission induced by Glauber exchange may become Reggeized, leading to additional evolution effects in the calculation of the non-contact medium-induced EEC. We leave a detailed investigation of this possibility to future work.

\subsection{Contact Energy-Energy Correlator at First Order in Opacity}
The contribution to contact EEC is uses quadratic energy weighting on each of the daughter parton, involving all four cases (RR, RV, VR, VV). The general formula is
\begin{align}
& \hspace*{-1cm}\frac{d\Sigma^i_{\rm contact}}{d\theta^2}(\theta, E) \nonumber\\
& = \sum_{(jk)}
\int_0^\infty dz^+ \int_0^1 dx \int d^2\bfk \int d^2\bfq \nonumber \\
&\times \left\{ \left(\frac{d\Sigma^j}{d\theta^2}(\theta, x_j E) + \frac{d\Sigma^k}{d\theta^2}(\theta, x_k E)\right)  \left[\frac{dN^{RR}_{i\rightarrow jk}}{dz^+ dx d^2\bfk d^2\bfq} + \delta^{(2)}(\bfq)\frac{dN^{RV}_{i\rightarrow jk}}{dz^+ dx d^2\bfk}
\right] \right. \nonumber\\
& \left. + \frac{d\Sigma^i}{d\theta^2}(\theta, E)\delta(1-x)\left[\delta^{(2)}(\bfk-\bfq) \frac{dN^{VR}_{i\rightarrow i}}{dz^+ d^2\bfq}  + \delta^{(2)}(\bfq)\delta^{(2)}(\bfk)\frac{dN^{VV}_{i\rightarrow i}}{dz^+}\right]
\right\}\,.
\end{align}
Remember that for the case where the RR and RV channel correspond to the splitting $g\rightarrow q+\bar{q}$, we define the VR and VV channel to be the $g\rightarrow g$ with an intermediate quark loops. Again, using the tree-level scaling law of EEC $\frac{d\Sigma}{d\theta^2}(\theta, x E) \approx x^2 \frac{d\Sigma}{d\theta^2}(\theta, E)$, we have the expression in the asymptotic region
\begin{align}
& \hspace*{-1cm}\frac{d\Sigma^i_{\rm contact}}{d\theta^2}(\theta, E) \nonumber\\
&=  \sum_{(jk)}
\int_0^\infty dz^+ \int_0^1 dx \int d^2\bfk \int d^2\bfq \nonumber \\
&\times \left\{ \left(x_j^2\frac{d\Sigma^j}{d\theta^2}(\theta, E) + x_k^2\frac{d\Sigma^k}{d\theta^2}(\theta, E)\right)  \left[\frac{dN^{RR}_{i\rightarrow jk}}{dz^+ dx d^2\bfk d^2\bfq} + \delta^{(2)}(\bfq)\frac{dN^{RV}_{i\rightarrow jk}}{dz^+ dx d^2\bfk}
\right] \right. \nonumber\\
& \left. + \frac{d\Sigma^i}{d\theta^2}(\theta, E)\delta(1-x)\left[\delta^{(2)}(\bfk-\bfq) \frac{dN^{VR}_{i\rightarrow i}}{dz^+ d^2\bfq}  + \delta^{(2)}(\bfq)\delta^{(2)}(\bfk)\frac{dN^{VV}_{i\rightarrow i}}{dz^+}\right]
\right\} \, .
\end{align}

In the following, we will demonstrator the calculate for $\Sigma^q$ due to the induced splittings $q\rightarrow q+g$, while all the other channels are obtained in a similar way.
First, we use the sum rules of the TMD $q\rightarrow q+g$ splittings at opacity one to reduce the general formula to only dependent on the RR and RV parts.
The flavor sum rules states that~\footnote{This resum rule is also verified explicitly in Ref.~\cite{Ke:2024ytw}, up to power-suppressed difference.}
\begin{align}
0 = &
\int_0^\infty dz^+ \int_0^1 dx \int d^2\bfk \int d^2\bfq \left\{ \frac{dN^{RR}_{q\rightarrow qg}}{dz^+ dx d^2\bfk d^2\bfq} + \delta^{(2)}(\bfq)\frac{dN^{RV}_{q\rightarrow qg}}{dz^+ dx d^2\bfk}
\right. \nonumber\\
& \left. + \delta(1-x)\delta^{(2)}(\bfk-\bfq) \frac{dN^{VR}_{q\rightarrow q}}{dz^+ dx d^2\bfq}  + \delta(1-x)\delta^{(2)}(\bfq)\delta^{(2)}(\bfk)\frac{dN^{VV}_{q\rightarrow q}}{dz^+}
\right\} \, .
\end{align}
Therefore, the last two terms in the non-contact EEC calculation can be replaced by the negative of the RR and RV, thus
\begin{align}
\frac{d\Sigma^q_{\rm contact}}{d\theta^2} =  &
\int_0^\infty dz^+ \int_0^1 dx \int d^2\bfk \int d^2\bfq \nonumber \\
&\times\left\{ \left[x^2 \frac{d\Sigma^q}{d\theta^2}(\theta, E) + (1-x)^2 \frac{d\Sigma^g}{d\theta^2}(\theta, E)\right] \left[\frac{dN^{RR}_{q\rightarrow qg}}{dz^+ dx d^2\bfk d^2\bfq} + \delta^{(2)}(\bfq)\frac{dN^{RV}_{q\rightarrow qg}}{dz^+ dx d^2\bfk}
\right] \nonumber\right.\\
&\quad \quad \left. - \frac{d\Sigma^q}{d\theta^2}(\theta, E)\left[\frac{dN^{RR}_{q\rightarrow qg}}{dz^+ dx d^2\bfk d^2\bfq} - \delta^{(2)}(\bfq)\frac{dN^{RV}_{q\rightarrow qg}}{dz^+ dx d^2\bfk}\right]
\right\}\nonumber\\
 = & \int_0^1 dx \left[x^2 \frac{d\Sigma^q}{d\theta^2}(\theta, E) + (1-x)^2 \frac{d\Sigma^g}{d\theta^2}(\theta, E)\right] \left[\frac{dN^{N=1}_{q\rightarrow q+g}}{dx}\right]_+\,, \\
\frac{dN^{N=1}_{q\rightarrow q+g}}{dx} = &\int_0^\infty dz^+ \int d^2\bfk \int d^2\bfq   \left(\frac{dN^{RR}_{q\rightarrow qg}}{dz^+ dx d^2\bfk d^2\bfq} + \delta^{(2)}(\bfq)\frac{dN^{RV}_{q\rightarrow qg}}{dz^+ dx d^2\bfk}\right).
\end{align}
The technical way to compute the in-meidum collinear real emission function $\frac{dN^{N=1}_{q\rightarrow q+g}}{dx}$ has been laid out in a previous work~\cite{Ke:2023ixa}. The key is to identify a region of scale hierarchy in thin medium that $P^+/L^+\gg m_{\rm eff}^2$. This way, once we shift $\bfk$ integrals under dimensional regularization in the rest of the terms we consider limit $P^+/L^+ \gg m_{\rm eff}^2$, we can study its asymptotic behavior at leading power of $m_{\rm eff}^2/(P^+/L^+)$, where the $\bfq$ integral can be computed directly with $m_{\rm eff}^2=0$.

\begin{align}
\frac{dN^{N=1}_{q\rightarrow q+g}}{dx} = & \frac{g^2 C_F}{(2\pi)^3}  P_{qq,\epsilon}(x) \int \frac{d^{2-2\epsilon}\bfk}{(2\pi)^{-2\epsilon}} \sum_T\int_0^\infty dz^+ \rho_T^-(z^+) \int \frac{d^{2-2\epsilon}\bfq}{(2\pi)^{2-2\epsilon}} \frac{ g^2 C_T/d_A}{(\bfq^2+m_{\rm eff}^2)^2} \nonumber\\
&  \times g^2\left\{
 (2C_F-C_A)\left[\frac{1}{\bfk^2}-\frac{\bfk\cdot(\bfk-(1-x)\bfq)}{\bfk^2 (\bfk-(1-x)\bfq)^2}\right]\Phi\left(\frac{\bfk^2 z^+}{2x(1-x)P^+}\right) \right.\nonumber 
\\ 
&+ C_A\left[\frac{1}{(\bfk-\bfq)^2}-\frac{(\bfk-\bfq)\cdot(\bfk-(1-x)\bfq)}{(\bfk-\bfq)^2 (\bfk-(1-x)\bfq)^2}\right]\Phi\left(\frac{(\bfk-\bfq)^2z^+}{2x(1-x)P^+}\right) \nonumber\\
&\left.+ C_A\left[\frac{1}{(\bfk-\bfq)^2}-\frac{\bfk\cdot(\bfk-\bfq)}{\bfk^2 (\bfk-\bfq)^2}\right]\Phi\left(\frac{(\bfk-\bfq)^2z^+}{2x(1-x)P^+}\right)\right\} \, .
\end{align}
Under dimensional regularization, a lot of the terms can be shown to cancel using a transverse momentum shift. The final result can be shown to be
\begin{align}
\frac{dN^{N=1}_{q\rightarrow q+g}}{dx} = & \frac{g^2 C_F}{(2\pi)^3} P_{qq,\epsilon}(x) \sum_T\int_0^\infty dz^+ \rho_T^-(z^+) g^2 C_T/d_A  \int \frac{d^{2-2\epsilon}\bfk}{(2\pi)^{-2\epsilon}} \frac{1}{\bfk^2} \Phi\left(\frac{\bfk^2 z^+}{2x(1-x)P^+}\right) \nonumber\\
&   \times \int \frac{d^{2-2\epsilon}\bfq}{(2\pi)^{2-2\epsilon}} \frac{g^2C_F }{(\bfq^2)^2} 
 \left\{ (2C_F-C_A)\left[\frac{-(1-x)\bfq\cdot(\bfk-(1-x)\bfq)}{(\bfk-(1-x)\bfq)^2}\right]\right. \nonumber \\ 
& \left. + C_A\left[\frac{x\bfq\cdot(\bfk+x\bfq)}{ (\bfk+x\bfq)^2}\right] + C_A\left[\frac{\bfq\cdot(\bfk+\bfq)}{ (\bfk+\bfq)^2}\right] \right\} \nonumber\\
=& F^{q\rightarrow qg}_{\epsilon}(x) \times \kappa_\epsilon^{\rm med} \times T_\epsilon\left(\frac{\mu^2}{2P^+/L_{\rm eff}^+}\right)\,.
\end{align}
Therefore, in the limit $p_T^2\gg P^+/L^+ \gg m_{\rm eff}^2$, we have factorize the collinear splitting function into an $x$-dependent process specific longitudinal emission function $F_\epsilon^{q\rightarrow q+g}$, a medium opacity integral $\kappa_\epsilon^{\rm med}$, and a transverse momentum integral $T_\epsilon$. Their expressions are 
\begin{align}
F^{q\rightarrow qg}_{\epsilon}(x) =& C_F P_{qq,\epsilon}(x) \frac{1}{[x(1-x)]^{1+2\epsilon}} \left[(1-x)^{2+2\epsilon}(2C_F-C_A) + x^{2+2\epsilon} C_A+ C_A \right] \, , \\
\kappa_\epsilon^{\rm med} =& \frac{1}{2P+/L_{\rm eff}^+}\sum_T \frac{g^2 C_T}{d_A} \int_0^\infty dz^+ \rho_T^-(z^+) \left(\frac{z^+}{L_{\rm eff}^+}\right)^{1+2\epsilon}  \, , \\
T_\epsilon\left(\frac{\mu^2}{2P^+/L_{\rm eff}^+}\right) =& \left[\frac{\mu^2}{2P^+/L_{\rm eff}^+}\right]^{2\epsilon}\frac{\alpha_s(\mu^2)}{2\pi^2}\int \frac{d^{2-2\epsilon}\tilde{\bfk}}{(2\pi)^{-2\epsilon}} \frac{\Phi\left(\tilde{\bfk}^2\right) }{\tilde{\bfk}^2} 
\nonumber\\
& \hspace*{2cm}\times \int \frac{d^{2-2\epsilon}\tilde{\bfq}}{(2\pi)^{-2\epsilon}} \frac{\alpha_s(\mu^2)}{\pi}\frac{1}{(\tilde{\bfq}^2)^2}\frac{\tilde{\bfq}\cdot(\tilde{\bfk}+\tilde{\bfq})}{ (\tilde{\bfk}+\tilde{\bfq})^2} \, ,
\end{align}
where in the last integral, tilded momentum are defined to be $\tilde{\bfp} = \bfp/\sqrt{2x(1-x)P^+/z^+}$ and are dimensionless.

\paragraph{The longitudinal function:} This integral carries all the process-specific information. It is proportional to the vacuum splitting function $P_{qq, \epsilon}(x)$ and the color factor $C_F$ of the radiation vertex. The second factor of $[x(1-x)]^{1+2\epsilon}$ is a result of the LPM effect that come from the interference of radiation from the position of the hard production point to the point of jet-medium scattering. As a result, is to makes the medium-induced collinear splitting function  much softer than its vacuum counter part, or relatively speaking, suppress the emission probability of energetic splittings. Finally the last factor $(1-x)^{2+2\epsilon}(2C_F-C_A) + x^{2+2\epsilon} C_A+ C_A$ can be interpreted as proportional to the effective color factor of the splitting system $q\rightarrow q+g$ as seen by the medium. For splittings involving a soft gluon $x\to 1$, the effective color factor is $ \propto C_A$, while in the flavor conversion limit $x\to 0$, the effective color factor becomes $ \propto C_F$.

\paragraph{The medium opacity integral:} This integral can be thought to measuring the order of magnitude of transverse momentum broadening relative to the semi-hard scale $2P^+/L_{\rm eff}^+$ in a thin medium, i.e., the higher-twist power suppression parameter. 
Here, we have introduced an ``arbitrary'' effective medium length $L_{\rm eff}^+$, which carries the dimension of a length. Consider an $\epsilon$ expansion of the density integral $D_\epsilon$
\begin{align}
\kappa_\epsilon^{\rm med} = \frac{\sum_T g^2 C_T/d_A}{2P^+/L_{\rm eff}^+}  \int_0^\infty dz^+\left[ \frac{z^+}{L_{\rm eff}^+} \rho_T^-(z^+) + 2\epsilon \frac{z^+}{L_{\rm eff}^+}\ln\frac{z^+}{L_{\rm eff}^+} \rho_T^-(z^+) + \mathcal{O}(\epsilon^2)\right] \, .
\end{align}
We can therefore always define the optimized effective medium length such that the linear term in $\epsilon$ vanish, i.e.,
\begin{align}
\label{eq:effective-med-param}
 0=\int_0^\infty dz^+ z^+ \ln\frac{z^+}{L_{\rm eff}^+} \rho_T^-(z^+)\,,\quad
\rho_{\rm eff}^- &\equiv \sum_T g^2 C_T/d_A \int_0^\infty \frac{dz^+}{L_{\rm eff}^+} \frac{z^+}{L_{\rm eff}^+} \rho_T^-(z^+)\,,
\end{align}
with $\rho_{\rm eff}^-$ being the effective density.
This way $\kappa^{\rm med} = \frac{\rho_{\rm eff}^-L_{\rm eff}^+}{2P^+/L_{\rm eff}^+}$. Note that these are general definitions for an arbitrary medium.

\paragraph{The transverse integral:} Note that this transverse integral acquires a very general form, independent of the color factors, and is actually the same for other channels. 
A direct evaluation gives
\begin{align}
T_\epsilon\left(\frac{\mu^2}{2P^+/L_{\rm eff}^+}\right) =& \alpha_s^2(\mu^2)\frac{1}{2} \left[\frac{\mu^2 e^{\gamma_E-1}}{2P^+/L_{\rm eff}^+}\right]^{2\epsilon} \left(1 + \mathcal{O}(\epsilon^2)\right)  \nonumber \\
=&\alpha_s^2(\mu^2)\left[\frac{1}{2}+\epsilon\ln \frac{\mu^2 e^{\gamma_E-1}}{2P^+/L_{\rm eff}^+}+\mathcal{O}(\epsilon^2)\right]\,.
\end{align}

With these definitions, we can summarize the result for all contact EEC's at first order in opacity
\begin{align}
\frac{d\Sigma^q_{\rm contact}}{d\theta^2}  = &
 \int_0^1 dx \left[x^2 \frac{d\Sigma^q}{d\theta^2}(\theta, E) + (1-x)^2 \frac{d\Sigma^g}{d\theta^2}(\theta, E)\right] \left(\frac{dN^{N=1}_{q\rightarrow qg}}{dx}\right)_+\,, \\
 \frac{d\Sigma^g_{\rm contact}}{d\theta^2}  = & \frac{d\Sigma^g}{d\theta^2}(\theta, E)\times \left[ \int_0^1 dx x^2  \frac{2}{x}\left(x\frac{dN^{N=1}_{g\rightarrow gg}}{dx}\right)_+ -2N_f \int_0^1 dx x \frac{dN^{N=1}_{g\rightarrow q\bar{q}}}{dx}\right]\nonumber\\
 &+ \frac{d\Sigma^q}{d\theta^2}(\theta, E)\times 2N_f\int_0^1 dx x^2 \frac{dN^{N=1}_{g\rightarrow q\bar{q}}}{dx}\, ,
\end{align}
where for the case of $g\rightarrow g+g$ and $g\rightarrow q+\bar{q}$, we have used the momentum sum rule to express the VR and VV contributions into real emission function that is incorporated in the plus prescription.
The medium-induced collinear splitting functions are
\begin{align}
\frac{dN^{N=1}_{i\rightarrow jk}}{dx} =&  F^{i\rightarrow jk}_{\epsilon}(x)   \alpha_s^2(\mu^2) \kappa^{\rm med}\left[\frac{1}{2}+\epsilon\ln \frac{\mu^2 e^{\gamma_E-1}}{2P^+/L_{\rm eff}^+}\right](1+\mathcal{O}(\epsilon^2))\,  , 
\end{align}
and for the specific partonic channels we have
\begin{align}
F^{q\rightarrow qg}_{\epsilon}(x) =& \frac{C_F P^q_{qq,\epsilon}(x) }{[x(1-x)]^{1+2\epsilon}} \left[(1-x)^{2+2\epsilon}(2C_F-C_A) + x^{2+2\epsilon} C_A+ C_A \right], \\
F^{g\rightarrow gg}_{\epsilon}(x) =& \frac{C_A P^g_{gg,\epsilon}(x) }{[x(1-x)]^{1+2\epsilon}} \left[(1-x)^{2+2\epsilon}C_A + x^{2+2\epsilon} C_A+ C_A \right], \\
F^{g\rightarrow q\bar{q}}_{\epsilon}(x) =& \frac{T_F P^g_{q\bar{q},\epsilon}(x) }{[x(1-x)]^{1+2\epsilon}} \left[(1-x)^{2+2\epsilon}C_A + x^{2+2\epsilon} C_A+ (2C_F-C_A) \right]\,,
\end{align}
where the last one is defined of one flavor of quark. 
\section{Medium-Modified Energy Correlator: Renormalization and Resummation}
\label{sec:Medium:resummation}

In this Section we study the renormalization of the medium-modified two point energy correlator, which connects the fixed-order calculation to its resumed form. We analyze how interactions with the medium modify the anomalous dimensions that govern the scale evolution of the correlator and, in turn, the structure of logarithmic terms that shape its angular dependence. This provides a systematic framework to quantify the deviation of the in-medium evolution from the vacuum case.

\begin{figure}
    \centering
    \includegraphics[width=0.7\linewidth]{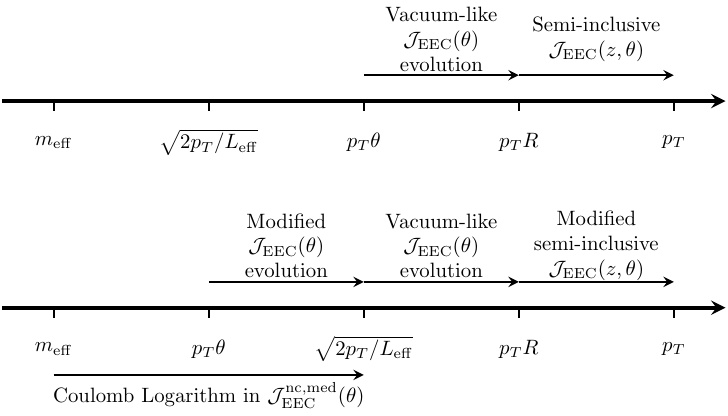}
    \caption{Schematic illustration of the scale hierarchies relevant to the Coulomb logarithm and EEC evolution. The upper panel shows the vacuum case, where the EEC evolution is governed by collinear logarithms of radiation. The lower panel shows the in-medium case, where Glauber interactions introduce an additional Coulomb logarithm regulated by the Debye mass, modifying the evolution of the jet and the EEC. }
    \label{fig:hierachy-of-scales}
\end{figure}

\subsection{Renormalization and Evolution with Medium Correction}
We start from the full expressions for the EEC, including both the vacuum result and opacity-one medium correction, each computed at NLO accuracy. To illustrate the derivation of the renormalization and the evolution effects we focus on the $q\rightarrow q+g$ case where
\begin{align}
\frac{d\Sigma_q}{d\theta^2} =& \frac{d\Sigma_q^{(0)}}{d\theta^2} + \frac{d\Sigma_{q, \textrm{non-contact}}^{\rm vac+med,(1)}}{d\theta^2} + \frac{d\Sigma_{q, \textrm{contact}}^{\rm vac+med,(1)}}{d\theta^2}  \,.
\end{align}
We have already discussed the nature of the medium non-contact term and showed in Section \ref{sec:Coulomb_log} that it is enhanced by a Coulomb logarithm that is screened by the medium Debye mass, not a phase-space logarithm. Therefore, we consider them to be already well defined at the considered level of accuracy and will no longer require renormalization.  

Now shifting attention to the contact term, we insert the expression for the contact vacuum and contact medium pieces
\begin{align}
& \frac{d\Sigma_{q, \rm contact}^{\rm  vac+med,(1)}}{d\theta^2}(\theta, p_T) \nonumber \\
=& \quad\frac{d\Sigma_q^{(0)}}{d\theta^2}(\theta, p_T)\times \left\{\frac{\alpha_s(\mu^2) }{4\pi}\left(\frac{1}{\epsilon}+\ln\frac{\mu^2}{(2p_T)^2\tan^2\frac{R}{2}}\right)\left[-2C_F\int_0^1 dx x^2 [P_{qq,\epsilon}(x)]_+ \right]\right. \nonumber\\
& \quad\quad\quad\quad\quad\quad\left. +\alpha_s^2(\mu^2) \kappa^{\rm med}\left[\frac{1}{2}+\epsilon\ln \frac{\mu^2 e^{\gamma_E-1}}{2P^+/L_{\rm eff}^+}\right] \int_0^1 dx x^2 \left[F^{q\rightarrow qg}_{\epsilon}(x)\right]_+ \right\} \nonumber\\
& + \frac{d\Sigma_g^{(0)}}{d\theta^2}(\theta, p_T)\times \left\{\frac{\alpha_s(\mu^2) }{4\pi}\left(\frac{1}{\epsilon}+\ln\frac{\mu^2}{(2p_T)^2\tan^2\frac{R}{2}}\right)\left[-2C_F\int_0^1 dx x^2 [P_{gq,\epsilon}(x)]_+\right] \right. \nonumber\\
& \quad\quad\quad\quad\quad\quad\left.+\alpha_s^2(\mu^2) \kappa^{\rm med}\left[\frac{1}{2}+\epsilon\ln \frac{\mu^2 e^{\gamma_E-1}}{2P^+/L_{\rm eff}^+}\right] \int_0^1 dx  (1-x)^2 \left[F^{q\rightarrow qg}_{\epsilon}(x)\right]_+ \right\} \,.
\end{align}
It is easy to see that the LPM effect, which accounts for the medium-induced splitting function being softer than the vacuum one, will lead to a $1/\epsilon$ pole. 
\begin{align}
f_{qq}^{\rm med}(N)=&\int_0^1 dx x^{N-1} \left[F^{q\rightarrow qg}_{\epsilon}(x)\right]_+ \nonumber \\
=& \frac{2(N-1)C_FC_A+C_F^2}{\epsilon} +  C_F(2C_F-C_A)\frac{-N^2+2N+1}{N^2-1}  \nonumber\\
& + C_FC_A \Big\{ 3+4N(\gamma_E-1)+2(N-2)\left[2\psi_0(N-2)-\psi_0(N-1)\right]\nonumber\\
& +N\left[3\psi_0(N)-\psi_0(N+1)\right]+(N+2)\psi_0(N+2)\Big\}
 + \mathcal{O}(\epsilon)\,, \\
f_{gq}^{\rm med}(N) = &\int_0^1 dx (1-x)^2 \left[F^{q\rightarrow qg}_{\epsilon}(x)\right]_+ \nonumber\\
=& -\frac{C_F^2}{\epsilon} + C_F\left\{C_A\frac{N^2+N+6}{(N+1)N(N-1)(N-2)}\right.\nonumber\\
&+C_A\left[\frac{1}{2}-\gamma_E+\frac{1}{(N-1)(N-2)}-\psi_0(N-2)\right]\nonumber\\
& \left.+(2C_F-C_A)\left[\frac{1}{2}-\gamma_E+\frac{1}{(N+1)N}-\psi_0(N)\right]\right\} + \mathcal{O}(\epsilon)\,,
\end{align}
\begin{align}
f_{gg}^{\rm med}(N) =&\int_0^1 dx x^2 \frac{2}{x}\left[xF^{g\rightarrow gg}_{\epsilon}(x)\right]_+ - 2N_f\int_0^1 dx x F^{g\rightarrow q\bar{q}}_{\epsilon}(x) \nonumber\\
=& \frac{2C_A^2(N-1)+2N_f T_F C_F}{\epsilon} 
+C_A^2\bigg\{\frac{N^5+6N^4-5N^3+6N^2+4N+96}{3N^5-15N^3+12N}\nonumber\\
& +(N-3)\left[-1+\gamma_E+2\psi_0(N-3)-\psi_0(N-2)\right] \nonumber\\
& +(N-1)\left[-1+\gamma_E+\psi_0(N-1)\right] \nonumber\\
&+(N+1)\left[-1+\gamma_E+2\psi_0(N+1)-\psi_0(N+2)\right] \nonumber\\
&+(N+3)\left[-1+\gamma_E+\psi_0(N+3)\right]\bigg\}\nonumber\\
& + 2N_f T_F\frac{2C_A+6C_F}{3}+\mathcal{O}(\epsilon)\,,\\
f_{qg}^{\rm med}(N) = & 2N_f \int_0^1 dx x^2 F^{g\rightarrow q\bar{q}}_{\epsilon}(x) \nonumber\\
=& -\frac{2N_f T_F C_F}{\epsilon} +2N_f T_F\left\{-C_F+C_A\frac{N^2-N+6}{(N+2)(N+1)N(N-1)}\right.\nonumber\\
&+(2C_F-C_A)\left[\frac{1}{2}-\gamma_E+\frac{1}{N(N-1)}-\psi_0(N+1)\right]\nonumber\\
&\left.+C_A\left[\frac{1}{2}-\gamma_E+\frac{1}{(N+2)(N+1)}-\psi_0(N+3)\right]\right\} + \mathcal{O}(\epsilon)\,.
\end{align}

From the $\overline{\rm MS}$ renoramlization with the medium correction, one can extract the modified EEC anomalous dimension with opacity-one medium correction included,
\begin{align}
\gamma_{ji}(N) = \begin{cases}
\gamma_{ji}^{\rm vac}(N) +  \Delta\gamma_{ji}^{\rm med}(N), & \Lambda_{\rm QCD}/p_T\ll \theta \ll \theta_{\rm LPM} \,,\\
\gamma_{ji}^{\rm vac}(N) , & \theta_{\rm LPM} \ll \theta \ll R\,.
\end{cases}
\end{align}
$\Delta\gamma_{ji}^{\rm med}(N)$ can be read off from the pole part of the $f_{ji}^{\rm med}$ function
\begin{align}
\label{eq:Delta_gamma_med}
\Delta\gamma_{qq}^{\rm med}(N) &= w_{\rm med} \times\left(2(N-1)C_FC_A+C_F^2\right)\, , \nonumber\\
\Delta\gamma_{gq}^{\rm med}(N) &= w_{\rm med} \times \left(-C_F^2\right) \, , \nonumber\\
\Delta\gamma_{gg}^{\rm med}(N) &= w_{\rm med}\times\left(2(N-1)C_A^2+2N_fT_FC_F\right)\, , \nonumber\\
\Delta\gamma_{qg}^{\rm N=1}(N) &= w_{\rm med} \times \left(-2N_f T_F C_F\right) \, ,
\end{align}
with the corrections all proportional to the small parameter
\begin{align}
w_{\rm med}\equiv 4\pi\alpha_s(\mu^2)\kappa^{\rm med} =  \frac{4\pi \alpha_s(\mu^2)\rho_{\rm eff}L_{\rm eff}}{2p_T/L_{\rm eff}} \ll 1 \, .
\end{align}
These are the key results regarding the scale evolution of EEC. $w_{\rm med}$ can be interpreted as the typical magnitude of monument broadening divided by the semi-hard scale due to LPM effect. The smallness of the parameter $w_{\rm med}$ is the indicator of whether we can consider the medium effect as a perturbative to the vacuum evolution, i.e., the validity region of the medium-modified EEC scale evolution is $w\ll 1$. In applications to small systems like $p$-Pb and O-O, we will see that this is very well satisfied. 

It is interesting to note that only the diagonal components linearly depend on $N$. This can be thought of as a direct consequence of the form of the opacity-one medium-induced splitting function.

For in-medium evolution, there are two possibilities as illustrated in Figure~\ref{fig:hierachy-of-scales}. First, $M_{\rm LPM} \ll p_T\theta \ll p_TR < p_T$. The medium-induced radiation does not lead to log-enhanced correction. One directly calculate the vacuum+medium non-contact term as initial condition, and then perform vacuum resummation from $p_T\theta$ to $p_T$. Second, $p_T\theta \ll M_{\rm LPM} \ll  p_TR < p_T$. The evolution is two-staged. First, construct vacuum at scale $p_T\theta$; then, perform vacuum+medium modified LL evolution from $p_T\theta$ to $M_{\rm LPM}$. At the scale $M_{\rm LPM}$, include medium fixed-order correction from contact and non-contact term. Then, perform vacuum evolution from $M_{\rm LPM}$ to $p_T R$.

This means that for angles parametrically smaller than $\theta_{\rm LPM}$
we first initialize the EEC at scale $\mu =\theta p_T$ with $\Sigma_{\rm vac}$, then perform the medium modified evolution from $\mu=\theta p_T$ to $\mu=M_{\rm LPM}$
\begin{align}
\frac{\partial}{\partial\ln\mu^2}\begin{bmatrix}
\mathcal{J}_{\rm EEC, q}^{\rm vac}(\theta; p_T, R, \mu) \\
\mathcal{J}_{\rm EEC, g}^{\rm vac}(\theta; p_T, R, \mu)
\end{bmatrix}
= 
- \frac{\alpha_s(\mu^2)}{4\pi}\begin{bmatrix}
\gamma_{qq}^{\rm vac} + \Delta\gamma_{qq}^{\rm med} & \gamma_{gq}^{\rm vac} + \Delta\gamma_{gq}^{\rm med} \\
\gamma_{qg}^{\rm vac} + \Delta\gamma_{qg}^{\rm med} & \gamma_{gg}^{\rm vac} + \Delta\gamma_{gg}^{\rm med}
\end{bmatrix}  
\begin{bmatrix}
\mathcal{J}_{\rm EEC, q}^{\rm vac}(\theta; p_T, R, \mu) \\
\mathcal{J}_{\rm EEC, g}^{\rm vac}(\theta; p_T, R, \mu)
\end{bmatrix}
\, .
\end{align}
Then at the scale $M_{\rm LPM}$, we include the medium non-contact correction $\mathcal{J}'(\mu=M_{\rm LPM}) = \mathcal{J}(\mu=M_{\rm LPM}) + \mathcal{J}_{\rm med}^{\rm contact}$. Then using $J'$ as the initial condition, we perform the vacuum-like evolution equation from scale $M_{\rm LPM}$ to the jet scale $\mu=ER$.
\begin{align}
\frac{\partial}{\partial\ln\mu^2}\begin{bmatrix}
\mathcal{J}_{\rm EEC, q}^{\rm vac}{}'(\theta; p_T, R, \mu) \\
\mathcal{J}_{\rm EEC, g}^{\rm vac}{}'(\theta; p_T, R, \mu)
\end{bmatrix}
= 
\frac{\alpha_s(\mu^2)}{2\pi}\begin{bmatrix}
\gamma_{qq}^{\rm vac} & \gamma_{gq}^{\rm vac} \\
\gamma_{qg}^{\rm vac} & \gamma_{gg}^{\rm vac}
\end{bmatrix}
\begin{bmatrix}
\mathcal{J}_{\rm EEC, q}^{\rm vac}{}'(\theta; p_T, R, \mu) \\
\mathcal{J}_{\rm EEC, g}^{\rm vac}{}'(\theta; p_T, R, \mu)
\end{bmatrix} \, .
\end{align}
If $\theta E$ is already greater than $M_{\rm LPM}$, then only the second stage of the evolution is needed.
\subsection{The Medium-Modified Semi-Inclusive Jet Function} \label{sec:medium:semi-inlcusive}
Finally, to get the correction of the semi-inclusive jet EEC, we further convolve the calculation with the medium-modified semi-inclusive jet function at NLO and to first order in opacity. The semi-inclusive jet cross-section in nuclear interaction $A$+$B$ is given by
\begin{align}
\frac{d\sigma_{AB\rightarrow \rm jet}}{dp_T^{\rm jet} dy} =& \sum_{ijkX}\int dx_1 \int dx_2  f_{i/A}(x_1, \mu^2)  f_{j/B}(x_2, \mu^2) \int dz \int dp_T \delta(p_T^{\rm jet}-zp_T) \nonumber\\
&\times  \left[ \frac{d\sigma_{ij\rightarrow kX}^{(0)}}{dp_T dy} \mathcal{J}_{j,AB}^{(0)}(z, R, \mu) + \frac{d\sigma_{ij\rightarrow kX}^{(1)}}{dp_T dy} \mathcal{J}_{j,AB}^{(0)}(z, R, \mu) + \frac{d\sigma_{ij\rightarrow kX}^{(0)}}{dp_T dy} \mathcal{J}_{j,AB}^{(1)}(z, R, \mu)  \right]\, ,
\end{align}
where the jet function labeled by subscript $AB$ denotes that it contains medium corrections in $A$+$B$ collisions.

The medium modified semi-inclusive jet function has been discussed in Ref.~\cite{Kang:2017frl}. The medium corrected semi-inclusive jet function where one parton is radiated outside of the jet cone is
\begin{align}
\mathcal{J}_{\rm EEC, c}^{\rm med, (1) >}(\theta; z_J, p_T, R, \mu) = &
 \sum_{(jk)}
\int_0^\infty dz^+ \int_0^1 dx \int d^2\bfk \int d^2\bfq \nonumber \\
&\times \left[\Theta_{\textrm{anti-}k_T}^{\rm >R} \delta(z-x_j)\mathcal{J}_j^{(0)}(\theta; p_T, R, \mu) \right. \nonumber\\ &\quad 
\left.+\Theta_{\textrm{anti-}k_T}^{\rm >R}\delta(z-x_k)\mathcal{J}_k^{(0)}(\theta; p_T, R, \mu)\right]  \nonumber\\
&\times \left[\frac{dN^{RR}_{i\rightarrow jk}}{dz^+ dx d^2\bfk d^2\bfq} + \delta^{(2)}(\bfq)\frac{dN^{RV}_{i\rightarrow jk}}{dz^+ dx d^2\bfk}
\right] \, .
\end{align}
The full expression can only be evaluated numerically, but using region analysis, we can obtain the most important leading contributions analytically. Because the scale $M_{\rm LPM}$  is smaller than the jet scale $p_T R$, the medium-induced collinear radiations outside of the cone will be power suppressed, and the relevant modes that causes jet energy loss is the collinear soft radiations.

To show this explicitly, start from the convolution of the semi-inclusive jet function with the hard parton spectrum,
\begin{align}
\mathcal{H}_{ab\rightarrow cX}\otimes \mathcal{J}^{(1)>} = \int \frac{dz_J}{z_J} \mathcal{H}_{ab\rightarrow cX}\left(\frac{p_T}{z_J}\right) \mathcal{J}_{\rm EEC, c}^{\rm med, (1)>}(\theta; z_J, p_T, R, \mu)\,,
\end{align}
first consider regions where $z_J=x_j=x$ or $z_J=x_k=1-x$ is not much smaller than unity. Then using the assumed hierarchy of energy scales $p_T R\gg M_{\rm LPM} = \sqrt{2 p_T/L}$, the radiative process outside of the cone must by power suppressed by additional power of $\frac{p_T/L}{(p_T R)^2}$. Including the opacity small parameter, the final correction is only of order $\frac{\rho_{\rm eff} L}{p_T/L} \times \frac{p_T/L}{(p_T R)^2} = \frac{\rho_{\rm eff} L}{(p_T R)^2}$. This is expected to be an negligible effect as it is suppressed by two powers of $p_T$. Nevertheless, in the soft-gluon emission region where $x_g\rightarrow 0$, there is a region that contributes at the power $\frac{\rho_{\rm eff} L}{p_T/L}$. To see this, we can start from the jet cone algorithm and express $\bfk^2$ with the LPM scaled dimensionless integration variable
\begin{align}
\Theta_{\textrm{anti-}k_T}^{\rm >R} =& \Theta\left(\frac{\bfk^2}{(1-z_J)^2(2p_T)^2} > \tan^2\frac{R}{2}\right) \nonumber \\
 =& \Theta\left(\frac{\tilde{\bfk}^2}{(1-z_J)^2(2p_T)^2} \frac{2z_J(1-z_J)p_T}{L} > \tan^2\frac{R}{2}\right) \, .
\end{align}

\begin{figure}
    \centering
    \includegraphics[scale=0.8]{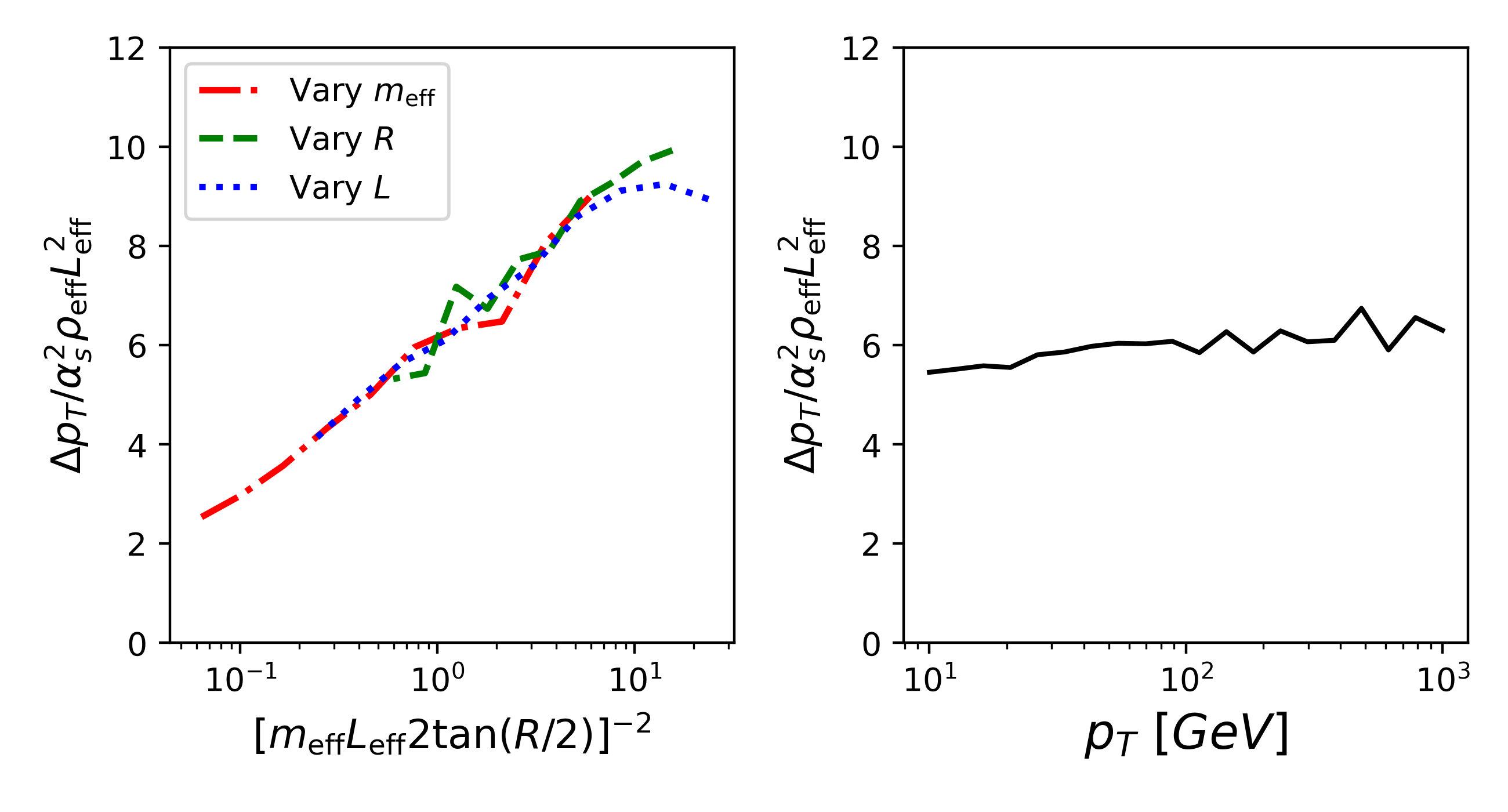}
    \caption{Left: the numerical calculation of jet energy loss with a screening mass $m_{\rm eff}$. A variations in the values of $R$, $m_{\rm eff}$, and $L_{\rm eff}$ in a constant brick medium confirms the scaling of the jet radiative energy loss as a function of $m_{\rm eff}L_{\rm eff}\tan\frac{R}{2}$. Right: for asymptotically high jet energy, the radiative energy loss of a jet is independent of jet $p_T$.}
    \label{fig:rad-eloss-jet}
\end{figure}

This way, we see that the cone condition is related to the power expansion parameter. In the $z_J\rightarrow 1$ limit, which is of phenomenological importance in the convolution with a steep falling spectrum, this requires
\begin{align}
1-z_J < \tilde{\bfk}^2\frac{2p_T/L}{\left(2 p_T\tan\frac{R}{2}\right)^2}\,.
\end{align}
Using this condition in the convolution along with the soft gluon approximation 
\begin{align}
\label{eq:jet-e-loss}
\mathcal{H}_{ab\rightarrow qX}\otimes \mathcal{J}_q^{(1)>}  
\approx & \int \frac{dx}{x} \mathcal{H}_{ab\rightarrow qX}\left(\frac{p_T}{x}\right) \int d^2\bfk \frac{dN^{N=1}_{q\rightarrow q+g}}{dx d^2\bfk}  \Theta\left(1-x < \tilde{\bfk}^2\frac{2p_T/L}{\left(2p_T\tan\frac{R}{2}\right)^2}\right) \nonumber\\
\approx & \int \frac{dx}{x} \mathcal{H}(p_T/x) \frac{\alpha_s^{(0)} C_F}{2\pi^2} P_{qq,\epsilon}(x) \sum_T\int_0^\infty dz^+ \rho_T^-(z^+) (g_s^{\rm med})^2 \frac{C_T}{d_A}    \nonumber\\
&   \times C_A \int \frac{d^{2-2\epsilon}\bfq}{(2\pi)^{-2\epsilon}} \frac{\alpha_s^{(0)}}{\pi}\frac{1}{(\bfq^2)^2} \int \frac{d^{2-2\epsilon}\bfk}{(2\pi)^{-2\epsilon}} \frac{2\bfk\cdot\bfq}{\bfk^2(\bfk-\bfq)^2}\Phi\left(\frac{(\bfk-\bfq)^2 z^+}{2x(1-x)P^+}\right) \nonumber\\
&   \times \
 \Theta\left((1-x)^2 < \frac{\bfk^2}{\left(2 p_T\tan\frac{R}{2}\right)^2}\right) \, .
\end{align}
Because the only scale other than the jet algorithm is the splitting-specific LPM scale squared $2x(1-x)p_T/L_{\rm eff}$, we approximate the $\bfk^2$ in the jet algorithm condition by $\bfk^2\sim 2(1-x)p_T/L_{\rm eff}$ under the soft gluon approximation. This way, the cone condition becomes a lower bound of $x$
\begin{align}
x > 1-\frac{2p_T/L_{\rm eff}}{\left(2p_T\tan\frac{R}{2}\right)^2} \equiv 1-\Delta \,.
\end{align}
The following analysis is very similar to the one in Ref.~\cite{Ke:2023ixa}, with a restricted range of the $x$ integration. The convolution with a splitting function with double pole at $x=1$ becomes
\begin{align}
\int_{1-\Delta }^1 dx \frac{\frac{1}{x}\mathcal{H}\left(\frac{z}{x}\right)-\mathcal{H}(z)}{(1-x)^{2+2\epsilon}} \approx& f'|_{x=1} \int_{1-\Delta }^1 dx \frac{x-1}{(1-x)^{2+2\epsilon}} 
= f'|_{x=1} \, \frac{\Delta^{-2\epsilon}}{2\epsilon} \, ,
\end{align}
which means that the logarithmic factor $\ln\frac{2 p_T/L}{\mu^2}$ to be resummed from Ref.~\cite{Ke:2023ixa} should is modified to 
\begin{align}
\ln\frac{2p_T/L_{\rm eff}}{\mu^2}\longrightarrow\ln\left(\frac{2p_T/L_{\rm eff}}{\mu^2}\frac{2p_T/L_{\rm eff}}{(2p_T\tan\frac{R}{2})^2}\right) \approx \ln\left(\frac{4}{\left[\mu L_{\rm eff} 2\tan\frac{R}{2}\right]^2}\right) \, .
\end{align}
The scale $\mu$ should be evolved to minimum invariant mass of the gluon inside the medium, which is also estimated by the medium screening mass $\mu=m_{\rm eff}$.
This is low energy radiation, but because medium-induced emission is wider than in the vacuum~\cite{Vitev:2005yg}, it still gives a sizable contribution to energy carried out of the jet cone. Such a logarithmic dependence on the combination $m_{\rm eff} L_{\rm eff} 2\tan\frac{R}{2}$ and the relative independence on $p_T$ can be checked by numerically computing the averaged energy loss as shown in Figure~\ref{fig:rad-eloss-jet}.

In the Appendix~\ref{app:jet-eloss}, we give a more careful derivation of this result using the method introduced in Ref.~\cite{Ke:2023ixa} and treat the medium-modification of the semi-inclusive jet function as an energy loss effect of the hard parton spectrum under the soft-gluon emission approximation. This lead to the following evolution equations
\begin{align}
\frac{\partial}{\partial \ln\mu^2} \left(p_T\mathcal{H}_{qq}(p_T,\mu^2)\right) = -\frac{[\alpha_s(\mu^2)]^2}{2\pi} 2C_F2C_A  \frac{\rho_{\rm eff} L_{\rm eff}^2}{2} \frac{s_0 \overline{\Phi}''(0)}{2} \frac{\partial }{\partial p_T}  \left(p_T\mathcal{H}_{qq}(p_T,\mu^2)\right)\,, \\
\frac{\partial}{\partial \ln\mu^2} \left(p_T\mathcal{H}_{gg}(p_T,\mu^2)\right) = -\frac{[\alpha_s(\mu^2)]^2}{2\pi} 2C_A2C_A  \frac{\rho_{\rm eff} L_{\rm eff}^2}{2} \frac{s_0 \overline{\Phi}''(0)}{2} \frac{\partial }{\partial p_T}  \left(p_T\mathcal{H}_{gg}(p_T,\mu^2)\right)\,.
\end{align}
These equations admits a traveling wave solution that correspond to a shift in parton energy $p_T\rightarrow p_T+\Delta p_{T, \rm rad}$.
\begin{align}
\Delta p_{T,\rm rad, R} =&\frac{2}{\beta_0} \left[\alpha_s\left(m_{\rm eff}^2\right)-\alpha_s\left(\frac{s_0^2/e^2}{L_{\rm eff}^2\tan^2\frac{R}{2}}\right)\right] \nonumber\\
&\times  4C_RC_A  \frac{\rho_{\rm eff} L_{\rm eff}^2}{2} \frac{s_0 \overline{\Phi}''(0)}{2}\,.
\end{align}
The first factor comes the interplay of the running coupling and the logarithm enhancement factor $\ln\frac{1}{(m_{\rm eff} L_{\rm eff} 2\tan\frac{R}{2})^2}$. The $C_FC_A$ is due to the color charge of the quark and the radiated gluon. $\frac{\rho_{\rm eff} L_{\rm eff}^2}{2}$ gives the scale of the typical energy loss, while $\frac{s_0 \overline{\Phi}''(0)}{2}$ encodes the information of the path-dependent medium profile and $s_0$ is an order-one number. For the detailed definition, please refer to the Appendix~\ref{app:jet-eloss}.

Finally, compared to Ref.~\cite{Kang:2017frl}, what we further include is the elastic energy loss (see Section~\ref{sec:coll-eloss}) of jet such that $\Delta p_{T,\rm rad}$ is replaced by $\Delta p_{T,\rm rad} + \Delta p_{T,\rm coll}$. Here we have only considered the effect of one parton losing energy inside the jet, which is consistent with our calculation to order $\alpha_s$. If collinear radiation inside the jet generates more partons as sources of energy loss, in principle the right-hand side will depend nonlinearly on the energy-loss distribution. Here we will not consider such complications.
\section{Phenomenological Applications}
\label{sec:pheno}
In this Section we apply the formalism developed above to small collision systems, focusing on $p$–Pb and O–O reactions where the medium is relatively short-lived but can still induce measurable modifications to jet substructure. These systems provide an ideal testing ground for the framework, as the first-order opacity expansion remains reliable while capturing essential medium-induced effects. Using the renormalized expressions for the EEC, we analyze how the medium parameters influence the angular distribution and assess the sensitivity of the observable to the path length and screening scale of the QGP.

\subsection{The Medium Path-Length Average}
\label{sec:results-L-avg}
In heavy-ion collisions, the produced quark--gluon plasma (QGP) is a highly dynamical medium undergoing rapid expansion and cooling. Its space--time evolution is well described by relativistic hydrodynamics, which provides the local energy density or effective temperature profile as a function of proper time and spatial coordinates. This framework has been successfully applied not only to nucleus--nucleus collisions but also to smaller systems such as proton--lead and other light-ion collisions, where collective QCD phenomena have been observed.

In this work, we model the medium evolution using the 2+1D viscous hydrodynamic framework VISHNew developed in Ref.~\cite{SHEN201661}, calibrated against soft-hadron observables in Refs.~\cite{Bernhard:2018hnz}, and further adapted for jet-quenching and small-system studies in Ref.~\cite{Ke:2022gkq}. The simulation provides the temperature field using an event-averaged initial condition for each centrality class, 
\begin{equation}
T = T(\tau, \vec{x}_\perp, \eta_s) \, ,
\end{equation}
which encodes both the transverse expansion and the longitudinal boost-invariant evolution of the plasma.

The initial production position of a hard scattering, denoted by $\vec{x}_0$, is sampled according to the binary-collision density in the transverse plane at $\tau = 0^+$,  
\begin{equation}
P_{\rm coll}(\vec{x}_0) \propto T_A(\vec{x}_0 - \tfrac{\vec{b}}{2}) \, T_B(\vec{x}_0 + \tfrac{\vec{b}}{2}) \, ,
\end{equation}
where $T_{A,B}$ (not to be confused with the temperature field) are the standard nuclear thickness functions for the colliding nuclei at impact parameter $\vec{b}$. $T_{A,B}$ are obtained from the same TRENTo model~\cite{Moreland:2014oya} that initializes hydrodynamic equations. The jet direction in the transverse plane is randomized with an azimuthal angle $\phi_{\rm jet}$, corresponding to a unit vector
\begin{equation}
\hat{n}_{\rm jet,\perp} = (\cos\phi_{\rm jet},\, \sin\phi_{\rm jet})\, .
\end{equation}
Assuming ultrarelativistic motion, the jet partons are taken to propagate at the speed of light along straight trajectories. Together with the sampled $\vec{x}_0$ and $\phi_{\rm jet}$, this defines a unique, event-by-event fluctuating temperature profile along the jet path,
\begin{align}
T_{\rm jet}(\tau;\, \vec{x}_0, \phi_{\rm jet}, \eta_{\rm jet})
&= T\big(\tau,\, \vec{x}_0 + \hat{n}_{\rm jet,\perp}\,\tau,\, \eta_{\rm jet}\big) \nonumber \\ 
&\simeq T\big(\tau,\, \vec{x}_0 + \hat{n}_{\rm jet,\perp}\,\tau,\, 0\big) \, ,
\end{align}
where the last approximation follows from the boost-invariance assumption of the hydrodynamic simulation around midrapidity.

For each individual jet trajectory, characterized by the local temperature field $T_{\rm jet}(\tau)$, we compute the medium-induced contributions to the energy-energy correlator using the following procedures:

\begin{enumerate}
  \item Evaluate the non-contact (medium-induced) term in the EEC at first order in opacity, integrating over the path-dependent temperature history;
  \item Solve Eq.~(\ref{eq:effective-med-param})
  to determine the path-averaged transport coefficients $\rho_{\rm eff}$ and $L_{\rm eff}$, which enter the definition of the medium-modified anomalous dimension.
\end{enumerate}

Finally, ensemble averages are performed over the sampled jet origins $\vec{x}_0$, propagation angles $\phi_{\rm jet}$, and pseudorapidities $\eta_{\rm jet}$, yielding the medium-averaged EEC correction and the corresponding in-medium anomalous dimension. The latter serves as input to the EEC evolution equation, through which resummed medium-induced effects are incorporated.

\subsection{Results and Discussion for the Energy Correlator in $p$-Pb and O-O}
\begin{figure}[t]
  \centering
  \includegraphics[scale=.8]{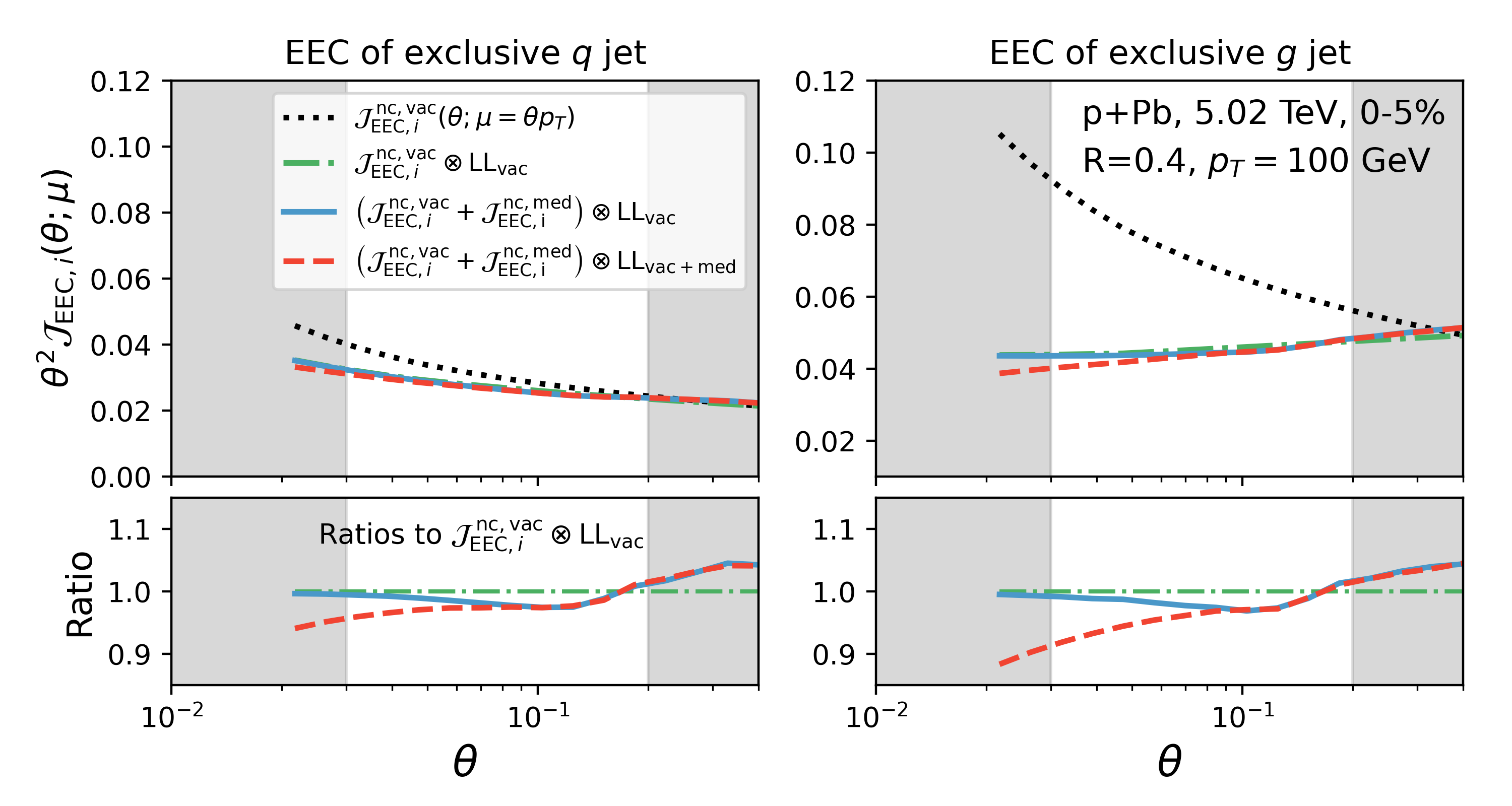}
  \caption{EEC of an exclusive quark jet (left panel) and a  gluon jet (right panel). Black dotted line: leading order expression in the vacuum under the small angle approximation. Blue solid line: including vacuum LL resummation. Green dashed line: further including the medium non-contact correction at opacity one. Red dash-dotted line: further including the medium-induced anomalous dimension to LL resummation. The effect of medium-induced resummation affects the small angle region. This calculation is performed for the centrality class 0-5\% in $p$-Pb and jet energy $100$~GeV (the corresponding energy of the charged jet is $2/3$ of the energy of the full jet).}
  \label{fig:pA_EEC}
\end{figure}

\begin{figure}[t]
  \centering
  \includegraphics[scale=.8]{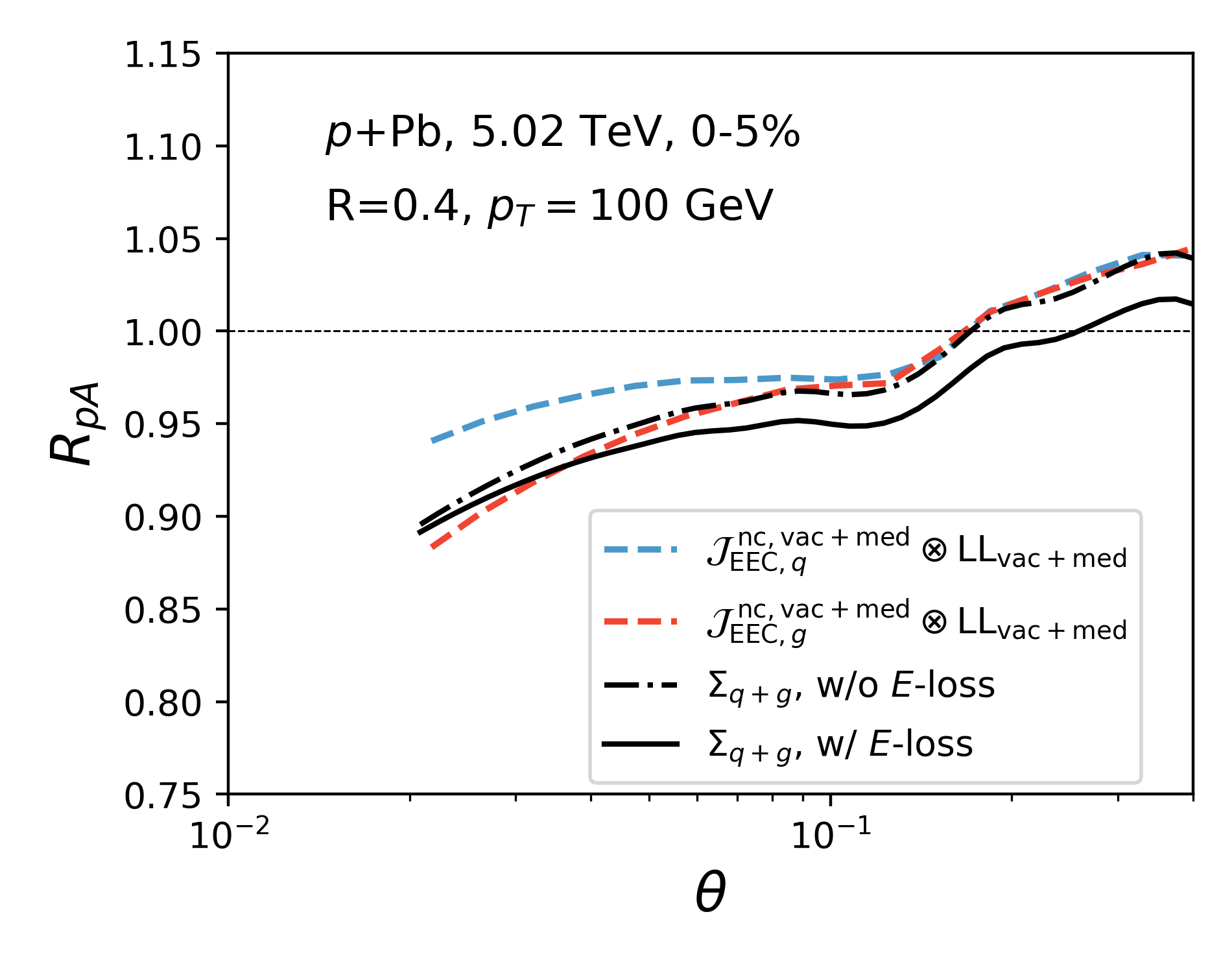}
  \caption{$R_{pA}$ of EEC (per-jet normalized) for inclusive jets (black) compared to those for pure quark (blue dashed) or gluon jets (red dashed). The black dot-dashed line does not include medium correction to the hard collinear function $\mathcal{H}_{ji}$. The black solid line includes medium energy loss correction to $\mathcal{H}_{ji}$, which can also also be recast as a modification to the hard function $\mathcal{H}_{ab\rightarrow cX}$. 
  }
  \label{fig:RpA_EEC_exclusive}
\end{figure}

Figure~\ref{fig:pA_EEC} demonstrates the exclusive jet function of the energy-energy correlator for quark jets (left panel) and gluon jets (right panel), i.e., the EEC before the convolution with the semi-inclusive jet function and the hard cross-section. The dotted lines show the initial condition of the EEC at scale $\mu_\theta = \theta p_T$ without any medium corrections, using Eq.~(\ref{eq:EEC-vac-initial-conditions}). Due to the running coupling $\alpha_s(\mu_\theta^2)$, the initial condition $\theta^2 \mathcal{J}_{\rm EEC}^{i}(\mu_\theta)$ increases at small angles. The green dash-dotted lines include the leadin-log resummation of the exclusive jet function in the vacuum by evolving the RG scale from $\mu_\theta$ to $\mu_R$. The evolution strongly suppresses the $\theta^2 \mathcal{J}_{\rm EEC}^{i}(\mu_R)$. Physically, this can be understood in the following way: allowing radiations within the phase space $\mu_\theta<\mu<\mu_R$ causes energy reduction in forming the EEC pair at scale $\mu_\theta$. The quark jet function after evolution still increases at small angle, while that for gluon jet slightly decreases, because the latter has a significantly larger anomalous dimension. 
The blue solid lines include the medium non-contact correction to the EEC jet function at first order in opacity. Its magnitude is rather small for 0-5\% $p$+Pb collisions. It's effect is more evident in the ratio plot with an excess at large angle and depletion at small angle with the transition happens near $ \theta_{\rm LPM}=\sqrt{8\pi/(p_T L_{\rm eff})}$. Finally, the red dashed lines includes medium-modified LL evolution between $\mu_\theta$ and $M_{\rm LPM}$. The effects of these medium-induced radiation is to further suppress the jet function below $\theta\sim \theta_{\rm LPM}$. The impact is larger for gluon jets due to the larger color charges and the resulting larger medium correction to the anomalous dimension in Eq.~(\ref{eq:Delta_gamma_med}).
The shaded areas indicate regions where non-perturbative effects may become large - hadronization at small angles and potential contamination from the medium at $\theta \sim R$. The latter is less of a problem for high $p_T \geq 100$~GeV jets, but can distort the measured EEC at lower transverse momenta.

Figure~\ref{fig:RpA_EEC_exclusive} shows the nuclear modification factor for the per-jet normalized energy-energy correlator in 0-5\% $p$-Pb collisions,
\begin{equation}
R_{pA}(\theta; p_T) = \frac{\left[\frac{\dd\sigma_{pA\rightarrow\rm jets}}{ p_T \dd p_T  \dd y}\right]^{-1}\frac{\dd\Sigma_{pA\rightarrow\rm jets}}{ p_T \dd p_T  \dd y\dd\theta^2}}{\left[\frac{\dd\sigma_{pp\rightarrow\rm jets}}{p_T \dd p_T  \dd y }\right]^{-1}\frac{\dd\Sigma_{pp\rightarrow\rm jets}}{p_T \dd p_T  \dd y \dd\theta^2}}
 \, .
\end{equation}
$\dd\sigma_{pA\rightarrow\rm jets}$ is the inclusive jet cross-section and $\dd\Sigma_{pA\rightarrow\rm jets}$ is the EEC for inclusive jets. The dash-dotted is the calculation without medium correction to the semi-inclusive jet function. It just reflects the effects of the mixture between quark and gluon jet EEC. The solid line is the result with medium modified semi-inclusive jet function, in the form of collisional and radiative jet energy loss. Because the gluon jet energy loss is larger than quark jet, the average after medium modification is slightly biased towards quark jet, and the quark-jet EEC is smaller in magnitude, the jet energy loss effects reduces the overall magnitude of the per-jet normalized EEC in $p$-Pb.

\begin{figure}[!t]
  \centering
  \includegraphics[scale=.8]{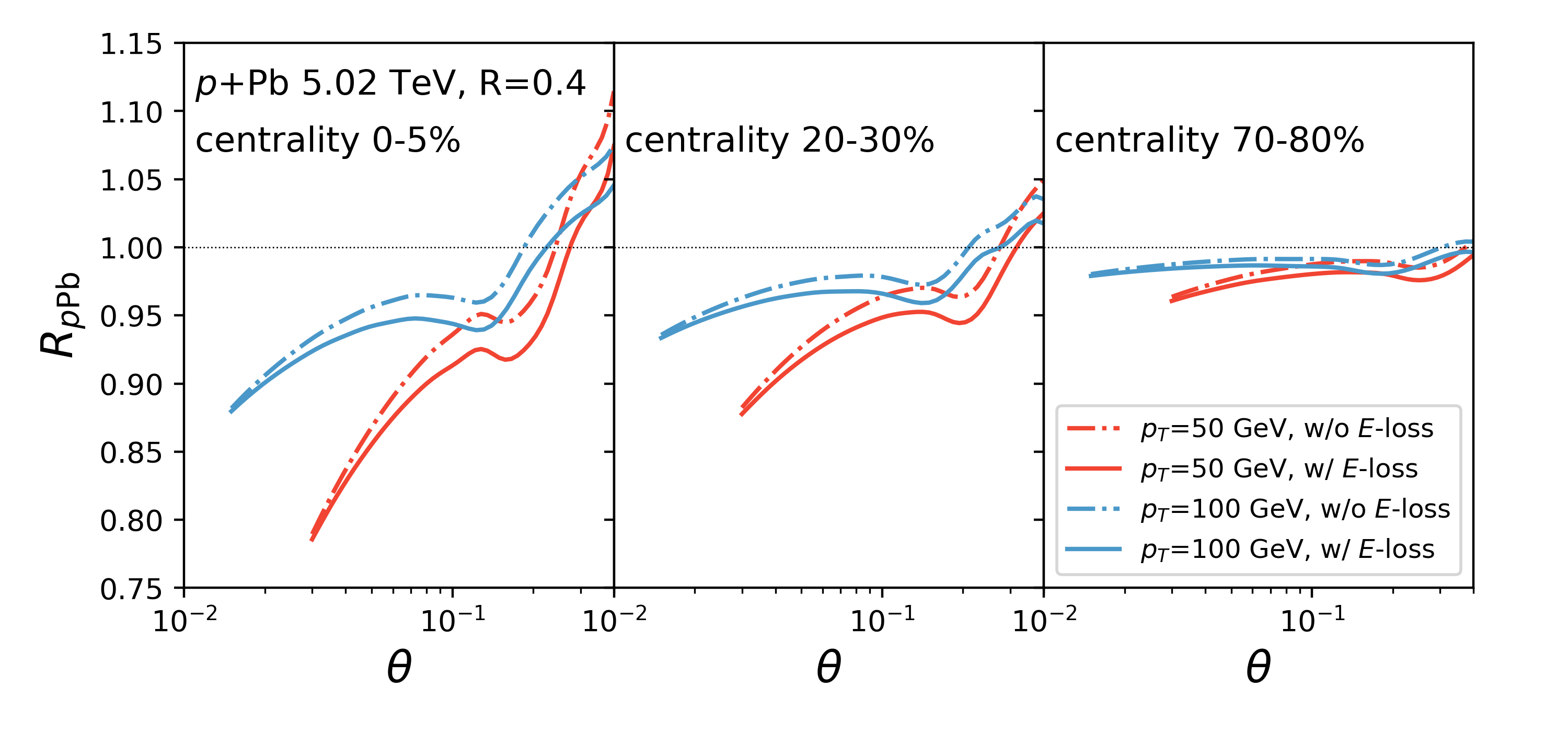}
  \caption{The centrality dependence of per-jet normalized $R_{p\rm Pb}$ of energy-energy correlators. Blue and red lines are results for full-jet $p_T=$ 50 GeV and 100 GeV, respectively. Dot-dashed and solid lines represent calculations without and with jet energy loss. }
  \label{fig:RpA_EEC_vs_centrality}
\end{figure}

To gain better insight into the transverse momentum and centrality dependence we apply the theoretical formalism to evaluate EEC is  $p$-Pb collisions for $p_T =50, \; 100$~GeV jets and 0-5\%, 20-30\%, and 70-80\% centrality classes.  We observe the expected growth in the nuclear modification 
$R_{p{\rm Pb}}$ in going from peripheral to central collisions. As the transverse momentum of the jet decreases, the relative
significance of medium-induced branching contributions grow. This is particularly pronounced for the out-of-cone energy-loss and evolution effects at intermediate values of $\theta$.  A much more subtle effect is the role of $ \theta_{\rm LPM} =\sqrt{8\pi/p_T L_{\rm eff}}$. It roughly separates the evolution and fixed order regions, the latter including the large angle enhancement from in-medium showers~\cite{Vitev:2005yg}. This can be observed through the migration of point where the curves intersect unity to larger $\theta$.

\begin{figure}[!t]
  \centering
  \includegraphics[scale=.8]{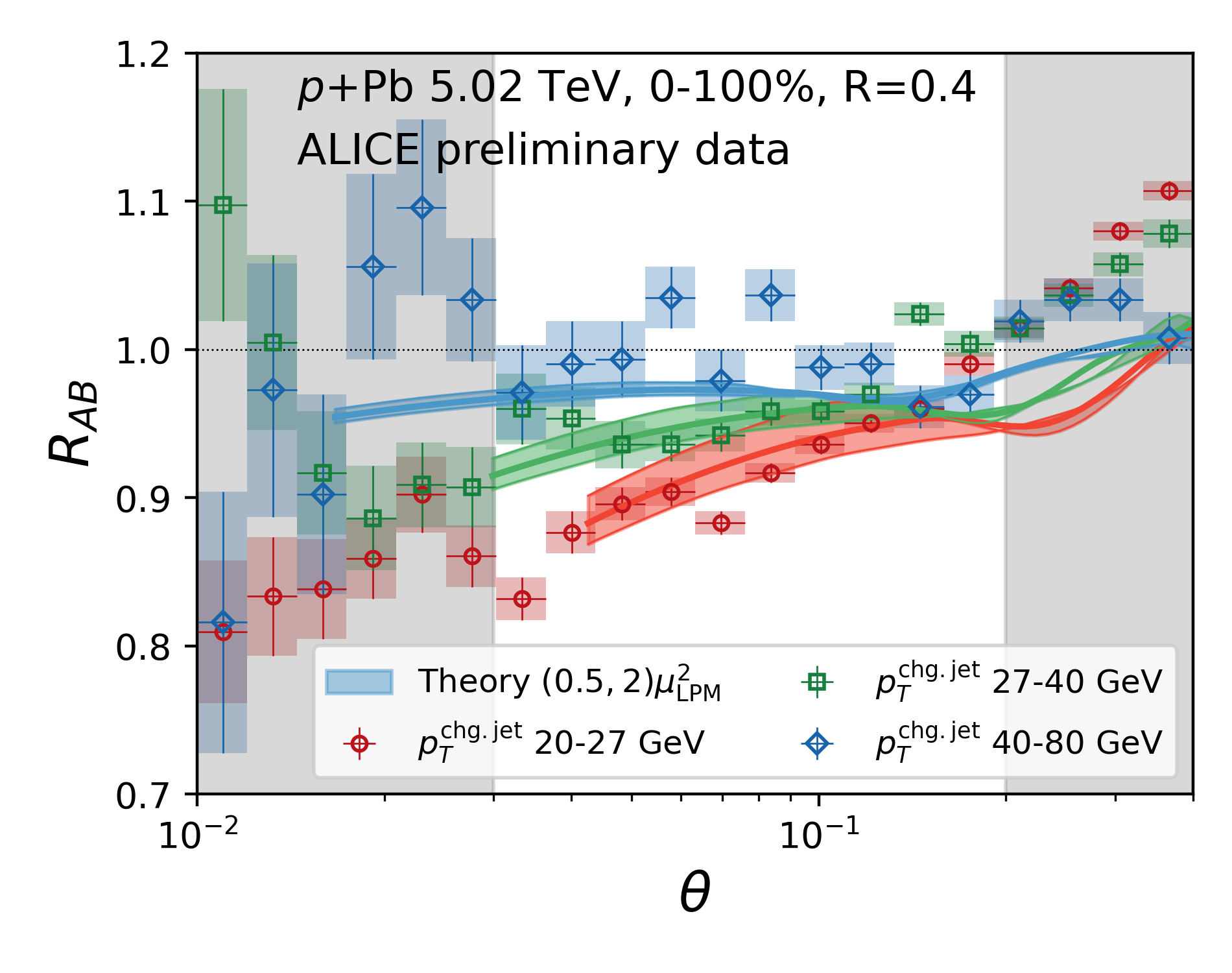}
  \caption{Nuclear modification factor $R_{pA}(\theta)$ of the energy--energy correlator
  for charged jets in different $p_T^{\rm ch.\,jet}$ intervals, compared with preliminary
  ALICE measurements~\cite{Liang-Gilman:2025gjl}. Bands correspond to the hydrodynamics-averaged EEC
  calculation including medium-induced corrections at first opacity order and in-medium
  resummation effects. The width of the band is obtained by varying the $\mu_{\rm LPM}$ scale by a factor of $1/2$ and $2$. Regions with potentially large non-perturbative contributions and expected deviations form a perturbative calculation are shaded. }
  \label{fig:RpA_EEC_inclusive_hydro}
\end{figure}

Finally, in Figure~\ref{fig:RpA_EEC_inclusive_hydro}, we average the jet spectra and EEC from 0 to 100\% centrality for $p$-Pb and compare the nuclear modification factor to the preliminary experimental data from ALICE Collaboration. 
The colored bands represent our theoretical calculation based on the hydrodynamic temperature profiles from the VISHNew framework and the in-medium EEC evolution described in the previous Section, while the data points denote the preliminary ALICE
measurements. Three representative jet transverse-momentum intervals,
$p_T^{\rm ch.\,jet}=20$--$27$~GeV (red), $27$--$40$~GeV (green), and
$40$--$80$~GeV (blue), are shown. It should be emphasized that the ALICE data is taken for charged jets. In principle, one should include the selection of charged tracks using the track function approach of EEC developed in Ref.~\cite{Jaarsma:2023ell}. This can certainly be improved in the future. Here we take the approximation that the charged jet $p_T$ is $2/3$ of that of a full jet, and compare the full-jet EEC with data. We note that that these are fairly low $p_T$ jets. 

Across all $p_T$ bins, $R_{pA}(\theta)$ exhibits a pronounced suppression below unity at small angles ($\theta \lesssim 0.1$) followed by a gradual recovery and mild enhancement toward large angles. This structure encodes the competing effects of medium-induced broadening and energy redistribution inside the jet. At small $\theta$, the reduction of correlation strength arises predominantly from the modified EEC anomalous dimension: multiple soft rescatterings in the QGP generate an in-medium resummation effect that depletes the energy flow near the jet core. This suppression is stronger for low-$p_T$ jets, consistent with the expected scaling of $\kappa^{\rm med}$ and the path-length dependence of the medium modification. In fact, if we take the LO running coupling and neglect the energy loss effects and the off-diagonal anomalous dimension, the suppression of EEC at $\theta_{\rm NP} \ll \theta\ll \theta_{\rm LPM}$ is approximately given by 
\begin{align}
\frac{\Sigma_{i}^{\rm vac+med}}{\Sigma_{i}^{\rm vac}} \approx \exp\left(-\frac{4\pi}{\beta_0}\Delta\gamma_{ii}^{\rm med}(N)\frac{\rho_{\rm eff} L_{\rm eff}^2}{p_T}\left[\alpha_s(p_T \theta) - \alpha_s(p_T R)\right]\right) \, .
\end{align}
Thus, for larger jet $p_T$ the suppression at small angle is reduced. This simple formula also predicts that higher point correlator are more suppressed in the region $\theta\ll \theta_{\rm LPM}$.

The significance of this finding is that for the first time we can connect medium properties directly to the anomalous dimension of an observable in its asymptotic region. This provides a very robust way of extracting the medium quantities of interest. The robustness is two fold: on the experimental side, extracting EEC slope in the per-jet normalized ratio can cancel many experimental uncertainties; on the theory side, such a behavior is derived from an RG analysis and other effects are power suppressed in this asymptotic region, which provide a reliable access to the medium properties, in this case, in the form of ensemble averaged $\rho_{\rm eff} L_{\rm eff}^2$ as defined in Eq.~(\ref{eq:effective-med-param}).

\begin{figure}[t]
  \centering
  \includegraphics[scale=.8]{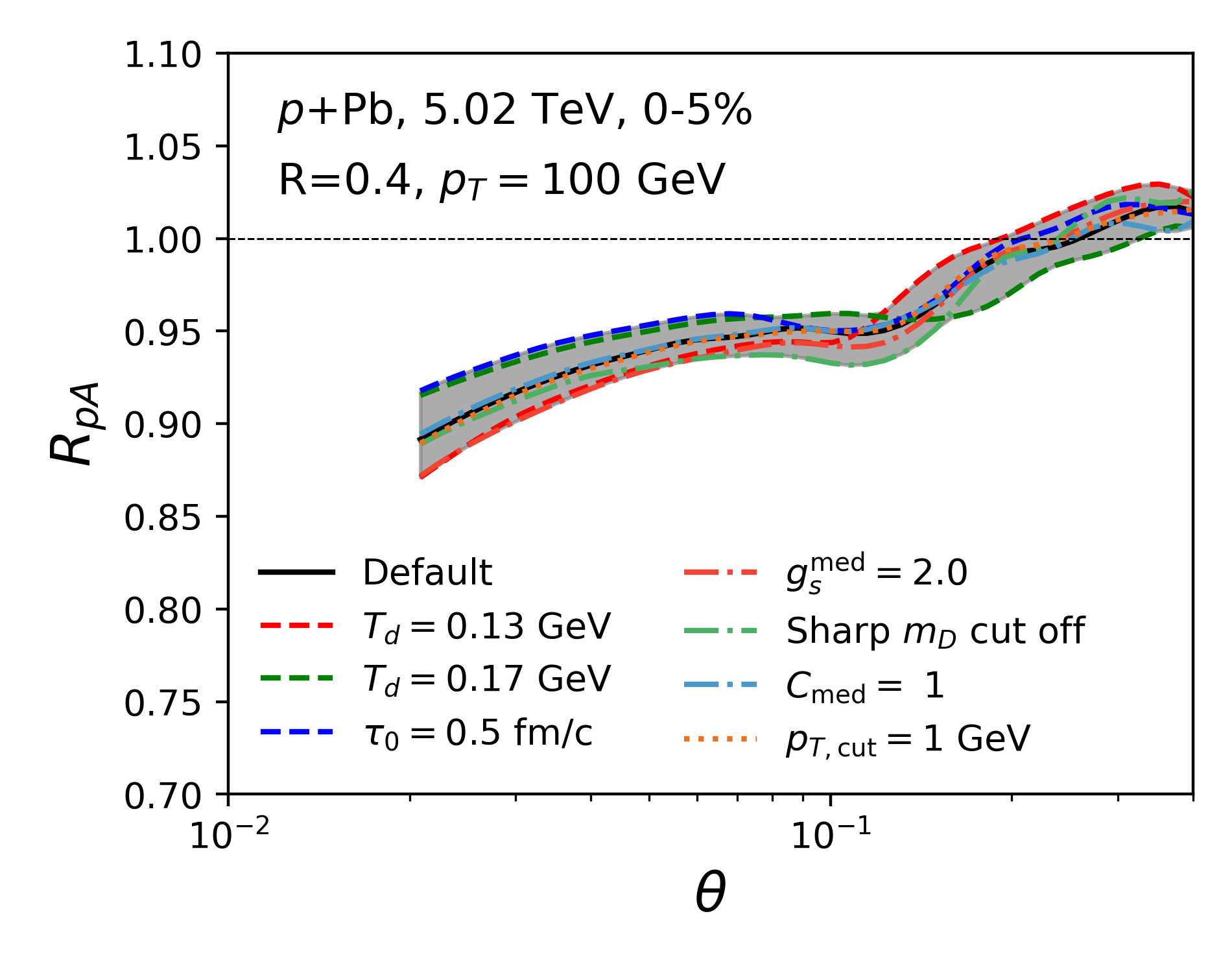}
  \caption{An estimation of the sensitivity of the EEC to the soft parameters. The default set of parameters has been summarized in table~\ref{tab:default-parameter}. Each of the line varies one of the parameters away from the default value. The shaded envelope of all the variations is a representation of the uncertainty from the soft parameters.
  }
  \label{fig:uncertainty}
\end{figure}

\paragraph{On the theoretical uncertainty:} We explore  the theoretical uncertainty on the example of $p_T =100$~GeV jets.  The band of Figure~\ref{fig:RpA_EEC_inclusive_hydro} corresponds to a variation of the the LPM scale up and down by a factor of 2. The uncertainty band from the variation of this semi-hard scale is reasonably under control.
Besides, there are many soft parameters used in above calculation related to the medium properties, including
\begin{itemize}
\item The initial time $\tau_0$ when the jet parton starts to couple to the medium. The uncertainty comes from the lack of knowledge of when the color charges of the QGP are fully produced from the initial scattering. 
It is varied from $0.1$ fm/$c$ to $0.5$ fm/$c$.
\item The final decoupling temperature $T_d$ between the jet and the quark-gluon plasma medium. This uncertainty is due to the cross-over nature of the QGP-hadron phase transition. The system between the pseudo-critical temperature $T_c$ is still a strongly-coupled system and may still impact the evolution of the jet. We vary $T_d$ around the default $T_c=155$ MeV from 130 MeV to 170 MeV.
\item The effective medium coupling parameter $g_s^{\rm med}$. This is the value of the strong coupling constant evaluated at the screening mass scale. However, at the temperatures reached in experiments, the Debye screening mass is not fully in the perturbative region. Therefore, we treat the coupling as a phenomenological parameter and vary it from $2.0$ to $2.2$.
\item The $p_T$ cut of the parton. Experimentally, not all hadrons are used to construct the EEC; there is a 1 GeV minimum $p_T$ cut. To estimate the impact of such a kinematic cut, we implement a parton-level $p_T$ cut from 0 to 1 GeV.
\item The type of IR screening used in the Glauber propagator. This is relevant in the numerical evaluation of the non-contact EEC at first order in opacity. The default is the soft cutoff where the $1/\mathbf{q}^2$ Glauber gluon propagator is replaced by $1/(\mathbf{q}^2+m_{\rm eff}^2)$. We also tested the results using the hard cutoff $\Theta(\mathbf{q}^2-m_{\rm eff}^2)/\mathbf{q}^2$.
\item The UV cutoff of the Glauber momentum integration. This is also used in the numerical integration. From the center-of-mass relation, the average center-of-mass energy between a jet parton of energy $E$ and a thermal parton in the medium is $\langle s\rangle = 6ET$, which is the kinematic upper bound for $|t|$. However, the Glauber gluon approximation requires $|t|\sim \mathbf{q}^2 \ll s$. Therefore, we choose the default UV cutoff as $\sqrt{3ET}$ and vary it down to $\sqrt{ET}$.
\end{itemize}
\begin{table}[h!]
    \centering
    \begin{tabular}{c|c|c|c|c|c}
    \hline
    Initial time $\tau_0$ & Decoupling $T_d$  & $g_{s}^{\rm med}$ & $p_{T, \rm cut}$ & Screening &  Glauber UV cut \\
    \hline
    0.1 fm/$c$ & 0.155 GeV  & 2.0 & 0.0 GeV & $\frac{1}{\bfq^2}\rightarrow \frac{1}{\bfq^2+m_{\rm eff}^2}$ & $\sqrt{3ET}$  \\
    \hline
    \end{tabular}
    \caption{Default values of the medium-related soft parameters and other kinematic regulators.}
    \label{tab:default-parameter}
\end{table}
All default parameters are listed in table~\ref{tab:default-parameter}. Figure~\ref{fig:uncertainty} shows the default result as the black solid line and the envelope of all other variations as the shaded bands. As one can see, even though each individual variation is small, the number of soft parameters means they can constitute a significant uncertainty, in addition to the uncertainty coming from the variation of the LPM energy scale alone. In the future, one may seek better ways to cancel such sensitivity to soft parameters, for example by considering ratios between E2C and E3C, which warrants further investigation.

\paragraph{A comment on data vs theory at large angle:}
We have shaded two regions $\theta\lesssim \theta_{\rm NP}$ and $\theta\sim R$ from the comparison. The reason for the first requirement is due to the perturbative nature of the calculation, it cannot describe the transition peak to the hadronic region. Now we discuss the discrepancy with data at angles comparable to $R$.
First, it has been found that in the large-angle area there can be differences between the opacity expansion and large number of soft scatterings energy loss approach~\cite{Andres:2023xwr}. This, however, should be applicable to large not small systems.  Second, Monte Carlo simulations also suggest the contribution of medium recoiled partons and possible medium collective response to the jet energy-momentum deposition to the large $R$ region~\cite{Yang:2023dwc}. If that is the case, it will contradict the argued insensitivity of the EEC to soft non-perturbative physics.   For these reasons, we primarily focus on the asymptotic region $\theta_{\rm NP}\ll \theta\ll \theta_{\rm LPM}$, where impacts from other large-angle energy transport mechanisms due to the medium are suppressed. The $\theta \approx R$ region merits further scrutiny and it will be useful to extend the EEC measurements in $p$-A to higher $p_T$.

\paragraph{Discussion of EECs in small and large systems:}
The first measurement of EEC with nuclear medium effect was  performed in Pb-Pb collisions~\cite{CMS:2025ydi}. It was found that the $R_{AA}$ for EEC (in CMS it is per particle-pair normalized, while in ALICE it is per-jet normalized) not only shows enhancement at large angles but also a significant upward trend at small angles. How can this be reconciled with the preliminary $p$-A data and  our evaluation of the modification in small system collisions? First, if the system is large, then the available phase space for the logarithmic term we considered, $\ln\frac{E/L_{\rm eff}}{m_{\rm eff}^2}$, will be significantly suppressed due to a much larger $L$. This is also due to stronger screening from a higher average temperature. Besides, if multiple scatterings are coherent over the medium-induced radiation formation time $\tau_f$,  $m_{\rm eff}^2$ might be replaced by the total transverse momentum broadening $\hat{q} \tau_f$. As a result, the resummation effect and its applicable angular domain are greatly reduced. It is possible that in Pb-Pb collisions, such medium-induced resummation effects are tiny unless for much larger jet energies than previously considered. Furthermore, jet energy loss becomes large when compared to the case in small systems. The large jet energy loss increases the quark-versus-gluon jet bias, which causes the enhancement at small angles in Pb-Pb collisions. This has been studied in MC simulations, e.g., in Ref.~\cite{}. In short, from small to large systems, we expect the medium-induced scale evolution and its related resummation effect on EEC to decrease, while the flavor bias due to energy loss increases, which eventually alters the behavior of EEC nuclear modification factors at small angles. For this reason, it is important to project EEC modifications for intermediate-sized collisions, such as the oxygen-oxygen system at the LHC center-of-mass energy.

\begin{figure}[t]
  \centering
  \includegraphics[scale=.8]{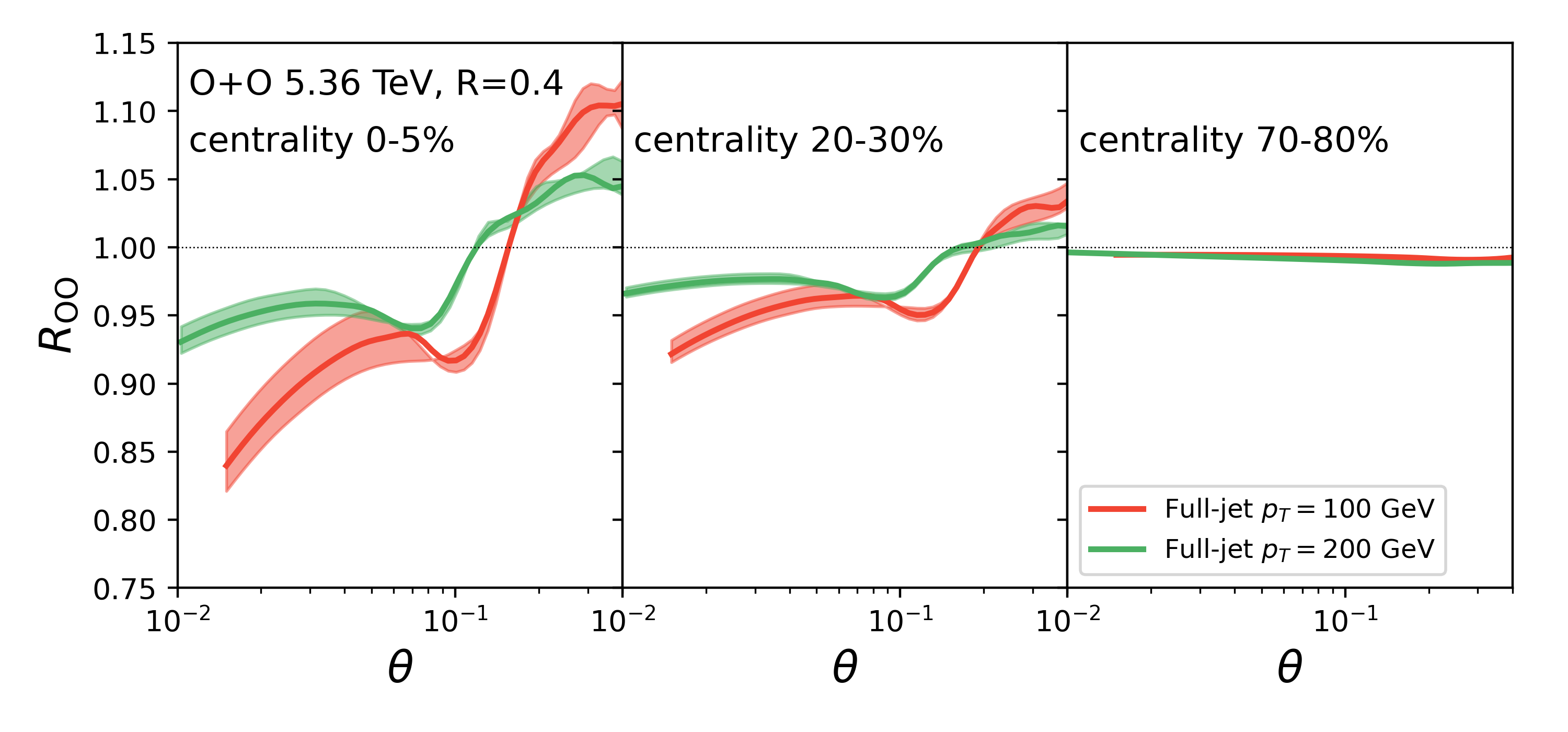}
  \caption{Predicted nuclear modification factor $R_{\rm OO}(\theta)$ of the energy--energy correlator for charged jets in two different full-jet $p_T$ intervals and three different centrality classes.
  }
  \label{fig:ROO_EEC}
\end{figure}

\paragraph{Projections for the EEC modification in Oxygen-Oxygen collisions:}
In Figure~\ref{fig:ROO_EEC}, we project the full-jet EEC for O-O collisions at $\sqrt{s_{NN}} = 5.36$ TeV~\cite{Brewer:2021kiv,Ke:2022gkq} for two different jet $p_T$. The left and right panels are projections for 0-5\% central and 20-30\% semi-central O-O collisions, respectively. A future measurement of the EEC in O-O may further differentiate if the effect of medium-induced resummation is needed for the asymptotic region and if the corresponding medium correction to the EEC anomalous dimension or opacity can be extracted from different systems.

\section{Conclusions and Outlook} 
\label{sec:conclude}
Jets are invaluable probes of the quark-gluon plasma and nuclear matter in general, offering a window into the medium’s microscopic structure and transport properties. In this work, we presented a first-principles study of the energy-energy correlation observable in heavy-ion collisions using SCET extended to include Glauber gluon interactions. In particular, this is the first calculation that presents a consistent derivation of both fixed order and resummed in-medium contributions, and obtains a renormalization group evolution for the EEC observable in matter. Furthermore, this framework enables a consistent QCD-based treatment of medium-induced effects on jet substructure, going beyond phenomenological models.

We incorporated medium interactions through a controlled opacity expansion that captures transverse momentum exchanges between hard partons and the QGP. Within this setup, we derived a factorized expression for the in-medium EEC and computed the quark and gluon jet functions at first order in opacity. These functions encode the leading effects of  hard scatterings on the collinear structure of the jet and provide a direct handle on medium-induced broadening and color decoherence. We further derived and analyzed the RG evolution of the in-medium EEC, identifying how Glauber exchanges modify the anomalous dimensions that govern its scale dependence. The evolution captures the medium-induced contributions associated with Coulomb logarithmic enhancements and their regulation by medium properties such as the Debye screening mass.

Using the resulting leading-order in opacity and leading-logarithmic in scale evolution expressions, we obtained numerical predictions for realistic jet energies and medium parameters. The comparison with $p$–Pb data demonstrates that the medium-modified EEC framework reproduces the characteristic suppression at small angles and enhancement at large angles, confirming that the observable is a sensitive probe of transverse momentum broadening and angular energy redistribution in QCD matter, and complementing traditional jet-shape and fragmentation measurements. It does suggest, however, that extending the measurements to higher jet $p_T$ could reduce the contamination of the observable with soft particles from the medium. Interestingly, the detailed EEC modification pattern was found to be consistent with the assumption of QGP formation in small systems. Predictions for O-O reactions were also presented.

This study establishes a rigorous baseline for interpreting energy flow observables in heavy-ion collisions. Beyond these first results, the framework offers a basis for systematically improving both precision and scope. Extensions to next-to-leading logarithmic accuracy and to higher orders in opacity will refine the quantitative extraction of medium parameters. Incorporating heavy-flavor jets will further enable studies of mass-dependent energy loss and color coherence, providing complementary sensitivity to the transport and diffusion properties of the QGP. In addition, generalizing the analysis to multi-point correlators will open a path toward probing correlations of collective excitations and nontrivial energy flow patterns in QCD matter. Such developments will deepen the connection between jet substructure measurements, heavy-quark dynamics, the properties of nuclear matter and space–time evolution of the quark–gluon plasma.

\section*{Acknowledgments} 
The work of B.M. and I.V. is supported by the US Department of Energy through the Los Alamos National Laboratory. Los Alamos National Laboratory is operated by Triad National Security, LLC, for the National Nuclear Security Administration of the U.S. Department of Energy (Contract No. 89233218CNA000001). The  research  performed by W.K, B.M., and I.V. is further funded by LANL’s Laboratory Directed Research and Development (LDRD) program.

\newpage
\appendix

\section{The non-contact EEC at first order in opacity}
We will show the detailed techniques to compute the non-contact EEC jet functions at first order in opacity for the $q\rightarrow q+g$ channel, and the way to identify the Coulomb logarithm. The derivations for the other two channels $g\rightarrow g+g$ and $g\rightarrow q+\bar{q}$ follow a similar approach. 

First, it is important to resolve the hierarchy between the different angular scales in order to identify the possible large logarithms contained in the non-contact term. There are two angles that are parametrically set by the medium itself and the jet energy (in terms of large light-cone components)
\begin{align}
\theta_{\rm NP} = \frac{m_{\rm eff}}{P^+},  \qquad
\theta_{\rm LPM} =& \sqrt{\frac{8\pi}{P^+L^+}} \, , 
\end{align}
where $\theta_{\rm NP}$ transforms as the minus component under a boost along the jet direction. For all high-energy scenarios, $\theta_{\rm LPM} \gg _{\rm NP}$, because $\theta_{\rm LPM}/\theta_{\rm NP} = \sqrt{\frac{8 \pi P^+}{m_{\rm eff}^2 L^+}} \gg 1$ for a jet.
The EEC measurement introduces anther scale $\theta P^+$ or simply the EEC angle $\theta$.
In this appendix, using the method of regions, we compute the non-contact EEC at opacity one for the case $\theta_{\rm NP} \ll \theta \ll \theta_{\rm LPM}$, and demonstrate the appearance of the Coulomb logarithm.

A direct numerical evaluation can demonstrate the angular regions of the two different power counting scenarios. Below we show details of the calculation for these two different kinematic regions, but first we list below some useful integrals and relations. In the evaluation of the integrals over the transverse momentum variables, we define the dimensionless variable {$\tilde{\bfq}$, such that the integral can be expressed in a compact dimensionally regularized form        
\begin{align}
A_{\epsilon; n,m} =& \int \frac{d^{2-2\epsilon}\tilde{\bfq}}{\pi^{1-\epsilon}} \frac{(\tilde{\bfq}^2)^n}{(\tilde{\bfq}^2+1)^{m}} = \frac{1}{\Gamma(1-\epsilon)} \frac{\Gamma(n+1-\epsilon)\Gamma(m-n-1+\epsilon)}{\Gamma(m)}\,.
\end{align}
The above integral satisfies the following useful property under shift in the exponent
\begin{align}
A_{\epsilon; n,m} =& \frac{n-\epsilon}{m-1}A_{\epsilon; n-1,m-1}\,.
\end{align}
We also encounter the following definite integral in the calculation 
\begin{align}
& \int_{-a}^a \frac{dt}{(it)^{2+\epsilon}} = \frac{2\sin\frac{\pi(1+\epsilon)}{2}}{1+\epsilon} a^{-1-\epsilon}\,.
\end{align}
Moreover for evaluating some of the double integrals it will prove useful to perform the following change of variables
\begin{align}
\int_0^{1} du \int_0^{1-u} dv =\int_0^{1/2} ds \int_{-2s}^{2s} dt,~~\textrm{for $s=\frac{u+v}{2}, t=u-v$} \,.
\end{align}

As shown in the main text, there are two diagrams that give rise to non-zero contributions in this case, the diagram with one real collinear splitting and a real Glauber interaction and the diagrams with a real collinear splitting and a virtual Glauber interaction as shown in Figure~\ref{fig:Jet-diagrams}. This   corresponds to integrating over phase space the corresponding TMD splitting functions as explicitly written down in Eq.~(\ref{eq:non-contact EEC}),
where the expressions for the corresponding contributions for the quark-to-quark channel are given in Eqs.~(\ref{eq:TMD-RR}), (\ref{eq:TMD-RV}). Using the relations Eqs.~(\ref{eq:RR-angle}), (\ref{eq:RV-angle}) we shift the transverse momenta $\bfk$ such that for the real-real and real-virtual diagrams respectively we have
\begin{align}
\textrm{RR: } \,\,\bfk \rightarrow & \bfw + (1-x)\bfq \,, \\
\textrm{RV: } \,\,  \bfk \rightarrow & \bfw\,,
\end{align}
where the new angular scale $\bfw$ is defined as 
\begin{align}
\bfw &=x(1-x)\frac{P^+}{2}\bft\, \\
Q_{\rm LPM} &= \sqrt{2x(1-x)P^+/L^+}\,.
\end{align}
This can also be seen as integrating out the real gluon momenta $\bfk$ using the angular delta functions in Eq.~(\ref{eq:non-contact EEC}).
Then the equation Eq.~(\ref{eq:non-contact EEC}) can be written in terms of sums of double integrals over the energy fraction $x$ and the Glauber gluon momenta $\bfq$ such that  
\begin{small}
\begin{align}
\frac{d\Sigma^{q}_{\rm non-contact}}{d^{2-2\epsilon}\bft} = & \frac{\sum_T g^2 \rho_T^-(0) L^+ C_T}{d_A} \frac{1}{\pi^{1-\epsilon}} \frac{\alpha_s(\mu^2) C_F}{2\pi} \mu^{2\epsilon}e^{\epsilon\gamma_E} \int_0^1 dx 
\bfw^{-2\epsilon} P
_{qq,\epsilon}(x) x(1-x)\nonumber\\
&\times \bigg\{ (2C_F-C_A) \mathfrak{I}_{1,\epsilon}(x, \bfw, Q_{\rm LPM}, m_{\rm eff}) + C_A \mathfrak{I}_{2,\epsilon}(x, \bfw, Q_{\rm LPM}, m_{\rm eff}) \nonumber\\
& \hspace{0.8cm} + C_A \mathfrak{I}_{3,\epsilon}(x, \bfw, Q_{\rm LPM}, m_{\rm eff})+ C_A  \mathfrak{I}_{4,\epsilon}(x, \bfw, Q_{\rm LPM}, m_{\rm eff})\nonumber\\
& \hspace{0.8cm} +  C_A \mathfrak{I}_{5,\epsilon}(x, \bfw, Q_{\rm LPM}, m_{\rm eff}) -C_A \mathfrak{I}_{6,\epsilon}(x, \bfw, Q_{\rm LPM}, m_{\rm eff}) \nonumber\\
& \hspace{0.8cm}  -C_A \mathfrak{I}_{7,\epsilon}(x, \bfw, Q_{\rm LPM}, m_{\rm eff}) \bigg\}\,,
\end{align}
\end{small}
where $\rho_{0}^- = \frac{\sum_T g^2 \rho_T^-(0)  C_T}{d_A}$ and the functions $\mathfrak{I}_{i,\epsilon}(x, \bfw, Q_{\rm LPM}, m_{\rm eff})$ represent the different integrals over $\bfq$ in dimensional regularization. The functions $\mathfrak{I}_{n=1\cdots 5,\epsilon}(x, \bfw, Q_{\rm LPM}, m_{\rm eff})$ are contributions from the RR diagram and the remaining $\mathfrak{I}_{n=6,7,\epsilon}(x, \bfw, Q_{\rm LPM}, m_{\rm eff})$ are contributions from the RV diagram. For completeness we show below explicit expressions for all $\mathfrak{I}_{n=1\cdots 7,\epsilon}(x, \bfw, Q_{\rm LPM}, m_{\rm eff})$
\begin{small}
\begin{align}
    \mathfrak{I}_{1,\epsilon}(x, \bfw, Q_{\rm LPM}, m_{\rm eff}) =&  \alpha_s(\mu^2)\mu^{2\epsilon}e^{\epsilon\gamma_E}\int \frac{d^{2-2\epsilon}\bfq }{\pi^{1-\epsilon}} \frac{1}{(\bfq^2+m_{\rm eff}^2)^2} \frac{-(1-x)\bfq\cdot\bfw (\bfw+(1-x)\bfq)^2}{[(\bfw+(1-x)\bfq)^2]^2+Q_{\rm LPM}^4}\,,
    \\
    \vspace{0.3cm}\nonumber\\
    \mathfrak{I}_{2,\epsilon}(x, \bfw, Q_{\rm LPM}, m_{\rm eff})  =& \mathfrak{I}_{1,\epsilon}(1-x, \bfw, Q_{\rm LPM}, m_{\rm eff}) \,, \\
    \vspace{0.3cm}\nonumber\\
    \label{eq:J3}
    \mathfrak{I}_{3,\epsilon}(x, \bfw, Q_{\rm LPM}, m_{\rm eff}) =&  \alpha_s(\mu^2)\mu^{2\epsilon}e^{\epsilon\gamma_E}\int \frac{d^{2-2\epsilon}\bfq }{\pi^{1-\epsilon}} \frac{\bfw^2}{(\bfq^2+m_{\rm eff}^2)^2} \frac{\bfq \cdot (\bfw+(1-x)\bfq)}{(\bfw+(1-x)\bfq)^2}\nonumber\\
    &\hspace{3.4cm}\times\frac{(\bfw-x\bfq)^2}{[(\bfw-x\bfq)^2]^2+Q_{\rm LPM}^4} \,,\\
    \vspace{0.3cm}\nonumber\\
    \mathfrak{I}_{4,\epsilon}(x, \bfw, Q_{\rm LPM}, m_{\rm eff}) =&  \mathfrak{I}_{3,\epsilon}(1-x, \bfw, Q_{\rm LPM}, m_{\rm eff}) \,,
    \label{eq:J5}
\end{align}
\begin{align}
    \mathfrak{I}_{5,\epsilon}(x, \bfw, Q_{\rm LPM}, m_{\rm eff}) =&  \alpha_s(\mu^2)\mu^{2\epsilon}e^{\epsilon\gamma_E}\int \frac{d^{2-2\epsilon}\bfq }{\pi^{1-\epsilon}} \frac{\bfw^2}{(\bfq^2+m_{\rm eff}^2)^2} \frac{(\bfw+(1-x)\bfq)\cdot(\bfw-x\bfq)}{(\bfw+(1-x)\bfq)^2(\bfw-x\bfq)^2} \nonumber\\
    & \hspace{3.cm}\times\frac{[(\bfw+(1-x)\bfq)^2-(\bfw-x\bfq)^2]^2}{[(\bfw+(1-x)\bfq)^2-(\bfw-x\bfq)^2]^2+Q_{\rm LPM}^4} \,, \\
    \vspace{0.3cm}\nonumber\\
    \mathfrak{I}_{6,\epsilon}(x, \bfw, Q_{\rm LPM}, m_{\rm eff}) =&  \alpha_s(\mu^2)\mu^{2\epsilon}e^{\epsilon\gamma_E}\int \frac{d^{2-2\epsilon}\bfq }{\pi^{1-\epsilon}} \frac{1}{(\bfq^2+m_{\rm eff}^2)^2} \frac{\bfq\cdot(\bfq-\bfw)}{(\bfw-\bfq)^2} \frac{\bfw^4}{\bfw^4+Q_{\rm LPM}^4}, \\
    \vspace{0.3cm}\nonumber\\
\mathfrak{I}_{7,\epsilon}(x, \bfw, Q_{\rm LPM}, m_{\rm eff}) =&  \alpha_s(\mu^2)\mu^{2\epsilon}e^{\epsilon\gamma_E}\int \frac{d^{2-2\epsilon}\bfq }{\pi^{1-\epsilon}} \frac{1}{(\bfq^2+m_{\rm eff}^2)^2} \vspace{0.3cm}\nonumber \\
&\hspace{3.2cm}\times\frac{\bfw\cdot(\bfw-\bfq)}{(\bfw-\bfq)^2} \frac{(\bfw^2-(\bfw-\bfq)^2)^2}{(\bfw^2-(\bfw-\bfq)^2)^2+Q_{\rm LPM}^4}\,.
\end{align}
\end{small}

Some of the integrals are related to each other by a simple change of variables from $x$ to $1-x$ as shown above. This means we only have to perform five integrations in total for the non-contact EEC. In the following subsections we present a detailed analysis for each of the $ \mathfrak{I}_{n=1\cdots 7,\epsilon}(x, \bfw, Q_{\rm LPM}, m_{\rm eff})$ integrals in the different kinematic regions. For simplicity, we will suppress the arguments of $ \mathfrak{I}_{n=1\cdots 7,\epsilon}(x, \bfw, Q_{\rm LPM}, m_{\rm eff})$ and will refer to these functions as simply $ \mathfrak{I}_{n=1\cdots 7,\epsilon}$ from now on. 
\subsection{$\mathfrak{I}_{1,\epsilon}$ and $\mathfrak{I}_{2,\epsilon}=\mathfrak{I}_{1,\epsilon}(x\rightarrow 1-x)$ contributions}

We consider the scenario where the EEC angle is much smaller than medium effect, where $\theta_{\rm NP}\ll \theta\ll \theta_{\rm LPM}$. Within this kinematic region, we further identify three different parametric regions. 
\begin{itemize}
\item The region $\bfw^2+(1-x)^2\bfq^2\sim Q_{\rm LPM}^2$, which is equivalent to $\bfq^2\sim M^2_{\rm LPM}$:
\begin{align}
\mathfrak{I}_{1,\epsilon} \approx&  \alpha_s(\mu^2)\mu^{2\epsilon}e^{\epsilon\gamma_E}\int \frac{d^{2-2\epsilon}\bfq }{\pi^{1-\epsilon}} \frac{1}{\bfq^4} \frac{-(1-x)^2 2(\bfq\cdot\bfw)^2}{(1-x)^2\bfq^4+Q_{\rm LPM}^4} \nonumber \\
=& \frac{\bfw^2}{(Q_{\rm LPM}^2)^{2}} \alpha_s(\mu^2)\left[\frac{\mu^2}{Q_{\rm LPM}^2}\right]^{\epsilon}e^{\epsilon\gamma_E}  (1-x)^{2+2\epsilon} 2A_{\epsilon; 0,3} \frac{\sin\frac{\pi(1+\epsilon)}{2}}{1+\epsilon} \frac{1}{\epsilon(1-\epsilon)} \nonumber \\
\approx& \frac{\bfw^2}{(Q_{\rm LPM}^2)^2}\alpha_s(\mu^2)(1-x)^{2} \left[\frac{1}{\epsilon}+\ln\frac{\mu^2}{Q_{\rm LPM}^2}+2\ln(1-x)+1\right] + \mathcal{O}(\epsilon) \, .
\end{align}
\item The region $\bfq^2\sim m^2_{\rm eff}$:
\begin{align}
\label{eq:q-m small region}
\mathfrak{I}_{1,\epsilon} \approx &  \alpha_s(\mu^2)\mu^{2\epsilon}e^{\epsilon\gamma_E}\int \frac{d^{2-2\epsilon}\bfq }{\pi^{1-\epsilon}} \frac{1}{(\bfq^2+m_{\rm eff}^2)^2} \frac{-(1-x)^2 2(\bfq\cdot\bfw)^2}{Q_{\rm LPM}^4} \nonumber \\
=& -\frac{\bfw^2}{(Q_{\rm LPM}^2)^2}\alpha_s(\mu^2)\left[\frac{\mu^2}{m_{\rm eff}^2}\right]^{\epsilon}e^{\epsilon\gamma_E} (1-x)^{2} \frac{A_{\epsilon; 1,2}}{1-\epsilon} \nonumber \\
\approx&\frac{\bfw^2}{(Q_{\rm LPM}^2)^2}\alpha_s(\mu^2)(1-x)^{2} \left[-\frac{1}{\epsilon}-\ln\frac{\mu^2}{m_{\rm eff}^2}\right]+ \mathcal{O}(\epsilon) \, .
\end{align}
\item The region $\bfq^2\sim \bfw^2 \sim \vec{\theta}^2 P^{+2}$ (scaleless integral) :
\begin{align}
\mathfrak{I}_{1,\epsilon} \approx&  \alpha_s(\mu^2)\mu^{2\epsilon}e^{\epsilon\gamma_E}\int \frac{d^{2-2\epsilon}\bfq }{\pi^{1-\epsilon}} \frac{1}{\bfq^4} \frac{-(1-x)^2 2(\bfq\cdot\bfw)^2}{Q_{\rm LPM}^4} = 0 \, .
\end{align}
\end{itemize}
We find that the pole from the region $\bfq^2\sim Q_{\rm LPM}^2$ cancels the pole from the region $\bfq^2\sim m^2_{\rm eff}$ and leading to a logarithm of the type $\ln\frac{Q_{\rm LPM}}{m_{\rm eff}}$. Therefore, the non-contact term can be sensitive to the infrared scale of QGP. This is different from the vacuum where the non-contact term is not sensitive to $\Lambda_{\rm QCD}$.


\subsection{$\mathfrak{I}_{3,\epsilon}$ and $\mathfrak{I}_{4,\epsilon}=\mathfrak{I}_{3,\epsilon}(x\rightarrow 1-x)$ contributions}
In this subsection we show the explicit integration of the function $\mathfrak{I}_{3,\epsilon}$, defined in Eqs.~(\ref{eq:J3}) for the different kinematic scenarios. We start with the small EEC angle region, where $\theta_{\rm NP}\ll\theta\ll\theta_{\rm LPM}$. We identify three different cases listed below. 
\begin{itemize}
\item The region $\bfq^2\sim \bfw^2$:
\begin{align}
\mathfrak{I}_{3,\epsilon} \approx& \alpha_s(\mu^2)\mu^{2\epsilon}e^{\epsilon\gamma_E}\frac{x^{2+2\epsilon}}{1-x}\bfw^2 \int \frac{d^{2-2\epsilon}\bfq }{\pi^{1-\epsilon}} \frac{1}{\bfq^4}\frac{1}{\bfq^4+Q_{\rm LPM}^4} \nonumber\\
=&\alpha_s(\mu^2)\left[\frac{\mu^2}{Q_{\rm LPM}^2}\right]^{\epsilon}e^{\epsilon\gamma_E}\frac{x^{2+2\epsilon}}{1-x}\frac{\bfw^2}{Q_{\rm LPM}^4} (-2)A_{\epsilon;0,3}\frac{\sin\frac{\pi(1+\epsilon)}{2}}{1+\epsilon}\frac{1}{\epsilon}\nonumber\\
\approx& \alpha_s(\mu^2)\frac{\bfw^2}{Q_{\rm LPM}^4} \left\{-\frac{x^2}{1-x}\left[\frac{1}{\epsilon}+\ln\frac{\mu^2}{Q_{\rm LPM}^2}\right]-\frac{2x^2\ln(x)}{1-x}+\mathcal{O}(\epsilon)
\right\} \, .
\end{align}
Here we have first performed the power expansion in $\bfw^2/Q_{\rm LPM}^2$ and then in $\bfq^2$. The order of the expansion matters in this case. 
\item The region $\bfq^2\sim m^2_{\rm eff}$:
\begin{align}
\mathfrak{I}_{3,\epsilon} \approx & \alpha_s(\mu^2)\mu^{2\epsilon}e^{\epsilon\gamma_E}\frac{x^2}{1-x}\frac{\bfw^2}{Q_{\rm LPM}^4} \left(1-x-\frac{1}{1-\epsilon}\right) \int \frac{d^{2-2\epsilon}\bfq }{\pi^{1-\epsilon}} \frac{\bfq^2}{(\bfq^2+m_{\rm eff}^2)^2} \nonumber\\
=& \alpha_s(\mu^2)\left[\frac{\mu^2}{m_{\rm eff}^2}\right]^{\epsilon}e^{\epsilon\gamma_E}\frac{x^2}{1-x}\frac{\bfw^2}{Q_{\rm LPM}^4} A_{\epsilon;1,2} \nonumber \\
\approx&  \alpha_s(\mu^2)\frac{\bfw^2}{Q_{\rm LPM}^4}\left\{-x\left[\frac{1}{\epsilon}+\ln\frac{\mu^2}{m_{\rm eff}^2}\right]+x-1\right\} \, .
\end{align}
Similarly to before, here we first perform power expansion in $\bfw^2/Q_{\rm LPM}^2$ and then in $\bfq^2$. 
\item The region $\bfq^2\sim \bfw^2$: 
\begin{small}
\begin{align}
\mathfrak{I}_{3,\epsilon} \approx& \alpha_s(\mu^2)\mu^{2\epsilon}e^{\epsilon\gamma_E}\frac{\bfw^2}{Q_{\rm LPM}^4}\int \frac{d^{2-2\epsilon}\bfq }{\pi^{1-\epsilon}} \frac{1}{\bfq^4} \frac{\bfq \cdot (\bfw+(1-x)\bfq)}{(\bfw+(1-x)\bfq)^2}(\bfw-x\bfq)^2\nonumber\\
=&\alpha_s(\mu^2)\left[\frac{\mu^2}{\bfw^2}\right]^{\epsilon}e^{\epsilon\gamma_E}\frac{\bfw^2}{Q_{\rm LPM}^4}(1-x)^{1+2\epsilon}\int_0^1 du(1-u)\left\{
A_{\epsilon;2,3}\left[u(1-u)\right]^{-\epsilon}\frac{x^2}{(1-x)^2}\right. \nonumber\\
& \left.-A_{\epsilon;0,3}\left[u(1-u)\right]^{-1-\epsilon}(1+u\frac{x}{1-x})^2
\right.\nonumber\\
&\left.
+A_{\epsilon;1,3}\left[u(1-u)\right]^{-1-\epsilon}\left[(1+u\frac{x}{1-x})^2-\frac{1-2u}{1-\epsilon}\frac{x}{1-x}\left(1+u\frac{x}{1-x}\right)-u(1-u)\frac{x^2}{(1-x)^2}\right]
\right\}\nonumber\\
=& \alpha_s(\mu^2)\left[\frac{\mu^2}{\bfw^2}\right]^{\epsilon}e^{\epsilon\gamma_E}\frac{\bfw^2}{Q_{\rm LPM}^4} \frac{4^\epsilon\pi^{3/2}(1-x)^{-1+2\epsilon}(x+2(1-x)\epsilon) \csc(\pi\epsilon)}{\Gamma(\frac{1}{2}-\epsilon)}\nonumber\\
\approx& \alpha_s(\mu^2)\frac{\bfw^2}{Q_{\rm LPM}^4}\left\{\frac{x}{1-x}\left[\frac{1}{\epsilon}+\ln\frac{\mu^2}{\bfw^2}\right] + \frac{-2x+2+2x\ln(1-x)}{1-x}+\mathcal{O}(\epsilon)
\right\} \, .
\end{align}
\end{small}
\end{itemize}
We note that again the poles here cancel among the three different regions.


\subsection{$\mathfrak{I}_{5,\epsilon}$ contribution}
Here we present the results for different integration regions of $\mathfrak{I}_{5,\epsilon}$ as defined in Eqs.~(\ref{eq:J5}). For $\theta_{\rm NP}\ll \theta\ll \theta_{\rm LPM}$, the three relevant regions are:
\begin{itemize}
\item Region $\bfq^2\sim m^2_{\rm eff}$:
\begin{align}
\mathfrak{I}_{5,\epsilon} \approx&
\alpha_s(\mu^2)\mu^{2\epsilon}e^{\epsilon\gamma_E}\frac{\bfw^2}{Q_{\rm LPM}^4}\frac{2}{1-\epsilon} \int \frac{d^{2-2\epsilon}\bfq }{\pi^{1-\epsilon}}\frac{\bfq^2}{(\bfq^2+m_{\rm eff}^2)^2} \nonumber\\
=& \alpha_s(\mu^2)\left[\frac{\mu^2}{m^2}\right]^\epsilon e^{\epsilon\gamma_E}\frac{\bfw^2}{Q_{\rm LPM}^4}\frac{2}{1-\epsilon} A_{\epsilon;1,2}\nonumber\\
\approx & \alpha_s(\mu^2)\frac{\bfw^2}{Q_{\rm LPM}^4}2\left[\frac{1}{\epsilon}+\ln\frac{\mu^2}{m_{\rm eff}^2}\right] \, .
\end{align}
\item Region $\bfq^2\sim M^2_{\rm LPM}$:
\begin{align}
\mathfrak{I}_{5,\epsilon} \approx&
\alpha_s(\mu^2)\mu^{2\epsilon} e^{\epsilon\gamma_E}\frac{-|1-2x|^{2+\epsilon}}{x(1-x)} \bfw^2 \int \frac{d^{2-2\epsilon}\bfq }{\pi^{1-\epsilon}} \frac{1}{\bfq^2}\frac{1}{\bfq^4 + Q_{\rm LPM}^4}\nonumber\\
=&
\alpha_s(\mu^2)\left[\frac{\mu^2}{M^2}\right]^\epsilon e^{\epsilon\gamma_E}\frac{|1-2x|^{2+\epsilon}}{x(1-x)} \frac{\bfw^2}{Q_{\rm LPM}^4} 2A_{\epsilon;0,3}\frac{\sin\frac{\pi(1+\epsilon)}{2}}{1+\epsilon}\frac{1}{\epsilon}\nonumber\\
\approx& \alpha_s(\mu^2)\frac{|1-2x|^2}{x(1-x)} \frac{\bfw^2}{Q_{\rm LPM}^4} \left\{\frac{1}{\epsilon}+\ln\frac{\mu^2}{Q_{\rm LPM}^2}+\ln|1-2x|\right\} \, .
\end{align}
\item Region $\bfq^2\sim\bfw^2$:
\begin{align}
\mathfrak{I}_{5,\epsilon}\approx & \alpha_s(\mu^2)\frac{\bfw^2}{Q_{\rm LPM}^4} \left[\frac{(1-2x)^2}{x(1-x)}+2\right]\left[-\frac{1}{\epsilon}-\ln\frac{\mu^2}{\bfw^2}+\mathcal{O}(1)\right] \, .
\end{align}
\end{itemize}

\subsection{$\mathfrak{I}_{6,\epsilon}$ contribution}
$\mathfrak{I}_{6}$ does not contain any large logs and the remaining integral over $x$ will not introduce new poles for $\mathfrak{I}_{6}$; therefore, we can evaluate it in the $\epsilon=0$ limit
\begin{align}
\lim_{\epsilon\rightarrow 0}\mathfrak{I}_{6,\epsilon}(x, \bfw, Q_{\rm LPM}, m_{\rm eff}) =&  \alpha_s(\mu^2)\int \frac{d\bfq^2}{2\pi} d\phi \frac{1}{(\bfq^2+m_{\rm eff}^2)^2} \frac{\bfq\cdot(\bfq-\bfw)}{(\bfw-\bfq)^2} \frac{\bfw^4}{\bfw^4+Q_{\rm LPM}^4} \nonumber\\
=&  \alpha_s(\mu^2)\int \frac{d\bfq^2}{2\pi}  2\pi\Theta(\bfq^2>\bfw^2)\frac{\bfw^4}{\bfw^4+Q_{\rm LPM}^4}\nonumber\\
=&  \alpha_s(\mu^2)\frac{\bfw^4}{\bfw^4+Q_{\rm LPM}^4} \frac{1}{\bfw^2+m^2} \, .
\end{align}
which does not contain large log in any of the two limits.

\subsection{$\mathfrak{I}_{7,\epsilon}$ contribution}
\begin{align}
\mathfrak{I}_{7,\epsilon} =&  \alpha_s(\mu^2)\mu^{2\epsilon}e^{\epsilon\gamma_E}\int \frac{d^{2-2\epsilon}\bfq }{\pi^{1-\epsilon}} \frac{1}{(\bfq^2+m_{\rm eff}^2)^2} \frac{\bfw\cdot(\bfw-\bfq)}{(\bfw-\bfq)^2} \frac{(\bfw^2-(\bfw-\bfq)^2)^2}{(\bfw^2-(\bfw-\bfq)^2)^2+Q_{\rm LPM}^4}
\end{align}
We look into the ordering $\theta_{\rm NP}\ll \theta\ll \theta_{\rm LPM}$ to identify
\begin{itemize}
\item Region $\bfq^2\sim m^2_{\rm eff}$:
\begin{small}
\begin{align}
\mathfrak{I}_{7,\epsilon} \approx&  \alpha_s(\mu^2)\mu^{2\epsilon}e^{\epsilon\gamma_E}\frac{\bfw^2}{M^4}\frac{2}{1-\epsilon}\int \frac{d^{2-2\epsilon}\bfq }{\pi^{1-\epsilon}} \frac{\bfq^2}{(\bfq^2+m_{\rm eff}^2)^2} \nonumber\\
=&\alpha_s(\mu^2)\left[\frac{\mu^2}{m^2}\right]^\epsilon e^{\epsilon\gamma_E}\frac{\bfw^2}{Q_{\rm LPM}^4}\frac{2}{1-\epsilon}A_{\epsilon;1,2}\nonumber\\
\approx &\alpha_s(\mu^2)\frac{\bfw^2}{Q_{\rm LPM}^4}2\left\{\frac{1}{\epsilon}+\ln\frac{\mu^2}{m_{\rm eff}^2}+\mathcal{O}(\epsilon)\right\} \, .
\end{align}
\item Region $\bfq^2\sim Q_{\rm LPM}^2$:
\begin{align}
\mathfrak{I}_{7,\epsilon} \approx&  \alpha_s(\mu^2)\mu^{2\epsilon}e^{\epsilon\gamma_E} \bfw^2 \int \frac{d^{2-2\epsilon}\bfq }{\pi^{1-\epsilon}} \left[ \frac{\left(1+\frac{1}{1-\epsilon}\right)Q_{\rm LPM}^4} {\bfq^2(\bfq^4+Q_{\rm LPM}^4)}
-\frac{2}{1-\epsilon}\frac{\bfq^2}{(\bfq^4+Q_{\rm LPM}^4)^2}\right]\nonumber\\
=&\alpha_s(\mu^2)\left[\frac{\mu^2}{Q_{\rm LPM}^2}\right]^{2\epsilon}e^{\epsilon\gamma_E} \frac{\bfw^2}{Q_{\rm LPM}^4} \left[-\left(1+\frac{1}{1-\epsilon}\right)
2A_{\epsilon;0,3}\frac{\sin\frac{\pi(1+\epsilon)}{2}}{1+\epsilon}\frac{1}{\epsilon}-\frac{1}{1-\epsilon}\frac{1}{\Gamma(1-\epsilon)}\frac{\pi\epsilon}{2}\csc\frac{\pi\epsilon}{2}\right]\nonumber\\
\approx& \alpha_s(\mu^2)\frac{\bfw^2}{Q_{\rm LPM}^4}2\left\{-\frac{1}{\epsilon}-\ln\frac{\mu^2}{Q_{\rm LPM}^2}-1\right\} \, .
\end{align}
\end{small}
Note that the divergence already cancels, so the last region should be finite.
\item Region $\bfq^2\sim \bfw^2$:
\begin{small}
\begin{align}
\mathfrak{I}_{7,\epsilon}
=& \alpha_s(\mu^2)\left[\frac{\mu^2}{\bfw^2}\right]^\epsilon e^{\epsilon\gamma_E} \frac{\bfw^2}{M^4}
2 \int_0^1 du (1-u)^2 \left\{ \left[1+\frac{2}{1-\epsilon}\right] A_{\epsilon;2,3}  [u(1-u)]^{-\epsilon} \right. \nonumber\\
&\left.+\left[2(-2+u)u+\frac{1-4u+2u^2}{1-\epsilon}\right]A_{\epsilon;1,3}[u(1-u)]^{-1-\epsilon} + (2-u)^2u^2A_{\epsilon;0,3}[u(1-u)]^{-2-\epsilon}
\right\}\nonumber\\
=& -\alpha_s(\mu^2)\left[\frac{\mu^2}{\bfw^2}\right]^\epsilon e^{\epsilon\gamma_E} \frac{\bfw^2}{Q_{\rm LPM}^4}\frac{2^{1+2\epsilon}\pi^{3/2}\epsilon\csc(\pi\epsilon)}{\Gamma(\frac{1}{2}-\epsilon)}\nonumber\\
\approx& -\alpha_s(\mu^2)\frac{\bfw^2}{Q_{\rm LPM}^4}\left\{1+\mathcal{O}(\epsilon)\right\} \, .
\end{align}
\end{small}
\end{itemize}

\subsection{Analytic expression for the non-contact EEC in exponential medium}
Combining all pieces and using the method of regions in the $\theta_{\rm NP} \ll \theta \ll \theta_{LPM}$ limit, we find
\begin{align}
\frac{d\Sigma^{RR+RV}_{\rm non-contact}}{d\theta^2} = &\frac{\left[\alpha_s(\mu^2)\right]^2C_F\rho_{\rm eff}^-{L^+}^3}{32\pi}\left\{\frac{3C_F+16C_A}{15}\left[\frac{1}{\epsilon}+\ln\frac{\mu^2}{m_{\rm eff}^2}-\frac{1}{\epsilon}-\ln\frac{\mu^2}{P^+/L^+}\right] \right. \nonumber \\
&\hspace{3.8cm}+ \textrm{non-log enhanced pieces}
\Big\}\, .
\end{align}
This formula agrees very well with the numerical evaluation. Note that the pole is completely canceled leaving the logarithmic factor $\ln\frac{P^+/L^+}{m_{\rm eff}^2}$. Because the radiation phase space has been restricted by the angle and the energy weighting, such an enhancement can only come from the Coulomb-like tail of the Glauber-gluon exchange, regulated by $m_{\rm eff}$. \\

\section{On the soft radiation outside of the jet cone}
\label{app:jet-eloss}
Following Eq.~\ref{eq:jet-e-loss}, the opacity-one correction to soft radiation outside of the jet cone is given by
\begin{align}
\mathcal{H}_{ab\rightarrow qX}\otimes \mathcal{J}_q^{(1)>}  \approx & \int \frac{dx}{x} \mathcal{H}_{ab\rightarrow qX}\left(\frac{p_T}{x}\right) \int d^2\bfk\frac{dN^{N=1}_{q\rightarrow q+g}}{dx d^2\bfk} \Theta_{{\rm anti}-k_T}^{>R}   \; ,\\
= & \int \frac{dx}{x} \mathcal{H}_{ab\rightarrow qX}\left(\frac{p_T}{x}\right) I_{qq, R}(x)   \; .
\end{align}
First, we use the soft gluon approximation for the splitting function and complete the transverse integral to arrive at the kernel $I_{qq, R}(x)$. In computing $I_{qq, R}(x)$, we shift $\bfq\rightarrow \bfq+\bfk$, define $t=|\bfq|^2/|\bfk|^2$ and reorganize the integral into
\begin{align}
I_{qq, R}(x) = & \frac{[\alpha_s(\mu^2)]^2C_F}{2\pi} 2C_A P_{qq,\epsilon}(x) \sum_T\int_0^\infty dz^+ \rho_T^-(z^+) (g_s^{\rm med})^2 \frac{C_T}{d_A}    \nonumber\\
&   \times \frac{\mu^{2\epsilon}e^{\epsilon\gamma_E}}{\Gamma(1-\epsilon)} \int  \frac{d\Omega_{\hat{q}}}{2\pi^{1-\epsilon}}\Gamma(1-\epsilon) \int_0^\infty d\bfq^2\frac{1}{(\bfq^2)^{2+2\epsilon}} \Phi\left(\frac{\bfq^2 z^+}{2x(1-x)P^+}\right) \nonumber\\
&\times  \frac{\mu^{2\epsilon}e^{\epsilon\gamma_E}}{\Gamma(1-\epsilon)} \int_0^\infty dt t^{\epsilon} \Theta\left(t < \frac{\bfq^2}{\left(2 (1-x)p_T\tan\frac{R}{2}\right)^2}\right) \nonumber\\
& \times \int \frac{d\Omega_{\hat{k},\hat{q}}}{2\pi^{1-\epsilon}}\Gamma(1-\epsilon)\frac{1+\sqrt{t}\cos\theta_{\hat{k},\hat{q}}}{\left(1+t+2\sqrt{t}\cos\theta_{\hat{k},\hat{q}}\right)^2} \, .
\label{eq:angint}
\end{align}
Here, $\Omega_{\hat{q}}$ is the solid angle of the $2-2\epsilon$ dimensional vector $\bfq$, and $\Omega_{\hat{q},\hat{\bfk}}$ is the solid angle of $\bfk$ relative to $\bfq$.
In $d=2$ dimension, the result of the angular integral of $\Omega_{\hat{q},\hat{\bfk}}$ is proportional to $1/(1-t)^2$, which is not integrable at the point $t=1$, i.e., $|\bfq|=|\bfk|$. In $2-2\epsilon$ dimension, however, this is regulated, and we can extract the asymptotic form of the regulated integral near $|t|=1$ as follows
\begin{align}
& \int \frac{d\Omega_{\hat{k},\hat{q}}}{2\pi^{1-\epsilon}}\Gamma(1-\epsilon)\frac{1+\sqrt{t}\cos\theta_{\hat{k},\hat{q}}}{\left(1+t+2\sqrt{t}\cos\theta_{\hat{k},\hat{q}}\right)^2} \nonumber\\
& \qquad \qquad =\frac{1+2\epsilon}{(1-t)^{2+2\epsilon}}\left(\Theta(t<1)-\Theta(t>1)\right)(1+\mathcal{O}(|1-t|))\,.
\end{align}
Carefully separating the integral into different regions depending on the $\Theta$ functions, it is possible to cast the Eq.~(\ref{eq:angint}) into the final form 
\begin{align}
I_{qq, R}(x) = & \frac{1}{(1-x)^{2+2\epsilon}} \frac{[\alpha_s(\mu^2)]^2}{\pi} 2C_FC_A  \frac{\rho_{\rm eff} L_{\rm eff}}{2p_T/L_{\rm eff}} \left(\frac{\mu^2}{2p_T/L_{\rm eff}}\right)^{2\epsilon} \left(\frac{e^{\epsilon\gamma_E}}{\Gamma(1-\epsilon)}\right)^2 \nonumber\\
&   \times  \int_0^\infty ds\frac{1}{s^{2+2\epsilon}} \overline{\Phi}\left(s\right) \mathrm{B}\left(\min\{\Delta,\Delta^{-1}\}; 1+\epsilon, -1-2\epsilon\right) \nonumber\, ,  \\
\Delta =& \frac{s}{2 (1-x) p_T L_{\rm eff} \tan^2\frac{R}{2}}\,.
\label{app:analytic}
\end{align}
$\mathrm{B}(z;a,b)$ is the incomplete Beta function, $s=\frac{\bfq^2 L_{\rm eff}^+}{2x(1-x)P^+}$ is the rescaled momentum square, and $\overline{\Phi}$ is the path-length-averaged LPM phase factor
\begin{align}
\overline{\Phi}\left(s\right) = \int_0^\infty du u \frac{\sum_T (g_s^{\rm med})^2 \frac{C_T}{d_A} \rho_T(uL_{\rm eff}) }{\rho_{\rm eff}} \Phi\left(su\right)\,.
\end{align}
At this point, we cannot proceed without a given form of the medium density profile. But several important conclusions can be drawn from the result Eq.~(\ref{app:analytic}).

First, we observe that the integral contains the power counting parameter $\frac{\rho_{\rm eff} L_{\rm eff}}{2p_T/L_{\rm eff}}$. Therefore, we only need the order one contribution from the remaining $s$ integral, as any appearance of small parameter will make itself irrelevant to our calculation at the current level of accuracy. To extract the dominant contribution, note that at small $\epsilon$, the incomplete Beta function is highly-peaked when its first argument approaches one. Furthermore, the integration over its first argument is
\begin{align}
\int_0^1 dx \mathrm{B}(z, 1+\epsilon, -1-2\epsilon) = -\frac{1}{2\epsilon}(1+\mathcal{O}(\epsilon^2)).
\end{align}
So, we can approximate by replacing $\mathrm{B}(z, 1+\epsilon, -1-2\epsilon)$ by $-\frac{1}{2\epsilon}\delta(1-z)$. Thus
\begin{align}
I_{qq, R}(x) \approx & \frac{1}{(1-x)^{2+2\epsilon}} \frac{[\alpha_s(\mu^2)]^2}{\pi} 2C_FC_A  \frac{\rho_{\rm eff} L_{\rm eff}}{2p_T/L_{\rm eff}} \left(\frac{\mu^2}{2p_T/L_{\rm eff}}\right)^{2\epsilon} \left(\frac{e^{\epsilon\gamma_E}}{\Gamma(1-\epsilon)}\right)^2 \nonumber\\
&   \times  \frac{1}{2\epsilon}\int_0^\infty ds\frac{1}{s^{2+2\epsilon}} \overline{\Phi}\left(s\right) \delta\left(1-\frac{s}{2 (1-x) p_T L_{\rm eff} \tan^2\frac{R}{2}}\right)\, \nonumber\\
\approx &  \frac{[\alpha_s(\mu^2)]^2}{\pi} 2C_FC_A  \frac{\rho_{\rm eff} L_{\rm eff}}{2p_T/L_{\rm eff}} \left(\frac{\mu^2}{2p_T/L_{\rm eff}}\right)^{2\epsilon} \left(\frac{e^{\epsilon\gamma_E}}{\Gamma(1-\epsilon)}\right)^2 \nonumber\\
&   \times \left(-\frac{1}{2\epsilon}\right)\frac{2p_T L_{\rm eff} \tan^2\frac{R}{2}}{(1-x)^{1+2\epsilon}} \left.\frac{\overline{\Phi}\left(s\right)}{s^{2+2\epsilon}} \right|_{s=2 (1-x) p_T L_{\rm eff} \tan^2\frac{R}{2}}\,.
\end{align}
Now, we can make a general assumption on the medium profile function $\overline{\Phi}(s)$: it is proportional to $s^2$ at small $s$, while approaches a constant at large $s$, as can be seen from the oscillating behavior of the LPM phase factor $\Phi$. Therefore, only when $s=2 (1-x) p_T L_{\rm eff} \tan^2\frac{R}{2}$ is smaller than an order-one value $s_0$ will the integral contributes to the considered order of the power counting. Defining $\eta = 2 p_T L_{\rm eff} \tan^2\frac{R}{2}$, this condition translates into
\begin{align}
(1-x) \eta < s_0\, .
\end{align}
So we further approximate 
\begin{align}
\left.\frac{\overline{\Phi}\left(s\right)}{s^{2+2\epsilon}} \right|_{s= (1-x) \eta} = \frac{\Theta\left((1-x) \eta<s_0\right)}{\left((1-x)\eta\right)^{2\epsilon}}\frac{\overline{\Phi}''(0)}{2} + \mathcal{O}\left(\frac{1}{p_T L_{\rm eff} \tan^2\frac{R}{2}}\right)\,.
\end{align}
Another approximation is to perform a gradient expansion of the hard function under the convolution, consistent with the use of the soft-gluon approximation
\begin{align}
\frac{1}{x}\mathcal{H}_{ab\rightarrow qX}\left(\frac{p_T}{x}\right)-\mathcal{H}_{ab\rightarrow qX}(p_T) \approx (1-x)\left(p_T\frac{d\mathcal{H}_{ab\rightarrow qX}}{dp_T}+\mathcal{H}_{ab\rightarrow qX}(p_T)\right)  \, .
\end{align}
Finally, we can proceed with the evaluation of the convolution of the opacity-one correction of the semi-inclusive and the hard function
\begin{align}
\mathcal{H}_{ab\rightarrow qX}\otimes \mathcal{J}_q^{(1)}  \approx & \frac{[\alpha_s(\mu^2)]^2}{\pi} 2C_FC_A  \frac{\rho_{\rm eff} L_{\rm eff}}{2p_T/L_{\rm eff}} \left(-\frac{1}{2\epsilon}\right)\left(\mu^2 L_{\rm eff}^2 \tan^2\frac{R}{2}\right)^{2\epsilon} \left(\frac{e^{\epsilon\gamma_E}}{\Gamma(1-\epsilon)}\right)^2 \frac{\overline{\Phi}''(0)}{2}\nonumber\\
& \times \left(p_T\frac{d\mathcal{H}_{ab\rightarrow qX}}{dp_T}+\mathcal{H}_{ab\rightarrow qX}(p_T)\right)  \eta^{1-4\epsilon} \int_0^1 \frac{dx}{(1-x)^{4\epsilon}} \Theta\left(1-x<\frac{s_0}{\eta}\right)\nonumber\\
\approx & \frac{[\alpha_s(\mu^2)]^2}{\pi} 2C_FC_A  \frac{\rho_{\rm eff} L_{\rm eff}}{2p_T/L_{\rm eff}} \frac{s_0 \overline{\Phi}''(0)}{2}\nonumber\\
& \times \left(p_T\frac{d\mathcal{H}_{ab\rightarrow qX}}{dp_T}+\mathcal{H}_{ab\rightarrow qX}(p_T)\right) \left[-\frac{1}{2\epsilon} -\ln\frac{\mu^2 e^2 L_{\rm eff}^2 \tan^2\frac{R}{2}}{s_0^2} + \mathcal{O}(\epsilon)\right] \, .
\end{align}
This result is similar to the qualitatively argument that we used in the main text, but it gives a practical way to estimate the prefactor and optimized scales for an arbitrary medium profile.

A renormalization lead to the energy-loss like medium-induced evolution equation for the hard parton momentum spectra $D(p_T)\equiv p_T \times (\mathcal{H}_{ab\rightarrow qX}\otimes \mathcal{J})$,
\begin{align}
\frac{\partial}{\partial \ln\mu^2} D(p_T,\mu^2) = -\frac{[\alpha_s(\mu^2)]^2}{2\pi} 2C_F2C_A  \frac{\rho_{\rm eff} L_{\rm eff}^2}{2} \frac{s_0 \overline{\Phi}''(0)}{2} \frac{\partial }{\partial p_T}  D_q(p_T, \mu^2)
\end{align}
which relates the hard parton spectrum between different scales. Choosing the other boundary as the medium screening scale $\mu^2=m_{\rm eff}^2$, where no medium-induced radiation is included, we have an analytic solution under the soft-gluon approximation
\begin{align}
D\left(p_T, \frac{s_0^2/e^2}{L_{\rm eff}^2\tan^2\frac{R}{2}}\right) = D(p_T+\Delta p_T, m_{\rm eff}^2) \, ,
\end{align}
with the radiative energy loss with running coupling improvement  given by 
\begin{align}
\Delta p_T = \frac{2}{\beta_0} \left[\alpha_s\left(m_{\rm eff}^2\right)-\alpha_s\left(\frac{s_0^2/e^2}{L_{\rm eff}^2\tan^2\frac{R}{2}}\right)\right] 4C_FC_A  \frac{\rho_{\rm eff} L_{\rm eff}^2}{2} \frac{s_0 \overline{\Phi}''(0)}{2}\,.
\end{align}
The first factor comes from running coupling, the $C_FC_A$ is due to the color charge of the quark and the radiated gluon. $\frac{\rho_{\rm eff} L_{\rm eff}^2}{2}$ gives the scale of the typical energy loss, while $\frac{s_0 \overline{\Phi}''(0)}{2}$ encodes the information of the path-dependent medium profile.

\bibliographystyle{jhep} 

\providecommand{\href}[2]{#2}\begingroup\raggedright\endgroup

\end{document}